\documentclass[usenatbib]{mn2e}
\usepackage{fixltx2e}
\usepackage{mathptmx}
\usepackage{color}
\usepackage{graphics,graphicx,amssymb,amsmath,fixltx2e,natbib,epsfig}
\usepackage{amsmath}

\include{defn}
\def\gtrsim{\lower 2pt \hbox{$\, \buildrel {\scriptstyle >}\over
{\scriptstyle \sim}\,$}}
\def\lesssim{\lower 2pt \hbox{$\, \buildrel {\scriptstyle <}\over
{\scriptstyle \sim}\,$}}

\begin{document}

\title{A Systematic Chandra study of Sgr A$^{\star}$: I. X-ray flare 
detection}

\author[]{Qiang Yuan\thanks{E-mail:yuanq@umass.edu} and Q. Daniel Wang\thanks{E-mail:wqd@astro.umass.edu}\\
Department of Astronomy, University of Massachusetts, 710 North Pleasant St., Amherst, MA, 01003, U.S.A.}

\maketitle

\label{firstpage}

\begin{abstract}
Daily X-ray flaring represents an enigmatic phenomenon of Sgr A$^{\star}$ 
--- the supermassive black hole at the center of our Galaxy. We report 
initial results from a systematic X-ray study of this phenomenon, based 
on extensive {\it Chandra} observations obtained from 1999 to 2012, 
totaling about 4.5 Ms. We detect flares, using a combination of the 
maximum likelihood and Markov Chain Monte Carlo methods, which allow for 
a direct accounting for the pile-up effect in the modeling of the flare 
lightcurves and an optimal use of the data, as well as the measurements 
of flare parameters, including their uncertainties. A total of 82 flares 
are detected. About one third of them are relatively faint, which were not 
detected previously. The observation-to-observation variation of the 
quiescent emission has an average root-mean-square of $6\%-14\%$, 
including the Poisson statistical fluctuation of faint flares below our 
detection limits. We find no significant long-term variation in the
quiescent emission and the flare rate over the 14 years. In particular, 
we see no evidence of changing quiescent emission and flare rate around 
the pericenter passage of the S2 star around 2002. We show clear evidence 
of a short-term clustering for the ACIS-S/HETG 0th-order flares on time 
scale of $20-70$ ks. We further conduct detailed simulations to characterize 
the detection incompleteness and bias, which is critical to a comprehensive 
follow-up statistical analysis of flare properties. These studies together 
will help to establish Sgr A$^{\star}$ as a unique laboratory to understand 
the astrophysics of prevailing low-luminosity black holes in the Universe.
\end{abstract}


\section{Introduction}

Sagittarius A$^\star$ (Sgr A$^\star$), the supermassive black hole (SMBH)
at the center of our own Galaxy, is an ideal and unique example to 
study the accretion of matter onto a black hole (BH). The proximity of
Sgr A$^\star$ enables us to have spatially resolved studies which can
hardly be achieved for other SMBHs \citep[e.g.,][]{2003ApJ...591..891B,
2013Sci...341..981W}. The pure hot phase of the accretion flow
\citep{2014ARA&A..52..529Y} makes the relevant physics simple and largely 
scalable, allowing for an in-depth study with reasonable certainty.  
Sgr A$^\star$ also represents a limiting case of low luminosity SMBHs. 
The study of Sgr A$^\star$ could then be very useful in understanding this 
entire class of SMBHs.

The X-ray emission provides very useful diagnostics of the gas inflow/outflow 
around an SMBH, as well as its interplay with the circum-nuclear environment. 
The X-ray emission from Sgr A$^\star$ consists of two components, the 
extended quiescent emission and hour-scale flares that occur on average
a couple of times a day, although the decomposition between the two is 
still somewhat uncertain \citep[e.g.,][]{2001Natur.413...45B,
2003ApJ...591..891B,2001ARA&A..39..309M,2013Sci...341..981W}.

Since the launch of the {\it Chandra} X-ray Observatory in 1999, Sgr A$^\star$ 
has been monitored frequently, and dozens of flares have been detected 
\citep[e.g.,][]{2001Natur.413...45B,2004A&A...427....1E,2006A&A...450..535E,
2008ApJ...682..373M,2008ApJ...682..361Y}.
In particular, a {\it Chandra} X-ray Visionary Project (XVP) was recently 
carried out for Sgr A$^\star$ with a total exposure of 3 Ms
\citep{2012ApJ...759...95N}. Spectral analyses of the data from the project
suggest that the quiescent component can be well explained by the primarily 
thermal emission from a radiatively inefficient accretion flow together 
with an outflow of a similar mass rate~\citep{2003ApJ...598..301Y,
2004ApJ...606..894Y,2013Sci...341..981W}. Using the data, \citet[][hereafter 
N13]{2013ApJ...774...42N} detect 39 flares. Flares have also been detected 
in observations made with other X-ray observatories such as
{\it XMM-Newton} \citep{2003A&A...407L..17P,2005ApJ...635.1095B,
2008A&A...488..549P}, {\it Swift} \citep{2013ApJ...769..155D},
and {\it NuSTAR} \citep{2014ApJ...786...46B}.
The suggested radiation mechanisms of X-ray flares include synchrotron 
\citep{2003ApJ...598..301Y,2004ApJ...606..894Y,2009ApJ...698..676D}, 
inverse Compton scattering \citep{2001A&A...379L..13M,2003ApJ...598..301Y,
2004A&A...427....1E,2006A&A...450..535E,2006ApJ...636..798L}, and 
bremsstrahlung \citep{2002ApJ...566L..77L}. In addition, flares have been 
observed in  near-infrared 
(NIR) observations~\citep{2003Natur.425..934G,2009ApJ...698..676D,
2011A&A...528A.140T,2012ApJS..203...18W}. The strong polarization of 
the NIR flare emission indicates a synchrotron origin, and hence 
the generation of non-thermal electrons~\citep{2006A&A...455....1E}. 

Two typical dynamical scenarios have been proposed to account for the 
production of flares. One is the production of hot spots or episodic 
mass ejections in the accretion flow and/or jets, due to the magnetic 
reconnection or other magnetohydrodynamical process \citep[e.g.,][]
{2001A&A...379L..13M,2004ApJ...606..894Y,2004ApJ...611L.101L,
2009MNRAS.395.2183Y,2010ApJ...725..450D}. The other is the tidal 
disruptions of approaching planetesimals by the SMBH \citep[e.g.,][]
{2009A&A...496..307K,2012MNRAS.421.1315Z}. In particular, the comparison 
of the flare statistics with the theoretical expectation of the so-called 
self-organized criticality (SOC) system \citep{1986JGR....9110412K,
1987PhRvL..59..381B} suggests that the spatial dimension responsible for 
the production of the flares is $S=3$, which is similar to that for solar 
flares and further implies a magnetic reconnection origin of the X-ray 
flares \citep{2015ApJS..216....8W,2015ApJ...810...19L}. Nevertheless, the 
exact physical nature of the Sgr A$^\star$ flares, both the radiation 
mechanism and the power source, remains unclear \citep{2014ARA&A..52..529Y}.

In this work, we present a systematic study of the X-ray variabilities 
of Sgr A$^\star$, based on all relevant {\it Chandra} observations taken  
from 1999 to 2012. The observations after 2012 are not included in the 
present study due to the strong confusion from the recently appeared 
bright magnetar SGR J1745-2900, which is only $2.4''$ away from Sgr 
A$^\star$ \citep{2013ApJ...770L..24K}. The data we use were also included 
in the very recent study by \citet[][hereafter P15]{2015MNRAS.454.1525P}. 
However, their emphasis, primarily on the rate of relatively bright flares 
and its connection to the pericenter passage of the G2 object 
\citep{2013ApJ...774...44G}, is quite different from ours. We here 
focus on both a systematic detection of the X-ray flares and a good 
characterization of the detection incompleteness and bias. 
This approach is very crucial to a statistical study of both 
flare and quiescent emissions for two main reasons. First, the analysis 
enables us to have a significant enlargement of the detected sample of the 
flares. In particular, the 2012 XVP observations of Sgr A$^\star$ were taken 
with the Advanced CCD Imaging Spectrometer - Spectroscopy array (ACIS-S) and
with the high energy transmission gratings (HETG) inserted. While allowing
for both a high spectral resolution view of Sgr A$^\star$ and a 
reduced pileup effect (\S~3.2) on bright flares, the 
instrument combination substantially decreased the effective collecting 
area and hence limited the sensitivity of the flare detection. All the 
observations prior to 2012 used the Advanced CCD Imaging Spectrometer - 
Imaging array (ACIS-I) without gratings, allowing for detection of flares 
to lower flare fluences. Second, the 1999-2012 coverage of the ACIS-I 
and -S/HETG observation combination enables us to explore the long-term 
evolution of the X-ray emission. The recent study of P15 shows an 
enhancement of the bright flare rate, but with a drop of the moderate 
flare rate, several months after the pericenter passage of G2 
\citep{2013ApJ...774...44G}. We extend this study by probing how the 
emission may be affected by the pericenter passages of the S2 star and 
G1 cloud in 2001-2002 \citep{2002Natur.419..694S,2015ApJ...798..111P}. 

Compared with previous works, we also have several technical improvements 
in the present analysis. We employ the unbinned photon events to maximize 
the use of information in the data. The Markov Chain Monte Carlo (MCMC) 
method is adopted for the likelihood fitting in order to improve the
characterization of the model parameters, especially their errors.
For example, the uncertainties in flare durations are so far not given, 
which prevents a rigorous statistical analysis involving this parameter. 
We also statistically test the long-term variability and short-term
clustering of the flare rate, which were not addressed in previous works.

\section{Observations}

Sgr A$^\star$ was observed 46 times (with a total exposure of 1.5 Ms) 
between 1999 and 2011, with the {\it Chandra} ACIS-I camera with no
grating. During the 2012 XVP campaign, 38 observations were performed 
using the ACIS-S camera combined with the HETG, reaching a total exposure
of 3 Ms. While the ACIS-S/HETG combination yielded high energy resolution 
data of Sgr A$^\star$ for the first time, it reduced the 0th order 
effective area greatly, compared with the ACIS-I instrument, for example,
150 cm$^2$ versus 320 cm$^2$ at 5 
keV\footnote{http://cxc.harvard.edu/caldb/prop\_plan/pimms/}, and
especially at energies $\lesssim4$ keV. The basic information of the
observations is compiled in Tables \ref{table:flare} and 
\ref{table:flare2012}. 

The data are reduced with the standard analysis tool CIAO (version 4.5),
which includes the exclusion of the time intervals of significant
background flaring. We extract the events within $1''.25$ circle region 
centered on Sgr A$^\star$, consistent with those used in other similar 
studies (N13, P15), to minimize the effect from the extended quiescent 
emission and the background \citep[][supplementary materials]
{2013Sci...341..981W}. The non-cosmic X-ray event background in 
such a small region is negligible.
We further filter the events within the $2-8$ keV channel energy range, 
as done in N13 and P15. For the 2012 ACIS-S/HETG observations, we adopt 
only the non-dispersed (0th order) events, different from N13 in which 
both the 0th and $\pm$1st order events were combined. The use of the 0th 
order data only (ACIS-S/HETG0 hereafter) makes the pileup correction 
(see below \S3.2) more straight-forward and reliable than that of the 
0th+1st order data, because the ratio of the non-dispersed to dispersed 
events depends on their spectra, which could vary from one flare to another 
and/or within a flare. The flare detection sensitivities of the 0th and 
0th+1st events are expected to be comparable\footnote{The 0th to 1st order 
count rate ratio of a typical flare is roughly $3:2$ (N13). However, the 
quiescent count rate (including the background) of the 0th order data 
used in this work is about $1.9$ ks$^{-1}$, which is about three times 
smaller than that of the 0th+1st order data \citep[$5.2-5.7$ ks$^{-1}$;]
[]{2015ApJ...799..199N}. To reach the same detection significance of a
flare, its count rate of the 0th+1st order data needs to be about 
$\sqrt{5.5/1.9}=1.7$ times higher than that of the 0th order data only, 
which is about the same as the flare count rate ratio.}.

\section{Analysis}

\subsection{Flare detection methodology}

We adopt a maximum likelihood fitting algorithm to detect flares. This 
algorithm allow us to use primarily the unbinned data to ensure the minimum 
loss of information, which is especially important for reliable measurements
of those flares whose durations are comparable to, or smaller than 300~s, 
the bin width used in N13. The use of the MCMC method in the fitting 
\citep{Neal1993,Gamerman1997,2003itil.book.....M} enables us to effectively 
survey the high-dimensional, correlated space of the model parameters and 
determine their posterior probability distributions (hence the uncertainties 
in the flare parameters). The Metropolis-Hastings algorithm is adopted 
to generate the Markov chains. 

The model used to describe the lightcurve consists of a 
quiescent\footnote{This ``quiescent'' component represents the sum of 
the background (X-ray and non-X-ray), as well as the truly quiescent 
emission and undetectable weak flares of Sgr A$^\star$. Throughout the 
paper we refer this steady component to ``quiescent emission''.}
emission plus a series of flares. We assume a Gaussian profile as an
approximation to the lightcurve of a typical flare. With the limited 
counting statistics of the data, only a few very bright ones show 
significant deviations from the profile \citep[e.g.,][]
{2012ApJ...759...95N}. Detailed treatment of the asymmetric profiles,
both individually and statistically, will be given in a future publication. 
The model lightcurve can then be expressed as
\begin{equation}
f(t)=\kappa+\sum_{i=1}^{n}G(t;\,A_i,\sigma_i,t_{0i}),
\label{model}
\end{equation}
where $\kappa$ is the quiescent count rate, and $G(t;\,A_i,\sigma_i,
t_{0i})=\frac{A_i}{\sqrt{2\pi}\sigma_i}\exp[-(t-t_{0i})^2/2\sigma_i^2]$ 
is the Gaussian profile with the total counts ($A_i$), the peak time 
($t_{0i}$), and the dispersion ($\sigma_i$) of the $i$th flare, and $n$ 
is the total number of flares in the lightcurve. The pileup effect, 
which will be described in \S~3.2, is applied directly to the model 
lightcurve. Then the (logarithmic) likelihood function of the unbinned 
Cash statistic is \citep{1979ApJ...228..939C}
\begin{equation}
C\equiv-2\ln{\mathcal L}=2\left(E-\sum_{j=1}^N\ln f^{\rm pi}(t_j)\right),
\label{C-stat}
\end{equation}
where $f^{\rm pi}(t)$ is the pileup affected lightcurve, $N$ is the total 
observed number of photons, $E$ is the expected number of photons according 
to Eq. (\ref{model}), $t_j$ is the arrival time of the $j$th event, and 
the summation is over all observed events. The likelihood fitting is 
performed for individual observation separately. 

The actual search procedure for flares in the lightcurve of an observation 
is as follows. We first bin the data with 300 s bins\footnote{We have 
tested that the final results are not sensitive to the start bin width 
adopted for the candidate searches only.}, and then start to search for 
flares from the highest count rate bins. For each bin with the maximum 
count rate in the observation, we fit the unbinned data with an initial 
flare centroid at the bin center. The width $\sigma_i$ of a flare is 
restricted to be larger than 100 s during the search for candidates (N13). 
This helps reduce false detections (see below). We define the {\it Test 
Statistic} of a flare, TS$=C_0-C$, where $C_0$ ($C$) is the $C$-statistic 
without (with) this flare \citep{1996ApJ...461..396M}. If the TS value for 
one flare is larger than $14$, which corresponds to the one-sided $3\sigma$ 
significance for three free parameters\footnote{The distribution of the 
TS value in the null hypothesis follows $\chi^2_n/2$ with degree of 
freedom $n$ \citep{1996ApJ...461..396M}. For $n=3$ and TS$=14$, the 
false detection probability is $p=0.0015$, which corresponds to $3\sigma$ 
significance for one-sided normal distribution.}, we have a detection. 
Then we remove the time interval $[t_{0i}-3\sigma_i,\,t_{0i}+3\sigma_i]$, 
and repeat the above analysis until that no flare has TS value larger than 
the threshold in the observed lightcurve. Note that this search 
procedure tends to miss some weak broad flares (e.g., with low peak count 
rates). However, it also reduces the number of false detections. We employ 
Monte Carlo simulations to quantify the probability of false detections 
due to the background fluctuation. Given the false-alarm probability of 
$p=0.0015$, we find that the average number of false detections is about 
$0.6$ ($1.2$) for the ACIS-I (-S/HETG0) data. The estimated number will 
be moderately higher if the search width of flare candidates is narrower.

The above flare detection, carried out independently from one observation 
to another, is not optimal for a few ACIS-I observations with short 
exposures. For such an observation, the quiescent emission level cannot 
be tightly constrained, which also limits the sensitivity of the flare 
detection. However, the quiescent level on average remains very steady 
over the 12 year period, as is shown in our later analysis (\S4.2). 
Therefore, we re-analyze the data with the quiescent count rate fixed to 
the average value of 4.86 (1.86) cts ks$^{-1}$ for the ACIS-I 
(ACIS-S/HETG0) observations whose quiescent fluxes deviate from the average 
ones, which are marked with the ``fixed'' $\kappa$ in Tables~\ref{table:flare} 
and \ref{table:flare2012}. Such a re-analysis leads to the detection of three
additional flares for the ACIS-I observations. These flares are removed 
in the re-calculation of the $\kappa$ values of the affected observations.

\begin{figure}
\centering
\includegraphics[width=\columnwidth]{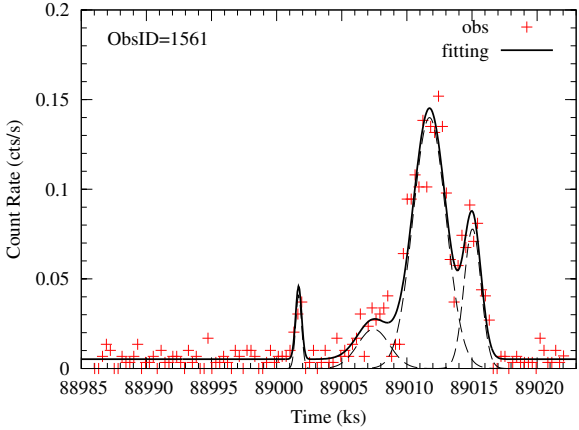}
\caption{Illustration of a lightcurve, constructed from the data collected
in the {\it Chandra} ACIS-I observation ID 1561 and with a bin size of 300 
s (red $+$; the binning is for the presentation only), compared with the 
best-fitting model (solid line). Two flares are detected, and the main one 
consists of three subflares, as shown by the dashed lines. 
}
\label{fig:lc}
\end{figure}

There may be significant substructures for some bright flares
\citep[][N13]{2012ApJ...759...95N}. We add additional Gaussian profiles 
to characterize such substructures or subflares. The 
criterion to detect a subflare is that its TS value is larger than $8$ 
($2\sigma$) and its $2\sigma$ width overlaps with the adjacent's. The 
detection is iterated until no more subflare is found. Finally we do a 
global fitting with all the flares and subflares to obtain their 
parameters as well as the quiescent count rate. The number of subflares
of each flare, $N_{\rm sub}$, is included in Tables~\ref{table:flare} and 
\ref{table:flare2012}. Fig. \ref{fig:lc} shows an example of a complex 
lightcurve with multi-flares. The lightcurves of all the detected flares 
together with the best-fitting results are given in the Appendix.

However, subflares could be due to chance overlapping of a flare with other 
independent ones. For each flare with detected subflares, we estimate the 
probability that all its subflares are chance overlapping of independent
flares as $P=\prod_{i=1}^{N_{\rm sub}-1}[1-\exp(-\Delta t_i/\overline
{\Delta t})]$, where $\Delta t_i$ is the time difference between two 
adjacent subflares and $\overline{\Delta t} = 44.6~(60.9)$ ks is the mean 
separation between two flares (assuming no correlation) averaged over all 
flares detected in the ACIS-I (ACIS-S/HETG0) data set. The estimated 
probability is listed in the last columns of Tables~\ref{table:flare} and 
\ref{table:flare2012}. We find that subflares in only about $1/3$ of such 
flares are expected to be due to the chance overlapping with a probability 
$\gtrsim 5\%$. Therefore, most of our detected subflares should represent 
intrinsic substructures of the flares and may be studied statistically. 
However, there is a complication that subflares are detected with different 
threshold compared with individual flares (lower significance but with 
higher background from overlapping flares). Therefore, the above probability 
estimate is somehow uncertain. More detailed analysis may be needed when 
subflares are studied.

The duration of a flare is defined to contain the fitting $95\%$ 
($2.5\%-97.5\%$) integrated counts, which includes subflares if present. 
The uncertainties in the fluences and durations are calculated from the
bootstrapping realized samples of the fitting model lightcurve, which 
accounts for the correlations among the parameters. 

\subsection{Pileup effect}

For bright flares, one also needs to consider the so-called pileup effect. 
If two or more photons arrived at the same detection pixel during a single 
frame integration time (3.2 s for the {\it Chandra} ACIS observations
considered here), they would be detected as a single event with a higher 
energy. This effect is significant only for the non-dispersed (0th order) 
counts when their incident rate at a pixel is sufficiently high (i.e.,
$\Lambda_{\rm in}\gtrsim\Lambda_{\rm in}^{\rm th}\sim0.02$ cts s$^{-1}$). 
We convert the fitting relation between $\Lambda_{\rm in}$ and the 
ACIS output count rate $\Lambda_{\rm out}$ given in P15 into the
following pileup functions:
\begin{eqnarray}
&&\Lambda_{\rm out}=\left(4.180\Lambda_{\rm in}^{-0.07387}+0.5381\Lambda_{\rm in}^{-1.160}\right)^{-1},\nonumber\\
&&~~~~~~~~~~~~~~~~~~~~~~~~~~~~~~\textrm{for~ACIS-I},\\
&&\Lambda_{\rm out}=\left(3.933\Lambda_{\rm in}^{-0.03541}+0.6564\Lambda_{\rm in}^{-1.107}\right)^{-1},\nonumber\\
&&~~~~~~~~~~~~~~~~~~~~~~~~~~~~~~\textrm{for~ACIS-S/HETG0}.
\end{eqnarray}

\begin{figure}
\centering
\includegraphics[width=\columnwidth]{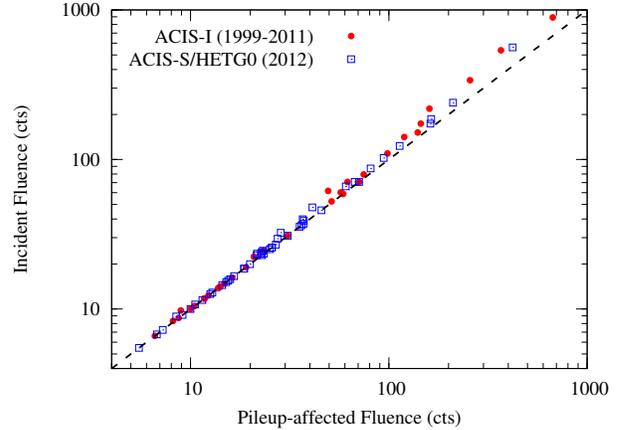}
\caption{Comparison of the pileup-affected fluences and the calculated
incident ones for our detected flares.
}
\label{fig:pileup}
\end{figure}

We apply the corresponding pileup function to the parts of the model 
lightcurve (Eq.(1)) with $\Lambda_{\rm in}>\Lambda_{\rm in}^{\rm th}$, 
before it is fitted to the observed one. The total incident and 
pileup-affected fluences for a flare are obtained from integrating its 
model lightcurve (after subtracting the quiescent emission) before and 
after the application of the pileup function. The largest pileup 
correction of the fluence is about $50\%$ for the ACIS-I flares, and 
about $35\%$ for the ACIS-S/HETG0 flares detected in this analysis 
(Fig. \ref{fig:pileup}; for the flare parameters, please refer to \S4.1). 
In comparison, the largest correction is $\sim20\%$ in N13 where the 
$\pm1$st order data (which suffered no pileup effect) are included, 
which is broadly consistent with ours taking into account the 0th to 
1st order count rate ratio. However, the pileup correction in N13 
is applied on an average ACIS-S/HETG 0th+1st order count rate, instead 
of on the (more directly relevant) ACIS-S/HETG 0th order lightcurve of 
a flare, as we have done in the present work. 


\subsection{Detection incompleteness, bias, and uncertainty}

With the limited counting statistics, the detection of weak flares can
be incomplete and even biased, making the measurements of parameters 
very uncertain \citep[e.g.,][]{2003ApJ...584.1016K,2004ApJ...612..159W}.
Such uncertainties need to be accounted for when studying the intrinsic 
properties (such as the flare fluence and duration distributions) of the 
sample. We characterize the  detection incompleteness and bias, as well as 
the uncertainties of the measured parameters with a redistribution matrix
\begin{equation}
P(F_{\rm det},\tau_{\rm det};F_{\rm int},\tau_{\rm int}),
\end{equation}
where the subscriptions ``$_{\rm det}$'' and ``$_{\rm int}$'' denote the 
detected and intrinsic parameters.

We use Monte Carlo simulations to construct the redistribution matrix.
Each simulation assumes a quiescent count rate $\kappa$ and a flare 
with a fluence $F_{\rm int}$ and a duration $\tau_{\rm int}$, in 
an observation with a typical exposure of $10^5$ s. This flare is 
randomly inserted in time and assumed to have a Gaussian profile. 
The arrival time of each photon is randomly generated following the
lightcurve Eq. (1). We apply the same flare detection and parameter 
measurement procedure to the simulated lightcurve as to the real data. 
If the flare is detected (i.e., with TS$\geq14$), then we measure the 
``detected'' parameters $F_{\rm det}$ and $\tau_{\rm det}$. This is 
repeated for 1000 simulations for each ($F_{\rm int},\tau_{\rm int}$).

\begin{figure*}
\centering
\includegraphics[width=\columnwidth]{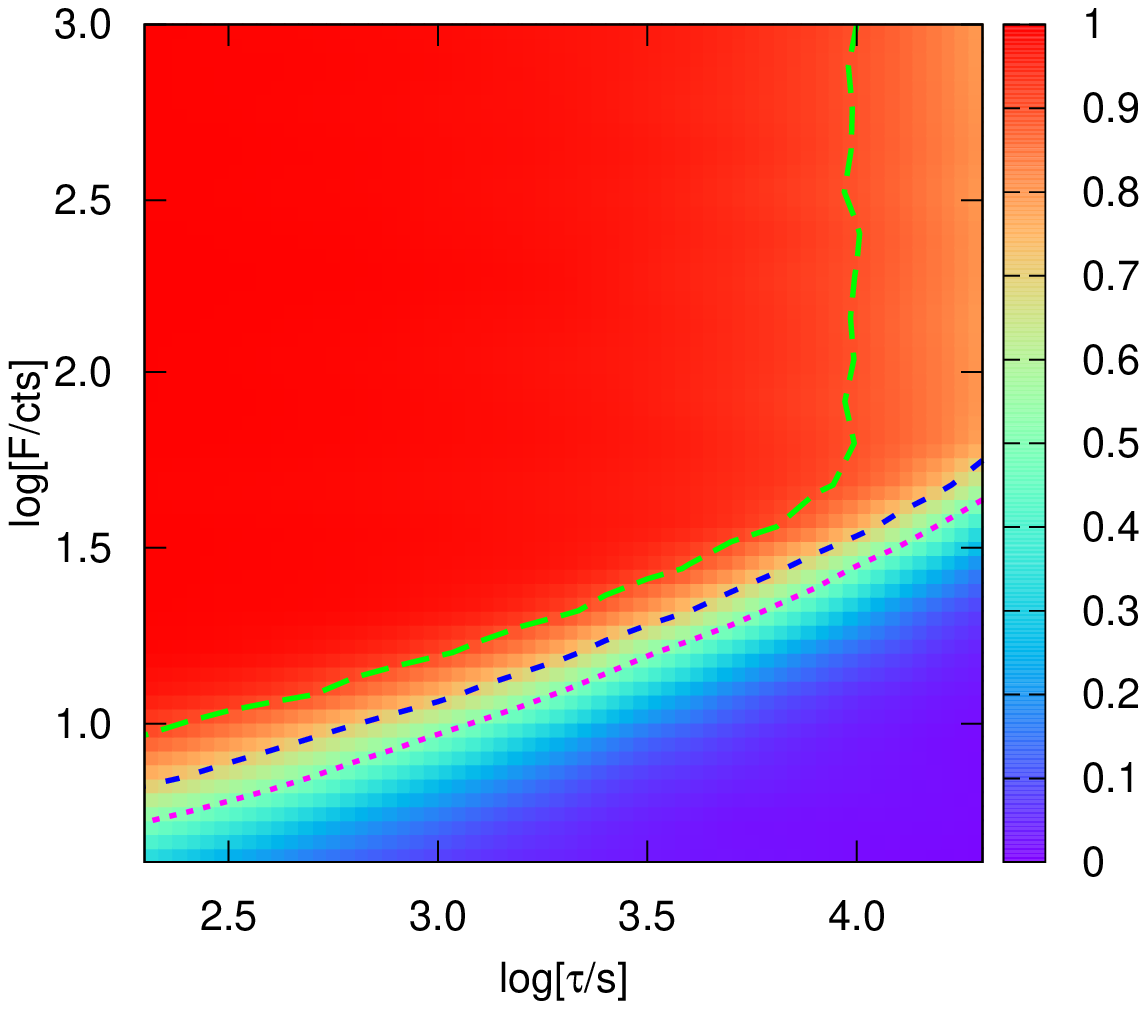}
\includegraphics[width=\columnwidth]{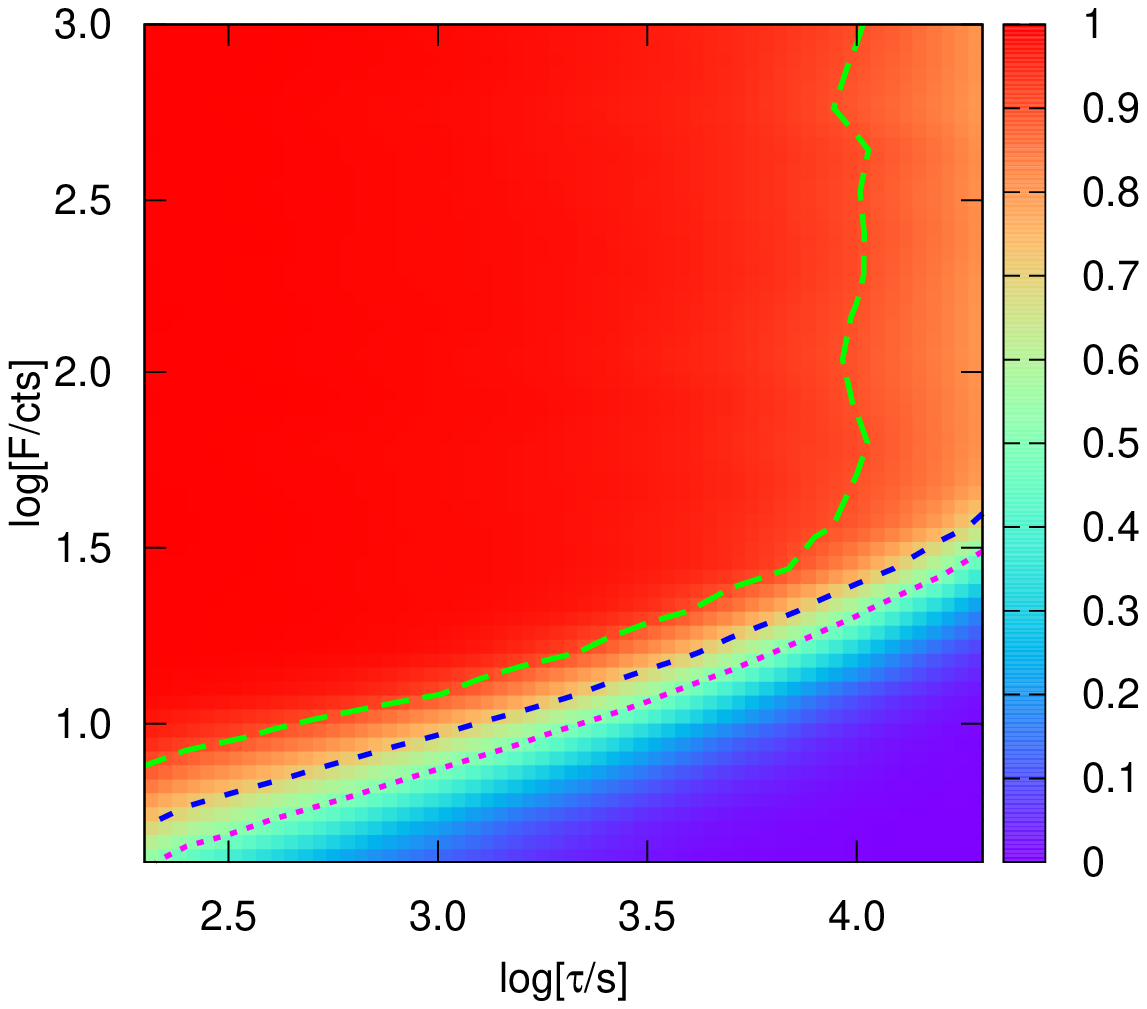}
\caption{Detection incompleteness as a function of the intrinsic flare 
fluence and duration. Lines show the $90\%$, $70\%$, and $50\%$ iso-contours 
of the detection fraction from top to bottom. The left panel is for an 
assumed quiescent rate of 4.9 cts ks$^{-1}$ (similar to ACIS-I data), 
while the right one for 1.9 cts ks$^{-1}$ (similar to ACIS-S/HETG0 data).
}
\label{fig:ratio}
\end{figure*}

Due to the statistical fluctuation, a flare may or may not be detectable 
above our defined threshold. This detection fraction of simulated flares 
as a function of the intrinsic fluence and duration is shown in Fig. 
\ref{fig:ratio}. The left and right panels are for different quiescent 
rates, which are chosen to mimic those in the ACIS-I and -S/HETG0 
observations, respectively. This figure shows that if the fluence is 
lower than a few tens of counts, the fraction could decrease considerably. 
The fraction also increases with the flare duration, $\tau_{\rm int}$, 
the increase of which decreases the signal-to-noise ratio. As illustrated 
by the $90\%$ fraction contour (long-dashed line in the figure), the 
incompleteness becomes important even for very high fluence flares when 
$\tau_{\rm int}\gtrsim10^4$ s. This is mainly because such a long-duration 
flare always has a good chance to be truncated by the start or end of 
an observation.

\begin{figure}
\centering
\includegraphics[width=\columnwidth]{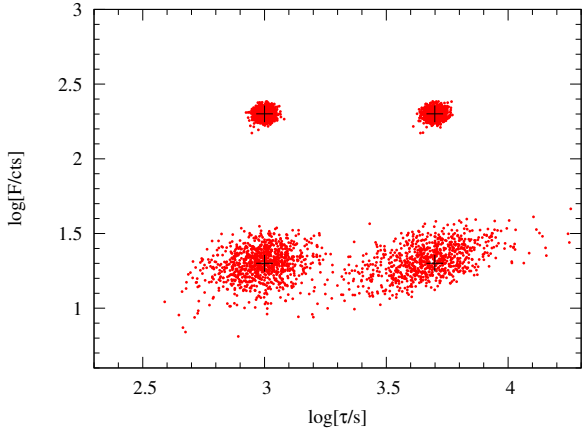}
\caption{Illustration of measured fluence and duration distributions 
(red dots) for four sets of simulated flares. The input flare parameters
of each set are marked by the black crosses.
}
\label{fig:F_tau_simu}
\end{figure}

The redistribution also determines the detection bias and the uncertainties 
in the parameter measurements. Fig. \ref{fig:F_tau_simu} illustrates how 
the parameters (red dots) are redistributed around the input ones
(black crosses). The scatters are larger if the fluence is lower and/or 
the duration is longer (e.g., bottom-right corner). The truncation from 
bottom-left to top-right for input parameters $(F_{\rm int},\tau_{\rm int})
=(20\,{\rm cts},\,5\,{\rm ks})$ is caused by realizations falling below
the detection threshold (see Fig. \ref{fig:ratio}). Conversely, a
considerable number of flares which are intrinsically weak could be realized 
to be above the detected limit. Depending mainly on the steepness of the 
fluence distribution, the redistribution leads to the so-called Eddington 
bias: many more flares scattered upwards than downwards \citep[e.g.,][]
{2004ApJ...612..159W}. The redistribution probabilities, each normalized to 
the detection fraction for a particular set of the input parameters, are 
then calculated on logarithmic grids of $\tau$ from 0.2 to 20 ks, and 
of $F$ from 4 to 1000 cts. The resulting 2-D redistribution matrix will 
enable us to account for the incompleteness and bias of the detection, 
as well as the measurement uncertainties of the parameters, critical for 
a rigorous statistical study of flare properties to be presented in a 
forthcoming paper.

\subsection{Observational gap effect}

Large gaps exist between consecutive observations, which need to be 
accounted for when the rate and waiting time of the flares are analyzed. 
Some of them were truncated at the start or end of an observation, 
which affects the detection of such flares (see Fig. \ref{fig:ratio}). 
The gap effect is especially important for the waiting time analysis
because a pair of flares with long enough waiting time will not be
detected with a shorter exposure. Under the null hypothesis in which
flares occur randomly, the distribution of the waiting times should
be exponential, ${\rm d}N_{\rm nogap}/{\rm d}(\Delta t)\propto 
\exp(-\Delta t/\overline{\Delta t})$. In case that there are observational 
gaps, this distribution needs to be modified by a detectable rate, which 
is an exposure-weighted probability
\begin{equation}
P(\Delta t)=\sum_i H(W_i-\Delta t)\cdot(W_i-\Delta t)/W_{\rm tot},
\end{equation}
where $W_i$ is the exposure of the $i$th observation, $W_{\rm tot}$
is the sum of all the exposures, $H(W_i-\Delta t)$ is the Heaviside step 
function, and the sum is over all observations. The term $(W_i-\Delta t)$ 
in the above equation means that an observation can only have an effective 
exposure of $(W_i-\Delta t)$ to detect a pair of flares with waiting 
time $\Delta t$. The cumulative distribution of the waiting time is then
\begin{equation}
N_{\rm gap}(>\Delta t)=\int_{\Delta t} \frac{{\rm d}N_{\rm nogap}}
{{\rm d}\Delta t'}\cdot P(\Delta t')~{\rm d}(\Delta t').
\end{equation}
We normalize both the expected and the corresponding detected distributions 
and then calculate the statistic $D$ (the maximum distance between the two 
distributions) to conduct the Kolmogorov-Smirnov (KS) null hypothesis test.

\subsection{Flux and luminosity conversions}

To facilitate the comparison of the flare detections based on the 
observations taken with the two different instrument setups, we convert 
the count rates of flares into their intrinsic fluxes. This conversion 
assumes the best-fit power-law model for the accumulated flare spectra 
presented in \citet[][supplementary materials]{2013Sci...341..981W}, 
accounting for the foreground absorption and dust scattering, as well 
as the pileup effect. The modeling was conducted with the spectral 
analysis package XSPEC \citep[version 12.8.0;][]{1996ASPC..101...17A}.
By removing the {\it multiplicative} pileup model component from this 
best-fit model, we obtain the conversion from the pileup-free count 
rate to the absorbed energy flux in the $2-8$ keV band as 
$2.81\,(7.22)\times10^{-11} {\rm~ergs~cm^{-2}~s^{-1}/[cts~s^{-1}}]$, 
for the ACIS-I (-S/HETG0) detected flares. The corresponding count
rate to the unabsorbed luminosity conversion is $0.73\,(1.88) \times 
10^{36} {\rm~ergs~s^{-1}/[cts\,s^{-1}}]$, assuming the distance of Sgr 
A$^\star$ to be 8 kpc. The ACIS-I to -S/HETG0 count rate ratio (hence 
effective area) is about 2.6 for the same incident flux.

\section{Results}

In total we detect 33 flares in the ACIS-I observations, and 49 flares 
in the ACIS-S/HETG0 observations. The fitting parameters of these flares and 
the quiescent emission count rates of individual observations are included
in Tables \ref{table:flare} and \ref{table:flare2012}. 

\subsection{Flare fluences and durations}

\begin{figure*}
\centering
\includegraphics[width=\columnwidth]{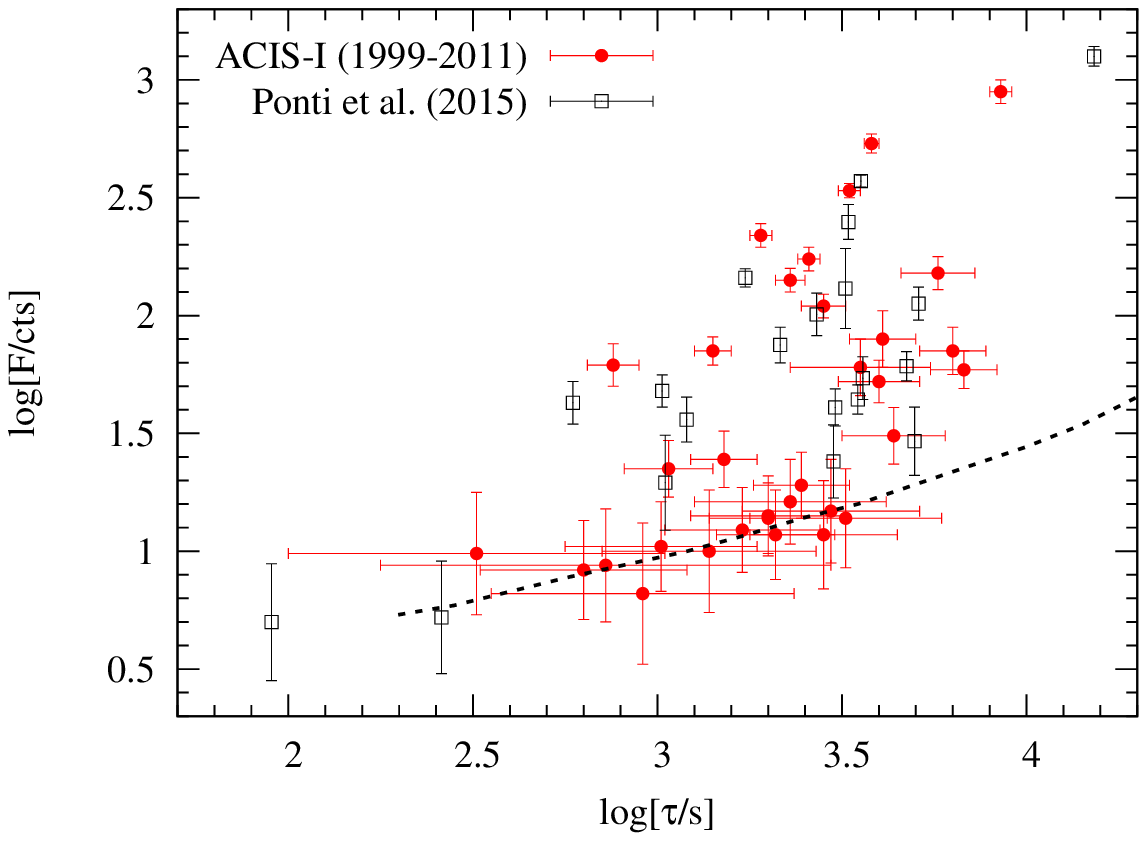}
\includegraphics[width=\columnwidth]{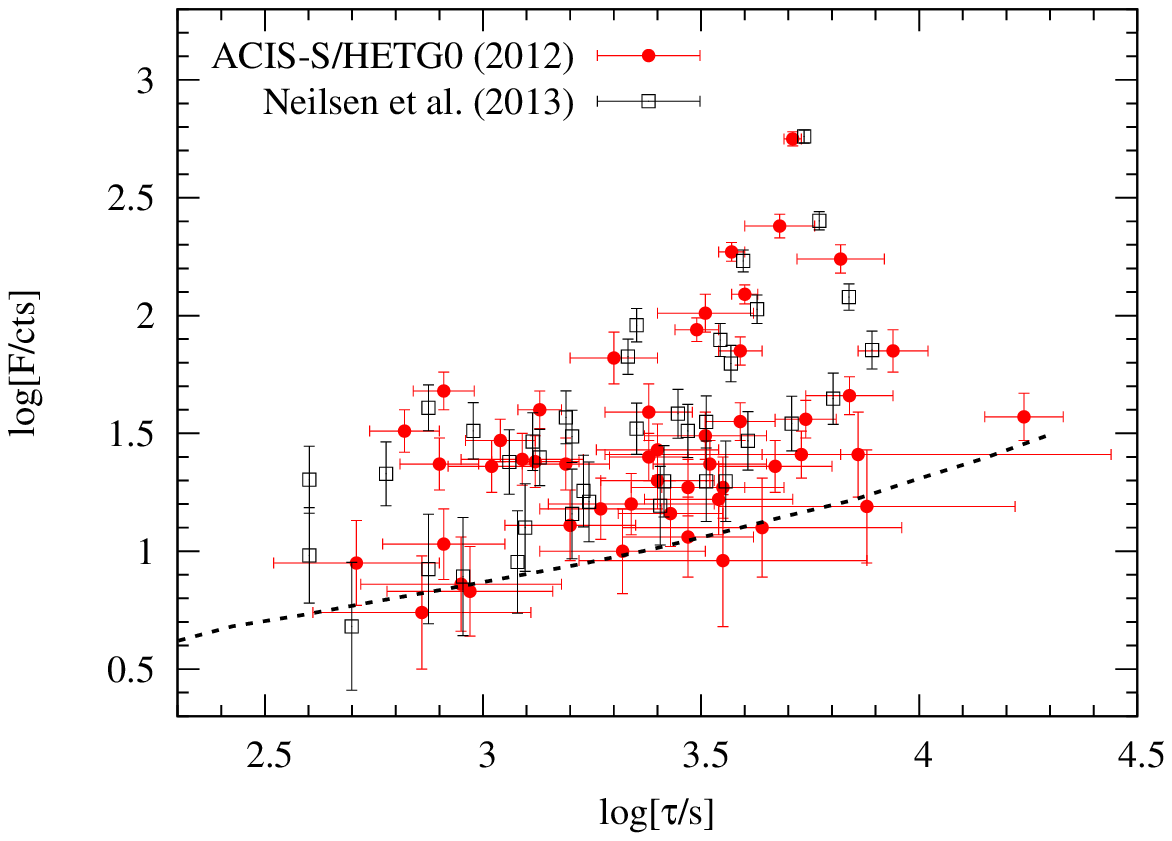}
\caption{Fluence versus duration distribution of the detected flares
for the ACIS-I (left) and -S/HETG0 (right) observations. The results of 
P15 and N13 are also shown for comparison. The dotted lines show the 
$50\%$ incompleteness limit of our flare detections (Fig. \ref{fig:ratio}).
}
\label{fig:F_tau}
\end{figure*}

\begin{figure*}
\centering
\includegraphics[width=\columnwidth]{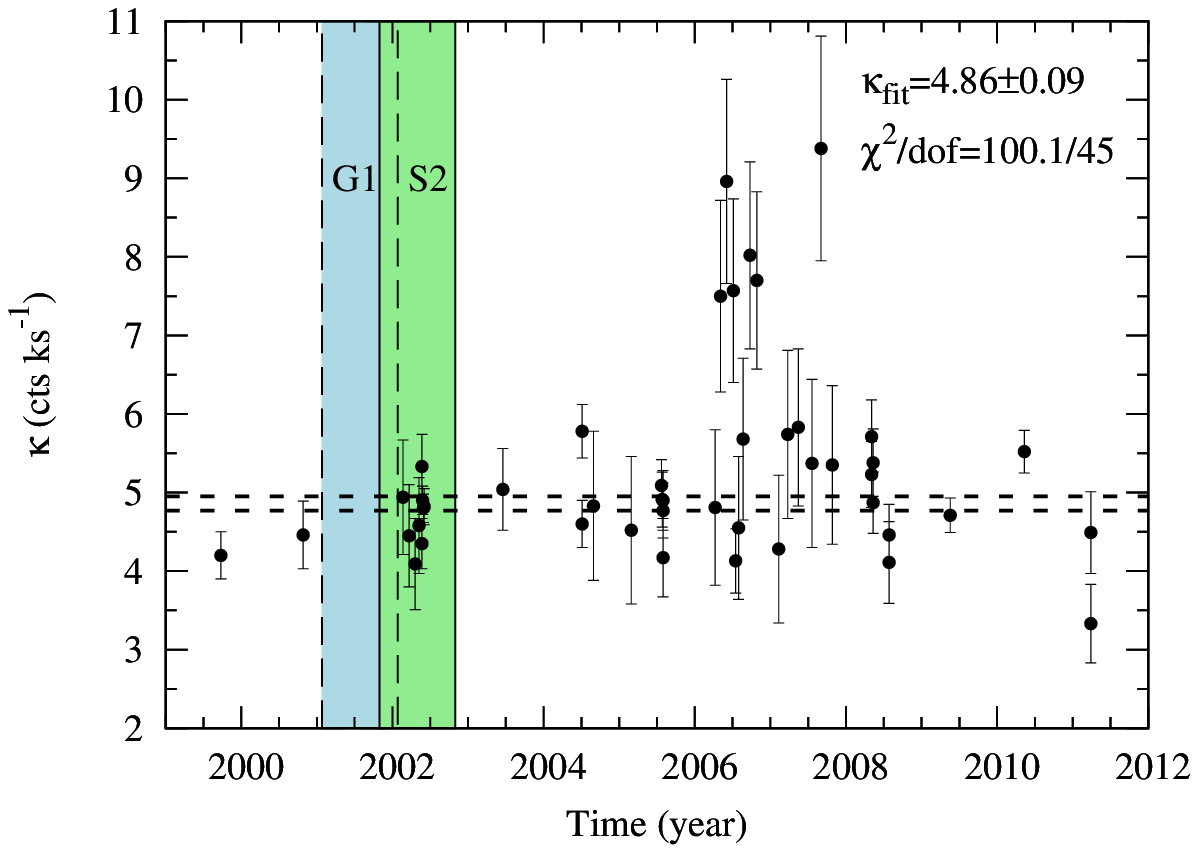}
\includegraphics[width=\columnwidth]{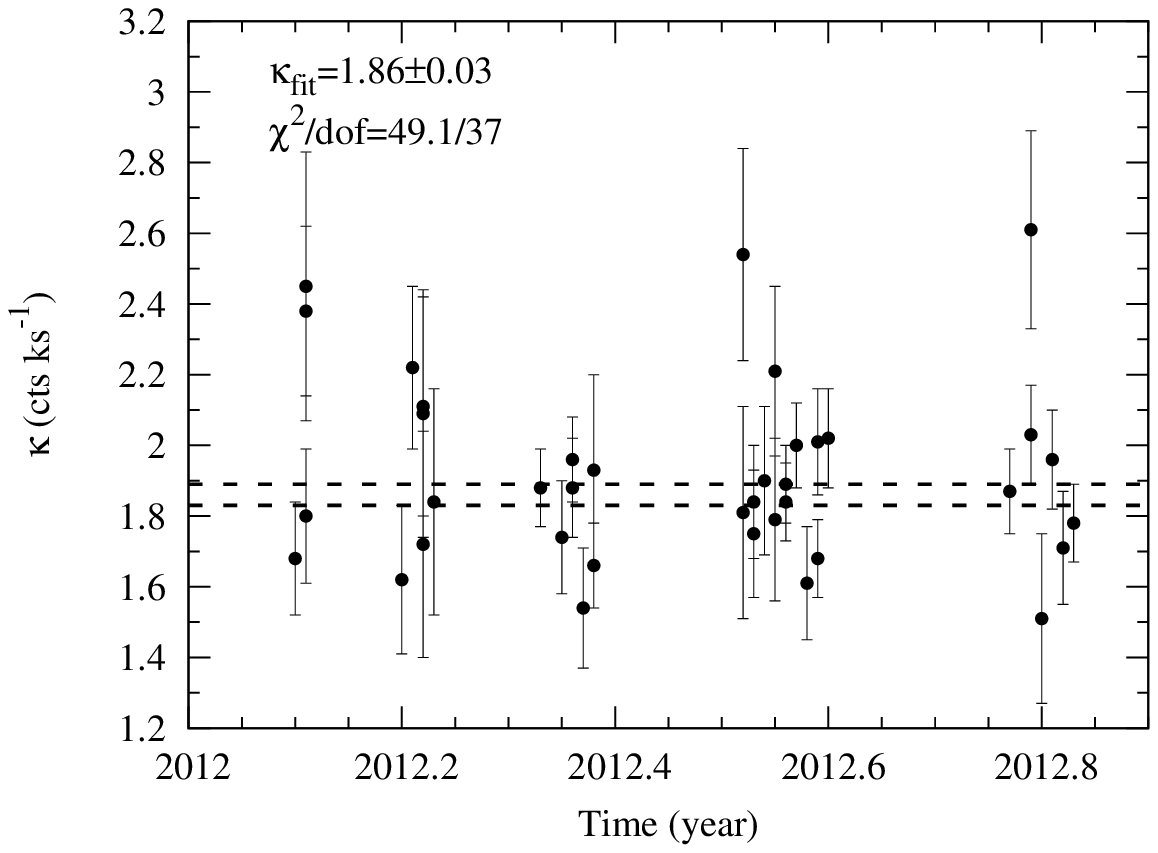}
\caption{The quiescent emission rate of Sgr A$^\star$ as a function of the 
date for the ACIS-I (left) and ACIS-S/HETG0 (right) observations. 
The horizontal lines mark the $\pm1\sigma$ range of the constant fittings. 
Shaded regions in the left panel show the $\pm0.5$ year around the 
best-fitting pericenter passages of the G1 cloud \citep{2015ApJ...798..111P} 
and S2 star \citep{2002Natur.419..694S}.
}
\label{fig:bkg}
\end{figure*}

Fig. \ref{fig:F_tau} shows the fluence-duration distributions of the 
detected flares, for the ACIS-I (left) and -S/HETG0 (right) observations, 
as well as those detected in P15 and N13 for comparison. The fluence 
(in cts) of a P15 flare is obtained through dividing the reported 
value (in erg cm$^{-2}$) by the conversion factor $4.2\times10^{-11}$ 
(erg cm$^{-2}$/cts). In order to have a direct comparison of our
sample with that of N13, we multiply the fluences of N13 flares by a 
factor of $\sim0.6$ to account for the 0th to 0th+1st order event 
ratio, and then apply the pileup function as presented in P15 
(the inverse of Eq. (4)) to the flare profiles to calculate the pileup 
corrected fluences. We find most of the flares that appear above or 
just around the 50\% incompleteness curve are missed in P15 and N13 
(see further discussion in \S5).

\begin{table*}
\scriptsize
\centering
\caption{Sgr A$^{\star}$ flares detected in the ACIS-I data}
\vspace{-2mm}
\begin{tabular}{cccccccccccccc}
\hline \hline
ObsID & Date & Start & End & Exp. & $\kappa$ & FlrID & $\log(F/{\rm cts})$ & Peak & $\log(\tau/{\rm ks})$ & $L_{2-8}^{\rm unabs}$ & $N_{\rm sub}$ & $P$ & Ref.\\
    & & (ks)  & (ks)& (ks)     & (cts/ks) & &  & (ks) & & ($10^{34}$ erg/s) & & & \\
\hline
242  & 1999-09-21 & 54270.275 & 54320.032 & 49.8 & $4.20\pm0.30$ &   &                &           &               &   &   &   & \\
     &            &           &           &      & fixed$^\ddagger$ & I1 & $1.07\pm0.23$ & 54272.168 & $0.45\pm0.20^\dagger$ & 0.30 & 1 & & (1)\\
1561 & 2000-10-26 & 88985.913 & 89022.169 & 36.3 & $4.46\pm0.43$ & I2 & $1.35\pm0.12$ & 89001.653 & $0.03\pm0.12$ & 1.53 & 1 & & \\
     &            &           &           &      &               & I3 & $2.95\pm0.05$ & 89011.674 & $0.93\pm0.03$ & 7.64 & 3 & 0.0067 & (2),(3)\\
2951 & 2002-02-19 & 130517.00 & 130529.54 & 12.5 & $4.94\pm0.73$ & I4 & $1.00\pm0.26$ & 130521.47 & $0.14\pm0.29$ & 0.53 & 1 & & \\
2952 & 2002-03-23 & 133274.58 & 133286.59 & 12.0 & $4.45\pm0.65$ & I5 & $0.82\pm0.30$ & 133278.21 & $-0.04\pm0.41$& 0.53 & 1 & & \\
2953 & 2002-04-19 & 135601.19 & 135612.93 & 11.7 & $4.09\pm0.58$ &   &                &           &               &   &   &   & \\
2954 & 2002-05-07 & 137151.56 & 137164.17 & 12.6 & $4.58\pm0.61$ &   &                &           &               &   &   &   & \\
2943 & 2002-05-22 & 138496.84 & 138535.02 & 38.2 & $4.35\pm0.32$ &   &                &           &               &   &   &   & \\
3663 & 2002-05-24 & 138629.52 & 138667.99 & 38.5 & $5.38\pm0.45$ & I6 & $1.85\pm0.10$ & 138656.20 & $0.80\pm0.09$ & 0.82 & 2 & 0.054 & (3) \\
3392 & 2002-05-25 & 138728.07 & 138896.97 & 168.9& $4.90\pm0.18$ & I7 & $1.72\pm0.09$ & 138774.36 & $0.60\pm0.11$ & 0.96 & 1 & & (3) \\
     &            &           &           &      &               & I8 & $1.02\pm0.19$ & 138782.73 & $0.01\pm0.26$ & 0.75 & 1 & & \\
     &            &           &           &      &               & I9 & $1.49\pm0.12$ & 138808.50 & $0.64\pm0.14$ & 0.52 & 1 & & (3) \\
     &            &           &           &      &               & I10& $1.39\pm0.12$ & 138865.34 & $0.18\pm0.09$ & 1.18 & 1 & & (3) \\
     &            &           &           &      &               & I11& $1.09\pm0.18$ & 138878.91 & $0.23\pm0.21$ & 0.53 & 1 & & \\
3393 & 2002-05-28 & 138952.39 & 139112.51 & 160.1& $4.80\pm0.18$ & I12& $2.53\pm0.03$ & 138987.68 & $0.52\pm0.03$ & 7.47 & 1 & & (3) \\
     &            &           &           &      &               & I13& $2.18\pm0.07$ & 139039.72 & $0.76\pm0.10$ & 1.92 & 2 & 0.059 & (3) \\
     &            &           &           &      &               & I14& $1.85\pm0.06$ & 139085.21 & $0.15\pm0.05$ & 3.66 & 1 & & (3) \\
3665 & 2002-06-03 & 139455.69 & 139546.81 & 91.1 & $4.82\pm0.23$ & I15& $0.94\pm0.24$ & 139465.08 & $-0.14\pm0.61$ & 0.88 & 1 & & \\
3549 & 2003-06-19 & 172435.60 & 172460.71 & 25.1 & $5.04\pm0.52$ & I16& $1.17\pm0.22$ & 172453.77 & $0.47\pm0.24$ & 0.37 & 1 & & (4) \\
4683 & 2004-07-05 & 205454.84 & 205505.02 & 50.2 & $4.60\pm0.30$ &    &               &           &               &   &   &    & \\
4684 & 2004-07-06 & 205541.03 & 205591.21 & 50.2 & $5.78\pm0.34$ &    &               &           &               &   &   &    & \\
     &            &           &           &      & fixed$^\ddagger$ & I17& $0.92\pm0.21$ & 205543.34 & $-0.20\pm0.28$  & 0.96 & 1 & & (5) \\
     &            &           &           &      &               & I18& $2.15\pm0.05$ & 205558.31 & $0.36\pm0.04$ & 4.50 & 1 & & (3),(5) \\
5360 & 2004-08-28 & 210082.80 & 210087.97 & 5.2  & $4.83\pm0.95$ &    &               &           &               &   &   &   & \\
6113 & 2005-02-27 & 225874.38 & 225879.30 & 4.9  & $4.52\pm0.94$ &    &               &           &               &   &   &   & \\
5950 & 2005-07-24 & 238623.22 & 238672.40 & 49.2 & $5.09\pm0.33$ &    &               &           &               &   &   &   & \\
5951 & 2005-07-27 & 238879.92 & 238925.10 & 45.2 & $4.91\pm0.35$ &    &               &           &               &   &   &   & \\
5952 & 2005-07-29 & 239054.88 & 239100.81 & 45.9 & $4.90\pm0.38$ & I19& $1.77\pm0.08$ & 239079.32 & $0.83\pm0.09$ & 0.64 & 1 & & (3),(6) \\
5953 & 2005-07-30 & 239141.35 & 239187.31 & 46.0 & $4.77\pm0.35$ & I20& $2.04\pm0.05$ & 239149.82 & $0.45\pm0.06$ & 2.84 & 1 & & (3),(7) \\
5954 & 2005-08-01 & 239314.38 & 239332.69 & 18.3 & $4.17\pm0.50$ &    &               &           &               &   &   &   & \\
6639 & 2006-04-11 & 261121.70 & 261126.25 & 4.5  & $4.81\pm0.99$ &    &               &           &               &   &   &   & \\
6640 & 2006-05-03 & 263083.36 & 263088.53 & 5.2  & $7.50\pm1.22$ &    &               &           &               &   &   &   & \\
6641 & 2006-06-01 & 265566.30 & 265571.42 & 5.1  & $8.96\pm1.30$ &    &               &           &               &   &   &   & \\
     &            &           &           &      & fixed$^\ddagger$ & I21& $1.15\pm0.17$ & 265567.89 & $0.30\pm0.21$ & 0.52 & 1 & & \\
6642 & 2006-07-04 & 268399.23 & 268404.41 & 5.2  & $7.57\pm1.17$ &    &               &           &               &   &   &   & \\
6363 & 2006-07-17 & 269496.92 & 269527.08 & 30.2 & $4.13\pm0.41$ &    &               &           &               &   &   &   & \\
     &            &           &           &      & fixed$^\ddagger$ & I22& $2.24\pm0.05$ & 269503.80 & $0.41\pm0.03$ & 4.94 & 1 & & (3),(7),(8) \\
6643 & 2006-07-30 & 270657.94 & 270662.98 & 5.0  & $4.55\pm0.91$ &    &               &           &               &   &   &   & \\
6644 & 2006-08-22 & 272614.20 & 272619.24 & 5.0  & $5.68\pm1.03$ &    &               &           &               &   &   &   & \\
6645 & 2006-09-25 & 275580.29 & 275585.47 & 5.2  & $8.02\pm1.19$ &    &               &           &               &   &   &   & \\
     &            &           &           &      & fixed$^\ddagger$ & I23& $1.07\pm0.19$ & 275581.21 & $0.32\pm0.16^\dagger$ & 0.41 & 1 & & \\
6646 & 2006-10-29 & 278480.67 & 278485.84 & 5.2  & $7.70\pm1.13$ &    &               &           &               &   &   &   & \\
7554 & 2007-02-11 & 287562.96 & 287568.11 & 5.1  & $4.28\pm0.94$ &    &               &           &               &   &   &   & \\
7555 & 2007-03-25 & 291251.43 & 291256.59 & 5.2  & $5.74\pm1.07$ &    &               &           &               &   &   &   & \\
7556 & 2007-05-17 & 295752.54 & 295757.58 & 5.0  & $5.83\pm1.00$ &    &               &           &               &   &   &   & \\
7557 & 2007-07-20 & 301287.05 & 301292.10 & 5.0  & $5.37\pm1.07$ &    &               &           &               &   &   &   & \\
7558 & 2007-09-02 & 305152.60 & 305157.65 & 5.0  & $9.38\pm1.43$ &    &               &           &               &   &   &   & \\
     &            &           &           &      & fixed$^\ddagger$ & I24& $1.14\pm0.15$ & 305153.70 & $0.30\pm0.16$ & 0.51 & 1 & & \\
7559 & 2007-10-26 & 309781.44 & 309786.51 & 5.1  & $5.35\pm1.01$ &    &               &           &               &   &   &   & \\
9169 & 2008-05-05 & 326348.29 & 326376.24 & 28.0 & $5.71\pm0.47$ &    &               &           &               &   &   &   & \\
     &            &           &           &      & fixed$^\ddagger$ & I25& $0.99\pm0.26$ & 326371.05 & $-0.49\pm0.51$ & 2.20 & 1 & & (3),(9) \\
9170 & 2008-05-06 & 326431.43 & 326458.58 & 27.1 & $5.23\pm0.42$ &    &               &           &               &   &   &   & \\
9171 & 2008-05-10 & 326777.71 & 326805.76 & 28.0 & $4.87\pm0.39$ &    &               &           &               &   &   &   & \\
9172 & 2008-05-11 & 326865.44 & 326893.23 & 27.8 & $5.38\pm0.43$ &    &               &           &               &   &   &   & \\
9174 & 2008-07-25 & 333410.76 & 333439.94 & 29.2 & $4.46\pm0.39$ &    &               &           &               &   &   &   & \\
9173 & 2008-07-26 & 333495.51 & 333523.64 & 28.1 & $4.11\pm0.52$ &    &               &           &               &   &   &   & \\ 
     &            &           &           &      & fixed$^\ddagger$ & I26& $1.14\pm0.21$ & 333499.49 & $0.51\pm0.26$ & 0.31 & 1 & & \\
     &            &           &           &      &               & I27& $1.21\pm0.18$ & 333504.75 & $0.36\pm0.26$ & 0.52 & 1 & & (9) \\
10556 & 2009-05-18 & 359001.57 & 359115.62 & 114.1& $4.71\pm0.22$ & I28& $1.78\pm0.12$ & 359002.25 & $0.55\pm0.19^\dagger$ & 1.24 & 2 & 0.048 & (3) \\
      &            &           &           &      &               & I29& $1.90\pm0.12$ & 359029.24 & $0.61\pm0.09$ & 1.42 & 2 & 0.049 & (3),(10) \\
      &            &           &           &      &               & I30& $2.34\pm0.05$ & 359075.91 & $0.28\pm0.03$ & 8.38 & 1 & & (3)\\
      &            &           &           &      &               & I31& $1.79\pm0.09$ & 359081.83 & $-0.12\pm0.07$ & 5.93 & 1 & & (3) \\
11843 & 2010-05-13 & 390104.87 & 390184.84 & 80.0 & $5.52\pm0.27$ &    &               &           &               &   &   &   & \\  
     &            &           &           &      & fixed$^\ddagger$ & I32& $2.73\pm0.04$ & 390110.65 & $0.58\pm0.02$ & 10.3 & 1 & & (3) \\
13016 & 2011-03-29 & 417782.70 & 417800.76 & 18.1 & $3.33\pm0.50$ &    &               &           &               &   &   &   & \\  
     &            &           &           &      & fixed$^\ddagger$ & I33& $1.28\pm0.14$ & 417783.64 & $0.39\pm0.13^\dagger$ & 0.57 & 1 & & (3) \\
13017 & 2011-03-31 & 417955.49 & 417973.56 & 18.1 & $4.49\pm0.52$ &    &               &           &               &   &   &   & \\
\hline \hline
\end{tabular}\\
Note: Columns from left to right are: observation ID, observing date, 
starting time and ending time from UT 1998-01-01 00:00:00, exposure, 
quiescent count rate, flare ID, logarithmic fluence, peak time, logarithmic 
duration, mean unabsorbed $2-8$~keV luminosity within the duration, number 
of subflares, the chance probability of uncorrelated flares overlapping 
with the main flare, and the references of previous works. 
Posterior marginalized $1\sigma$ errors are included 
for the fluence and the duration.\\
$^\dagger$Flare truncated by the starting or ending of an observation. \\
$^\ddagger$Quiescent emission is fixed at 4.86 cts ks$^{-1}$ --- the mean 
rate of all the ACIS-I observations.\\
References: (1) \citet{2003ApJ...591..891B}; (2) \citet{2001Natur.413...45B};
(3) \citet{2015MNRAS.454.1525P}; (4) \citet{2004A&A...427....1E};
(5) \citet{2006A&A...450..535E}; (6) \citet{2008A&A...479..625E};
(7) \citet{2007ApJ...667..900H}; (8) \citet{2008ApJ...682..373M};
(9) \citet{2012AJ....144....1Y}; (10) \citet{2012A&A...537A..52E}.
\label{table:flare}
\end{table*}

\begin{table*}
\scriptsize
\centering
\caption{Sgr A$^{\star}$ flares detected in the ACIS-S/HETG0 data}
\begin{tabular}{cccccccccccccc}
\hline \hline
ObsID & Date & Start & End & Exp. & $\kappa$ & FlrID & $\log(F/{\rm cts})$ & Peak & $\log(\tau/{\rm ks})$ & $L_{2-8}^{\rm unabs}$ & $N_{\rm sub}$ & $P$ & Ref.\\
    & & (ks)  & (ks)& (ks)     & (cts/ks) & &  & (ks) & & ($10^{34}$ erg/s) & & & \\
\hline
13850 & 2012-02-06 & 444877.14 & 444937.14 & 60.0 & $1.68\pm0.16$ &    &               &           &               &   &   &   & \\
14392 & 2012-02-09 & 445156.63 & 445215.88 & 59.2 & $1.80\pm0.19$ & S1 & $1.18\pm0.13$ & 445171.72 & $0.27\pm0.14$ & 1.53 & 1 & & (1),(2)\\
      &            &           &           &      &               & S2 & $2.75\pm0.03$ & 445187.73 & $0.71\pm0.02$ & 20.6 & 1 & & (1),(2)\\
14394 & 2012-02-10 & 445232.14 & 445250.21 & 18.1 & $2.45\pm0.38$ &    &               &           &               &   &   &   & \\
14393 & 2012-02-11 & 445343.58 & 445385.13 & 41.5 & $2.38\pm0.24$ &    &               &           &               &   &   &   & \\
13856 & 2012-03-15 & 448189.56 & 448229.62 & 40.1 & $1.62\pm0.21$ &    &               &           &               &   &   &   & \\
13857 & 2012-03-17 & 448363.06 & 448402.61 & 39.6 & $2.22\pm0.23$ &    &               &           &               &   &   &   & \\
13854 & 2012-03-20 & 448626.68 & 448649.75 & 23.1 & $2.09\pm0.35$ & S3 & $1.37\pm0.11$ & 448631.25 & $-0.10\pm0.09$ & 5.55 & 1 & & (2)\\
      &            &           &           &      &               & S4 & $1.38\pm0.11$ & 448634.81 & $0.12\pm0.10$ & 3.42 & 1 & & (2)\\
      &            &           &           &      &               & S5 & $1.37\pm0.11$ & 448639.76 & $0.19\pm0.10$ & 2.85 & 1 & & (2)\\
      &            &           &           &      &               & S6 & $1.51\pm0.09$ & 448647.98 & $-0.18\pm0.08$ & 9.21 & 1 & & (2)\\
14413 & 2012-03-21 & 448700.54 & 448715.26 & 14.7 & $1.72\pm0.32$ &    &               &           &               &   &   &   & \\
13855 & 2012-03-22 & 448803.91 & 448823.97 & 20.1 & $2.11\pm0.31$ &    &               &           &               &   &   &   & \\
14414 & 2012-03-23 & 448913.21 & 448933.27 & 20.1 & $1.84\pm0.32$ &    &               &           &               &   &   &   & \\
13847 & 2012-04-30 & 452191.00 & 452345.06 & 154.1& $1.88\pm0.11$ & S7 & $1.55\pm0.08$ & 452263.05 & $0.59\pm0.08$ & 1.71 & 1 & & (2)\\
14427 & 2012-05-06 & 452722.87 & 452802.93 & 80.1 & $1.74\pm0.16$ & S8 & $1.43\pm0.11$ & 452747.93 & $0.40\pm0.14$ & 2.01 & 1 & & (2)\\
      &            &           &           &      &               & S9 & $1.36\pm0.11$ & 452777.65 & $0.67\pm0.13$ & 0.92 & 1 & & \\
13848 & 2012-05-09 & 452953.42 & 453051.58 & 98.2 & $1.88\pm0.14$ &    &               &           &               &   &   &   & \\
13849 & 2012-05-11 & 453095.24 & 453273.98 & 178.7& $1.96\pm0.12$ & S10& $1.41\pm0.18$ & 453141.81 & $0.86\pm0.58$ & 0.67 & 2 & 0.058 & (2)\\
      &            &           &           &      &               & S11& $1.27\pm0.12$ & 453170.73 & $0.47\pm0.13$ & 1.19 & 1 & & (2)\\
      &            &           &           &      &               & S12& $1.10\pm0.21$ & 453195.75 & $0.64\pm0.32$ & 0.54 & 1 & & (2)\\
      &            &           &           &      &               & S13& $1.94\pm0.05$ & 453267.26 & $0.49\pm0.05$ & 5.30 & 1 & & (2)\\
13846 & 2012-05-16 & 453552.19 & 453607.19 & 55.0 & $1.54\pm0.17$ &    &               &           &               &   &   &   & \\
14438 & 2012-05-18 & 453703.94 & 453729.73 & 25.8 & $1.93\pm0.27$ &    &               &           &               &   &   &   & \\
13845 & 2012-05-19 & 453812.55 & 453947.86 & 135.3& $1.66\pm0.12$ &    &               &           &               &   &   &   & \\
      &            &           &           &      & fixed$^\ddagger$ & S14& $1.00\pm0.18$ & 453823.23 & $0.32\pm0.19$ & 0.90 & 1 & & \\
      &            &           &           &      &               & S15& $1.82\pm0.11$ & 453935.16 & $0.30\pm0.10$ & 6.23 & 2 & 0.012 & (2)\\
      &            &           &           &      &               & S16& $1.11\pm0.15$ & 453938.91 & $0.20\pm0.15$ & 1.53 & 1 & & \\
      &            &           &           &      &               & S17& $1.06\pm0.17$ & 453946.43 & $0.47\pm0.15^\dagger$ & 0.73 & 1 & & \\
14460 & 2012-07-09 & 458261.60 & 458285.66 & 24.1 & $2.54\pm0.30$ &    &               &           &               &   &   &   & \\
13844 & 2012-07-10 & 458351.01 & 458371.07 & 20.1 & $1.81\pm0.30$ &    &               &           &               &   &   &   & \\
14461 & 2012-07-12 & 458459.40 & 458510.37 & 51.0 & $1.75\pm0.18$ &    &               &           &               &   &   &   & \\
13853 & 2012-07-14 & 458613.56 & 458687.22 & 73.7 & $1.84\pm0.16$ &    &               &           &               &   &   &   & \\
13841 & 2012-07-17 & 458947.94 & 458993.01 & 45.1 & $1.90\pm0.21$ &    &               &           &               &   &   &   & \\
14465 & 2012-07-18 & 459041.14 & 459085.48 & 44.3 & $1.79\pm0.23$ & S18& $1.56\pm0.08$ & 459043.46 & $0.74\pm0.07^\dagger$ & 1.24 & 1 & & (2)\\
      &            &           &           &      &               & S19& $1.16\pm0.14$ & 459059.71 & $0.43\pm0.12$ & 1.01 & 1 & & (2)\\
14466 & 2012-07-20 & 459176.36 & 459221.44 & 45.1 & $2.21\pm0.24$ &    &               &           &               &   &   &   & \\
      &            &           &           &      & fixed$^\ddagger$ & S20& $1.39\pm0.11$ & 459177.21 & $0.09\pm0.14$ & 3.75 & 1 & & (2)\\
      &            &           &           &      &               & S21& $1.03\pm0.15$ & 459217.78 & $-0.09\pm0.14$ & 2.48 & 1 & & \\
13842 & 2012-07-21 & 459260.69 & 459452.44 & 191.8& $1.89\pm0.11$ & S22& $1.85\pm0.06$ & 459320.36 & $0.59\pm0.05$ & 3.42 & 1 & & (2)\\
      &            &           &           &      &               & S23& $1.36\pm0.11$ & 459381.31 & $0.02\pm0.10$ & 4.11 & 1 & & (2)\\
      &            &           &           &      &               & S24& $1.85\pm0.09$ & 459437.62 & $0.94\pm0.08$ & 1.53 & 2 & 0.039 & (2)\\
13839 & 2012-07-24 & 459501.75 & 459678.00 & 176.3& $1.84\pm0.11$ & S25& $1.47\pm0.09$ & 459509.11 & $0.04\pm0.08$ & 5.06 & 1 & & (2)\\
      &            &           &           &      &               & S26& $0.95\pm0.18$ & 459606.20 & $-0.29\pm0.19$ & 3.27 & 1 & & (2)\\
      &            &           &           &      &               & S27& $2.27\pm0.04$ & 459649.70 & $0.57\pm0.03$ & 9.42 & 1 & & (2)\\
13840 & 2012-07-26 & 459721.42 & 459883.94 & 162.5& $2.00\pm0.12$ & S28& $1.27\pm0.13$ & 459863.37 & $0.55\pm0.14$ & 0.99 & 1 & & \\
      &            &           &           &      &               & S29& $0.96\pm0.28$ & 459876.27 & $0.55\pm0.33$ & 0.48 & 1 & & \\
14432 & 2012-07-30 & 460041.72 & 460115.99 & 74.3 & $1.61\pm0.16$ &    &               &           &               &   &   &   & \\
      &            &           &           &      & fixed$^\ddagger$ & S30& $1.41\pm0.10$ & 460044.94 & $0.73\pm0.09^\dagger$ & 0.90 & 1 & & \\
      &            &           &           &      &               & S31& $2.09\pm0.04$ & 460113.72 & $0.60\pm0.03^\dagger$ & 5.81 & 1 & & (2)\\
13838 & 2012-08-01 & 460230.82 & 460330.38 & 99.6 & $2.01\pm0.15$ & S32& $2.01\pm0.08$ & 460254.74 & $0.51\pm0.11$ & 5.95 & 2 & 0.034 & (2)\\
      &            &           &           &      &               & S33& $0.86\pm0.20$ & 460269.49 & $-0.05\pm0.23$ & 1.53 & 1 & & \\
13852 & 2012-08-04 & 460436.77 & 460593.34 & 156.6& $1.68\pm0.11$ &    &               &           &               &   &   &   & \\
      &            &           &           &      & fixed$^\ddagger$ & S34& $1.60\pm0.08$ & 460453.54 & $0.13\pm0.05$ & 5.55 & 1 & & (2)\\
      &            &           &           &      &               & S35& $1.57\pm0.10$ & 460495.90 & $1.24\pm0.09$ & 0.40 & 1 & & \\
      &            &           &           &      &               & S36& $1.37\pm0.10$ & 460541.81 & $0.52\pm0.13$ & 1.33 & 1 & & (2)\\
14439 & 2012-08-06 & 460679.74 & 460791.48 & 111.7& $2.02\pm0.14$ & S37& $1.30\pm0.12$ & 460783.48 & $0.40\pm0.13$ & 1.49 & 1 & & (2)\\
14462 & 2012-10-06 & 465929.41 & 466063.49 & 134.1& $1.87\pm0.12$ & S38& $1.22\pm0.15$ & 465971.27 & $0.54\pm0.17$ & 0.90 & 1 & & (2)\\
      &            &           &           &      &               & S39& $1.40\pm0.10$ & 466058.86 & $0.38\pm0.09$ & 1.97 & 1 & & (2)\\
14463 & 2012-10-16 & 466737.09 & 466767.86 & 30.8 & $2.61\pm0.28$ &    &               &           &               &   &   &   & \\
      &            &           &           &      & fixed$^\ddagger$ & S40& $1.68\pm0.08$ & 466753.89 & $-0.09\pm0.07$ & 11.1 & 1 & & (2)\\
13851 & 2012-10-16 & 466802.39 & 466909.45 & 107.1& $2.03\pm0.14$ & S41& $0.74\pm0.24$ & 466827.54 & $-0.14\pm0.25$ & 1.43 & 1 & & (2)\\
      &            &           &           &      &               & S42& $2.38\pm0.05$ & 466891.41 & $0.68\pm0.08$ & 9.42 & 2 & 0.043 & (2)\\
15568 & 2012-10-18 & 466938.99 & 466975.06 & 36.1 & $1.51\pm0.24$ &    &               &           &               &   &   &   & \\ 
      &            &           &           &      & fixed$^\ddagger$ & S43& $1.20\pm0.13$ & 466972.41 & $0.34\pm0.19$ & 1.36 & 1 & & \\
13843 & 2012-10-22 & 467310.03 & 467430.71 & 120.7& $1.96\pm0.14$ & S44& $2.24\pm0.06$ & 467372.68 & $0.82\pm0.10$ & 4.94 & 2 & 0.072 & (2)\\
      &            &           &           &      &               & S45& $1.19\pm0.24$ & 467418.81 & $0.88\pm0.34$ & 0.38 & 1 & & \\
15570 & 2012-10-25 & 467524.25 & 467592.96 & 68.7 & $1.71\pm0.16$ & S46& $1.59\pm0.12$ & 467532.67 & $0.38\pm0.10$ & 3.05 & 2 & 0.025 & (2)\\
14468 & 2012-10-29 & 467943.30 & 468089.36 & 146.1& $1.78\pm0.11$ & S47& $1.66\pm0.08$ & 467970.68 & $0.84\pm0.10$ & 1.24 & 1 & & (2)\\
      &            &           &           &      &               & S48& $0.83\pm0.19$ & 468005.49 & $-0.03\pm0.19$ & 1.36 & 1 & & (2)\\
      &            &           &           &      &               & S49& $1.49\pm0.10$ & 468079.07 & $0.51\pm0.14$ & 1.80 & 1 & & (2)\\
\hline \hline
\end{tabular}
\label{table:flare2012}\\
Note: Same as Table \ref{table:flare}.\\
$^\dagger$Flare is truncated by the start or end of an observation. \\
$^\ddagger$Quiescent emission is fixed to be 1.86 cts ks$^{-1}$.\\
References: (1) \citet{2012ApJ...759...95N}; (2) \citet{2013ApJ...774...42N}.
\end{table*}

A correlation between the flare fluences and durations is apparent in Fig.
\ref{fig:F_tau}. We characterize this correlation with a linear function
\begin{equation}
\log(F/{\rm cts})=\log(\alpha)+\beta\log(\tau/{\rm ks}).
\end{equation}
The fitting parameters and the $\chi^2$ values over the number of 
degree-of-freedom (dof) are given in Table \ref{table:fit}. We obtain
a correlation slope of $\beta\sim2.5-2.8$. The large scatter of the
data around this relation indicates an intrinsic dispersion around the
function. We estimate this dispersion by quadratically adding an intrinsic 
error, $\sigma_{\log(\tau)}^{\rm int}$, to the statistical ones. Solving 
$\chi^2/{\rm dof}\approx1$, we obtain $\sigma_{\log(\tau)}^{\rm int}
\approx0.20$ and 0.26 for the ACIS-I and -S/HETG0 data sets, respectively. 
However, we note that these estimates and fittings, obtained without 
proper accounting for the incompleteness and bias of the flare detection 
(\S3.3), are for crude characterizations only. More rigorous treatments 
will be presented in a forthcoming work.

\begin{table}
\centering
\caption{Fitting results of the flare fluence-duration correlations}
\begin{tabular}{cccccc}
\hline \hline
    & $\log(\alpha)$ & $\beta$ & $\chi^2/{\rm dof}$ & $\sigma_{\log(\tau)}^{\rm int}$ \\
\hline
ACIS-I & $1.14\pm0.08$ & $2.45\pm0.16$ & $239.8/31$ & 0.20 \\
P15    & $1.70\pm0.03$ & $1.05\pm0.04$ & $609.0/18$ &      \\
\hline
ACIS-S/HETG0 & $0.56\pm0.09$ & $2.78\pm0.17$ & $392.9/47$ & 0.26 \\
N13    & $1.54\pm0.03$ & $1.26\pm0.05$ & $562.1/37$ &      \\
\hline
\hline
\end{tabular}
\label{table:fit}
\end{table}

\subsection{Quiescent emission}

Fig. \ref{fig:bkg} presents the fitting count rates of the quiescent 
component for all the observations. This component includes the background 
emission, as well as the truly quiescent emission and contribution from 
undetected weak flares of Sgr A$^\star$. The difference between the mean 
ACIS-I and -S/HETG0 rates of the component is consistent with the effective 
area ratio of the two instrument setups ($\sim2.6$; see \S~3.5). We estimate 
the background count rates from nearby regions around Sgr A$^\star$, scaled 
to our aperture for the flare detection. The rates are about $1.00$ cts 
ks$^{-1}$ and $0.38$ cts ks$^{-1}$ for the ACIS-I and -S/HETG0 observations. 
Therefore, most of the fitting quiescent component actually comes from 
Sgr A$^\star$ itself or other positionally overlapping sources. 
About $10\%-15\%$ of this emission could be attributed to undetected weak 
flares assuming a simple fluence distribution extrapolation from detected 
ones \citep{2015ApJ...799..199N}. Thus the truly quiescent emission accounts 
for the largest portion of this quiescent component. 

We now examine the variation of the quiescent emission from Sgr A$^\star$
on various time scales. First we check whether or not there is any significant 
change which might be attributed to the pericenter passages of the G1 
cloud around the time of the year 2001.57 \citep{2015ApJ...798..111P} 
and S2 star around 2002.33 \citep{2002Natur.419..694S}. The $\pm0.5$ 
year intervals about the passages are shown in Fig. \ref{fig:bkg}. 
While there was no observation during the interval around the G1 passage, 
9 observations were around the S2 passage. No significant variation in 
the quiescent count rate during this latter interval is found compared 
with the 12-year average. 

We also do not find any significant systematic trend in the quiescent 
count rate history, for either ACIS-I or -S/HETG0 observations. However, 
we do find significant rate variation among different observations
(judging from the values of the best-fitting $\chi^2$/dof). The intrinsic 
root-mean-square (RMS) fluctuation of the quiescent component, added
quadratically to the Poisson errors, is estimated to be $\sim14\%$ (6\%) 
for the ACIS-I (-S/HETG0) data via setting $\chi^2/{\rm dof}\sim1$. 
To characterize the intrinsic non-flare variation, we need to account 
for various statistical fluctuations. The error bars in Fig. \ref{fig:bkg} 
only account for the counting statistics of the observed photon events.
Additional fluctuations are expected from the limited number statistics 
of weak flares below our detection threshold in individual observations, 
which are particularly important in those with short exposures (e.g., 
the ACIS-I observations with ObsID 6640, 6641, 6642, 6645, 6646, and 7558). 
Therefore, the above RMSs represent only the upper limit to the true
variation of the quiescent rate among the observations. 
Because the exposures of the ACIS-S/HETG observations are long, the 
fluctuation due to the number statistics of undetected weak flares is 
smaller, as shown by the $\sim6\%$ intrinsic RMS.

\subsection{Flare rate}

The flare rate is another important statistical quantity related to 
the nature of the flares. The rate is estimated to be about $1.9\pm0.3$ 
or $1.4\pm0.2$ day$^{-1}$ for the ACIS-I or -S/HETG0 observations. 
The intrinsic fluence (defined as fluence divided by area) detection 
limit of the ACIS-I flares is expected to be about 1.6 times lower 
than that of the -S/HETG0 ones\footnote{The detected lowest intrinsic 
fluence of the ACIS-I flares is about 2.2 times lower than that of the 
-S/HETG0 ones, which is subject to uncertainty from the detection of 
faint flares (\S~3.3).}. According to the fluence distribution 
$N(>F)\propto F^{-0.5}$ (N13), the ACIS-I to -S/HETG0 flare rate ratio 
is expected to be $\sim1.3$, which is roughly consistent with the above 
detected rate difference. On the other hand, if taking the lowest 
intrinsic fluence of the ACIS-S/HETG0 flares as threshold, we find an 
ACIS-I to -S/HETG0 flare rate ratio of $0.9\pm0.2$. The flare rates 
between these two data sets are consistent with each other.

\begin{figure*}
\centering
\includegraphics[width=\columnwidth]{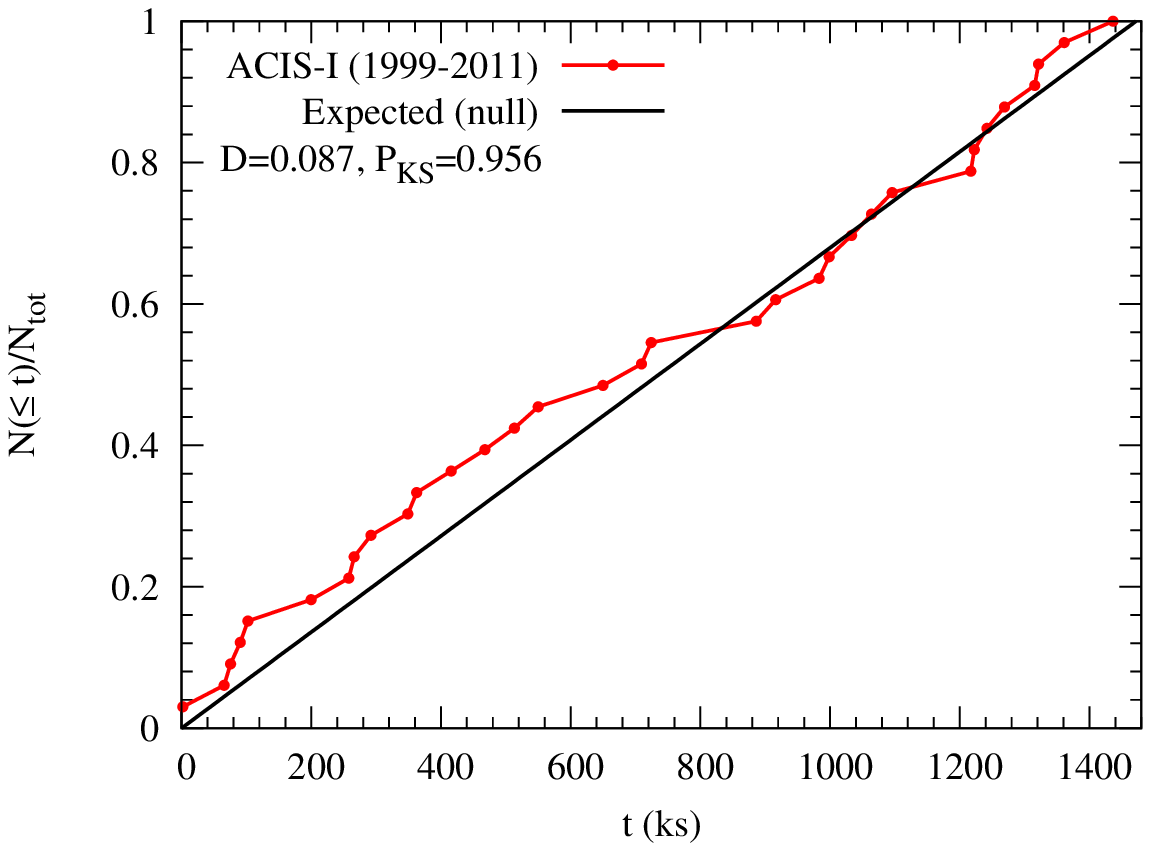}
\includegraphics[width=\columnwidth]{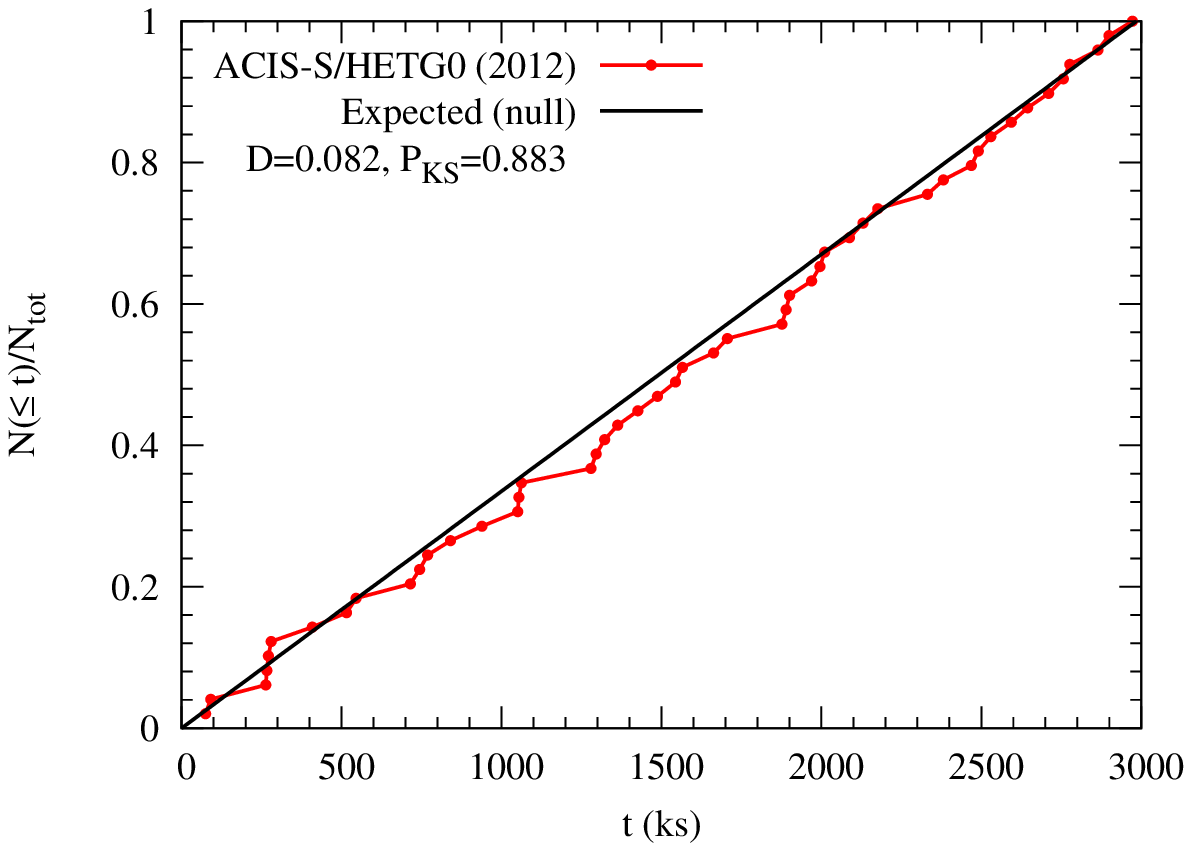}
\includegraphics[width=\columnwidth]{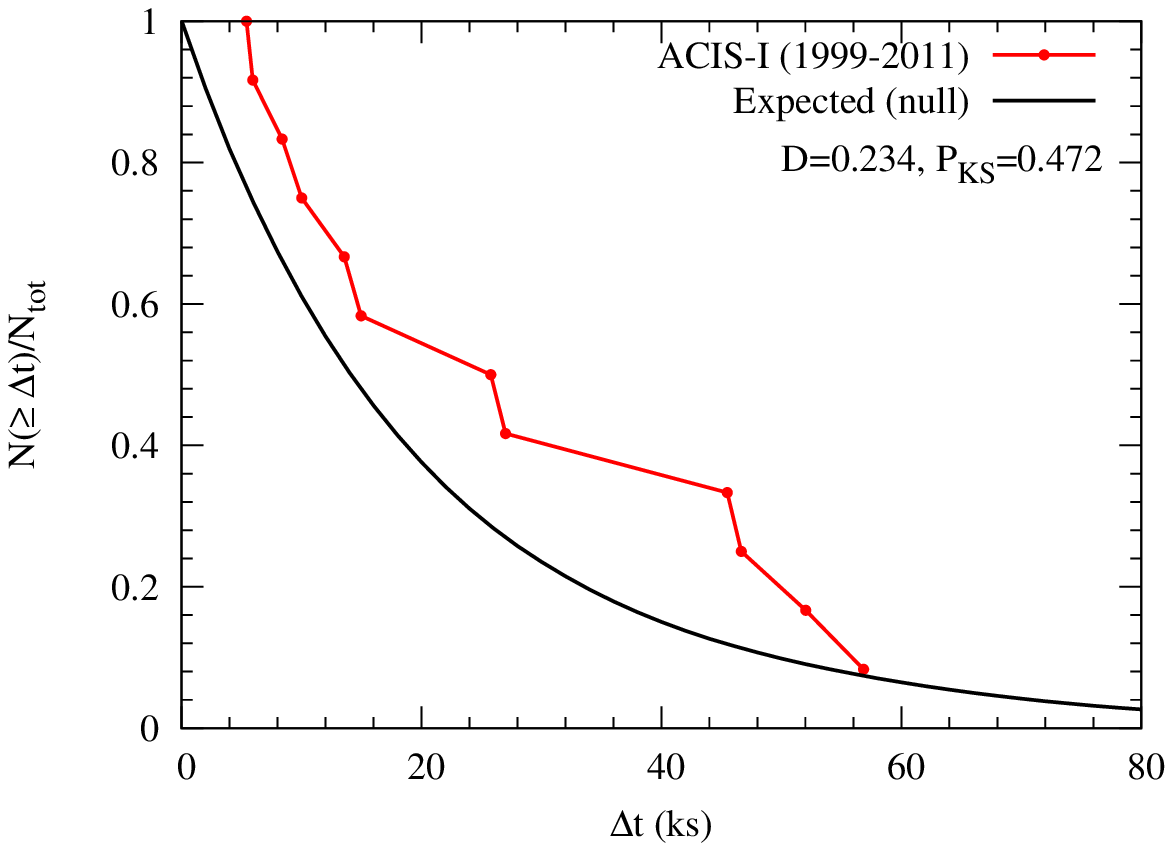}
\includegraphics[width=\columnwidth]{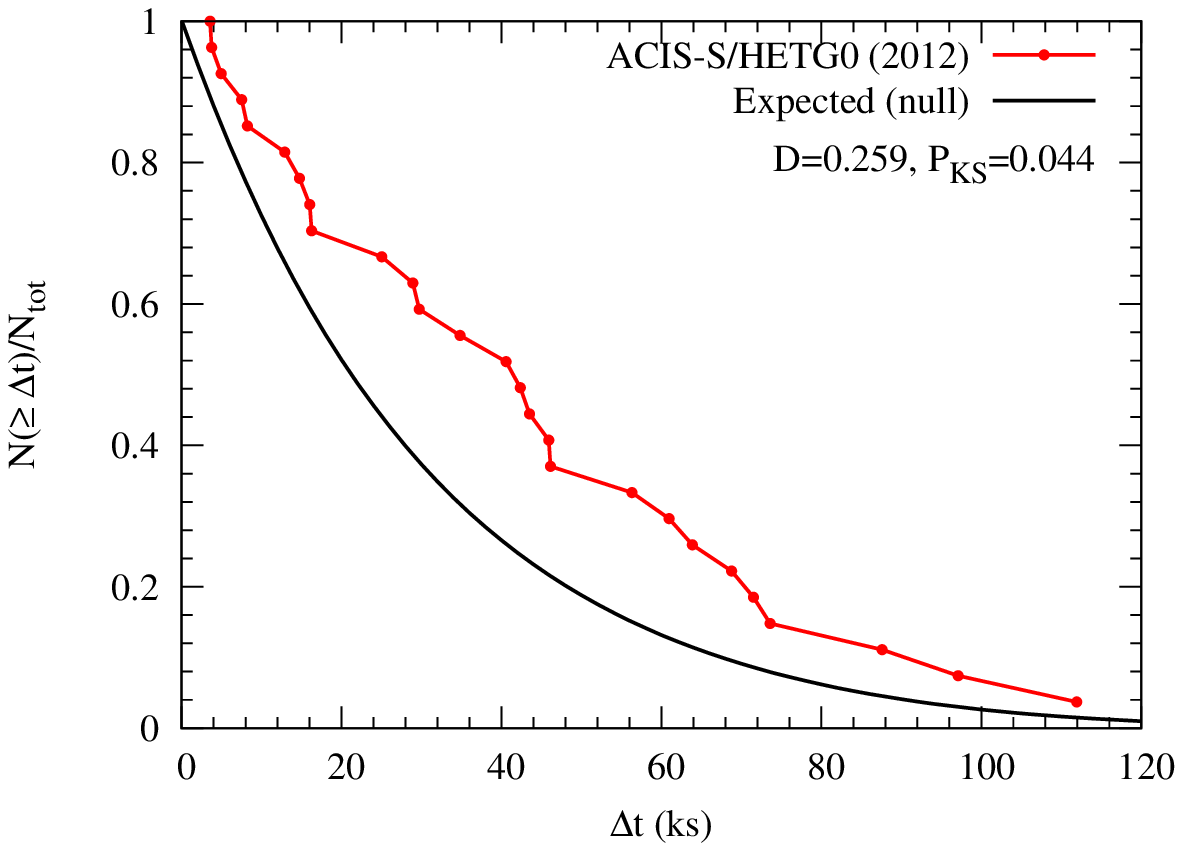}
\caption{Top panels: normalized backward cumulative number $N(\leq t)/
N_{\rm tot}$ of flares versus the accumulated exposure time. The red 
thick lines with dots represent the results from the real observational 
data, while the black lines show linear behaviors expected from random 
occurrence of flares (null hypothesis) with constant flare rates: 
$f\approx1.9$ day$^{-1}$ for the ACIS-I observations (left) and 
$f\approx1.4$ day$^{-1}$ for the ACIS-S/HETG0 observations (right). 
Bottom panels: normalized forward cumulative number $N(\geq\Delta t)/
N_{\rm tot}$ of flares versus the waiting time since the preceding one. 
The expected distributions for null hypothesis (black solid lines) are 
calculated according to Eq. (7).
}
\label{fig:Nflr}
\end{figure*}

To explore the systematic long-term change of the flare rate, we calculate 
the cumulative number of flares as a function of exposure time, as shown 
in the top panels of Fig. \ref{fig:Nflr}. The nearly linear increase 
of the number with time suggests that the flare rate is approximately 
constant during these observations. The KS tests show that the observations 
are consistent with the null hypothesis with high probabilities (see the 
statistic $D$ and ${\rm P_{KS}}$ values inserted in the upper panels of 
Fig. \ref{fig:Nflr}).

The bottom panels of Fig. \ref{fig:Nflr} show the (cumulative) waiting 
time distributions of the consecutive flares in individual observations,
observed and expected for non-clustering hypothesis (see Eq. (7) in \S~3.4). 
Deviations from the expectations of the null hypothesis, consistently 
occurring on time scales of $\sim40$ ks for both the ACIS-I and -S/HETG0 
flare samples, can be seen in these plots, although only the deviation 
in the latter sample is significant at the 96\% confidence, according to 
the KS test. The clustering is less significant for the ACIS-I flare 
sample, apparently due to its smaller size\footnote{Note that the $D$ 
values are comparable for the two samples (Fig. \ref{fig:Nflr}).}. 
The presence of this short-term clustering may suggest that X-ray flares 
of Sgr A$^\star$ mimic the foreshock or aftershock of earthquakes, which 
can be described with a piecewise-deterministic Markov process 
\citep{Davis1984}.

\section{Comparison with previous works}

\begin{figure*}
\centering
\includegraphics[width=\columnwidth]{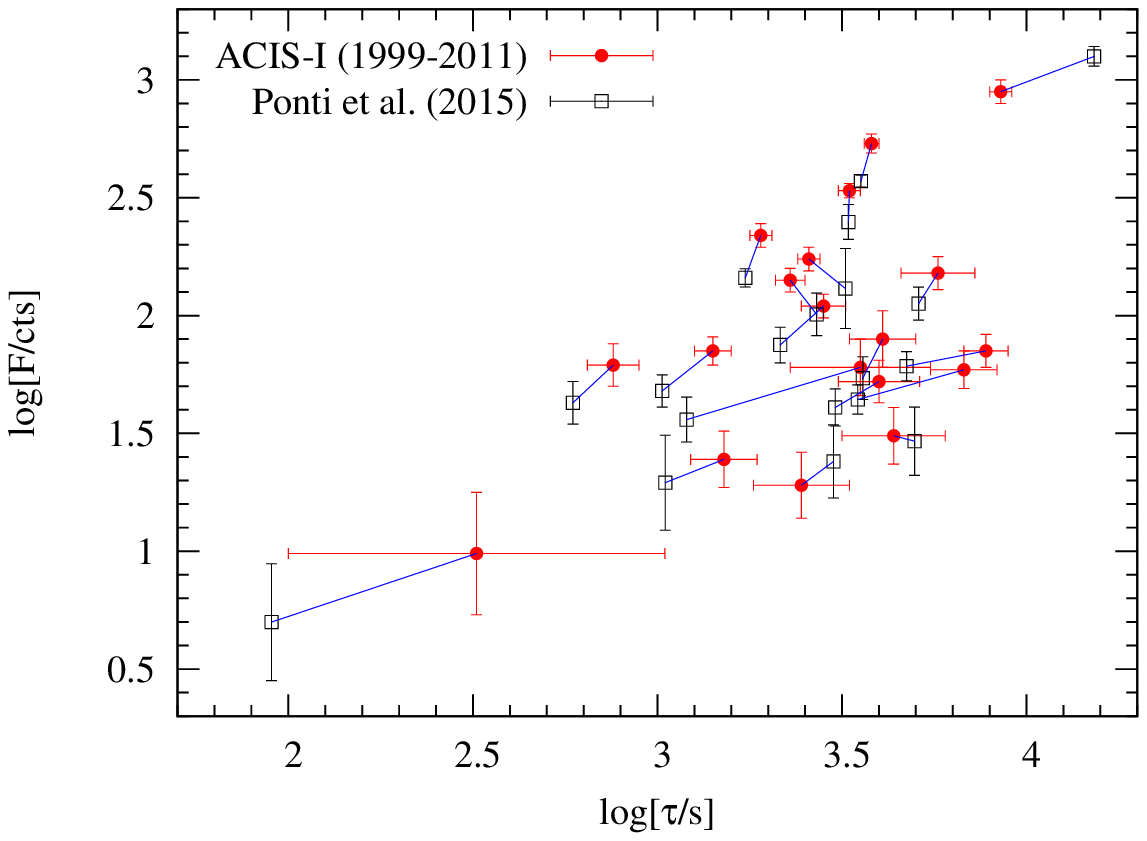}
\includegraphics[width=\columnwidth]{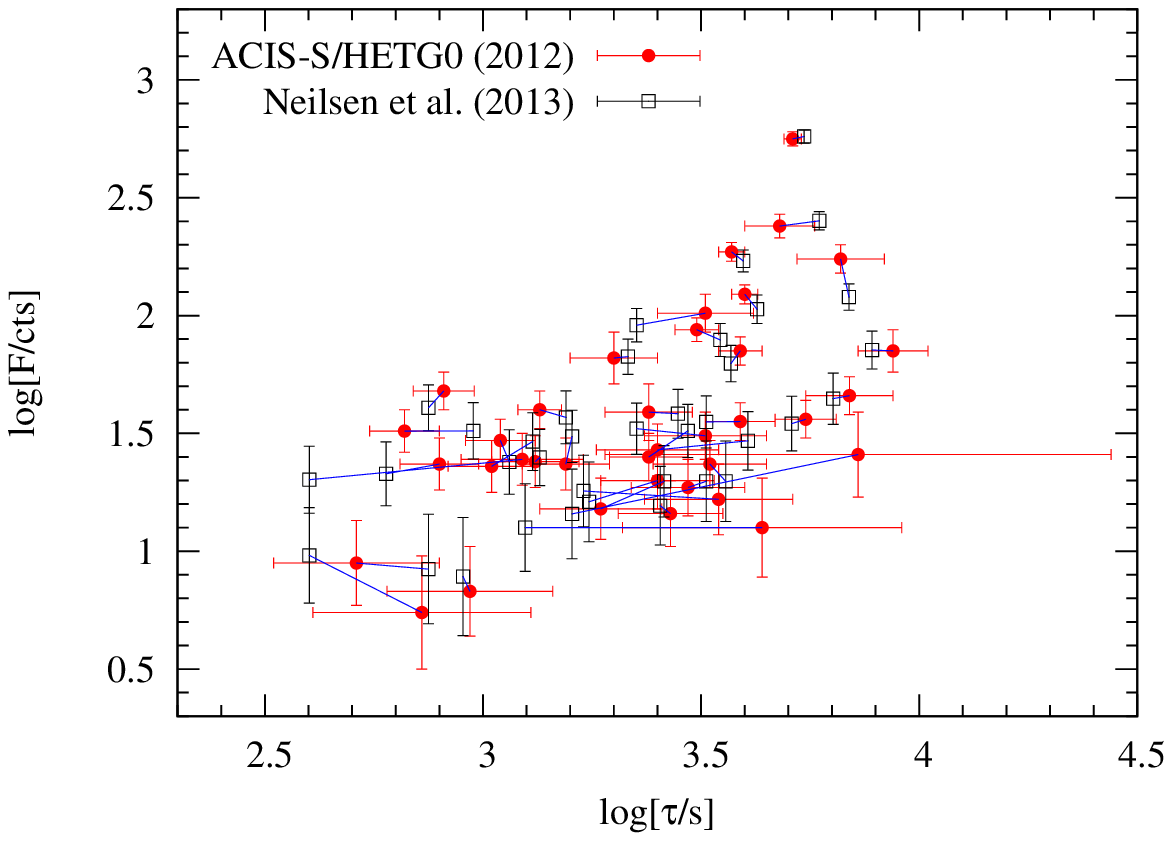}
\includegraphics[width=\columnwidth]{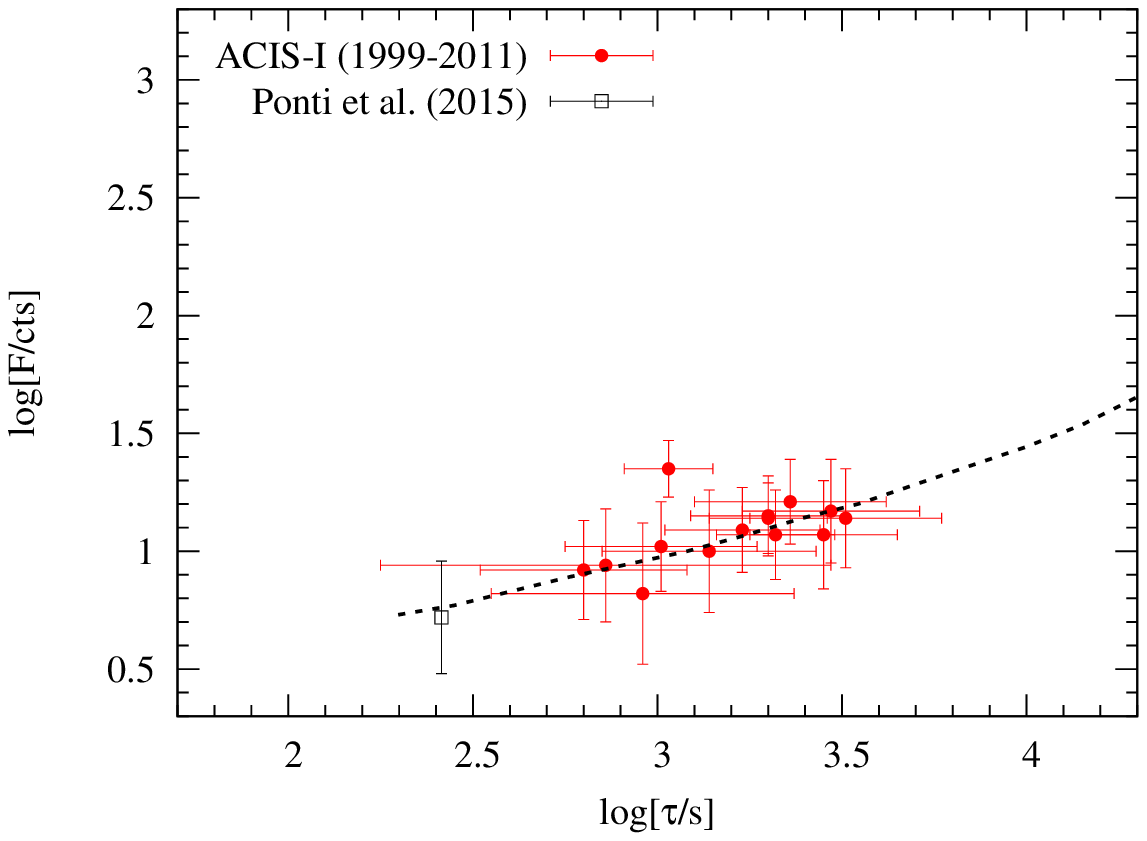}
\includegraphics[width=\columnwidth]{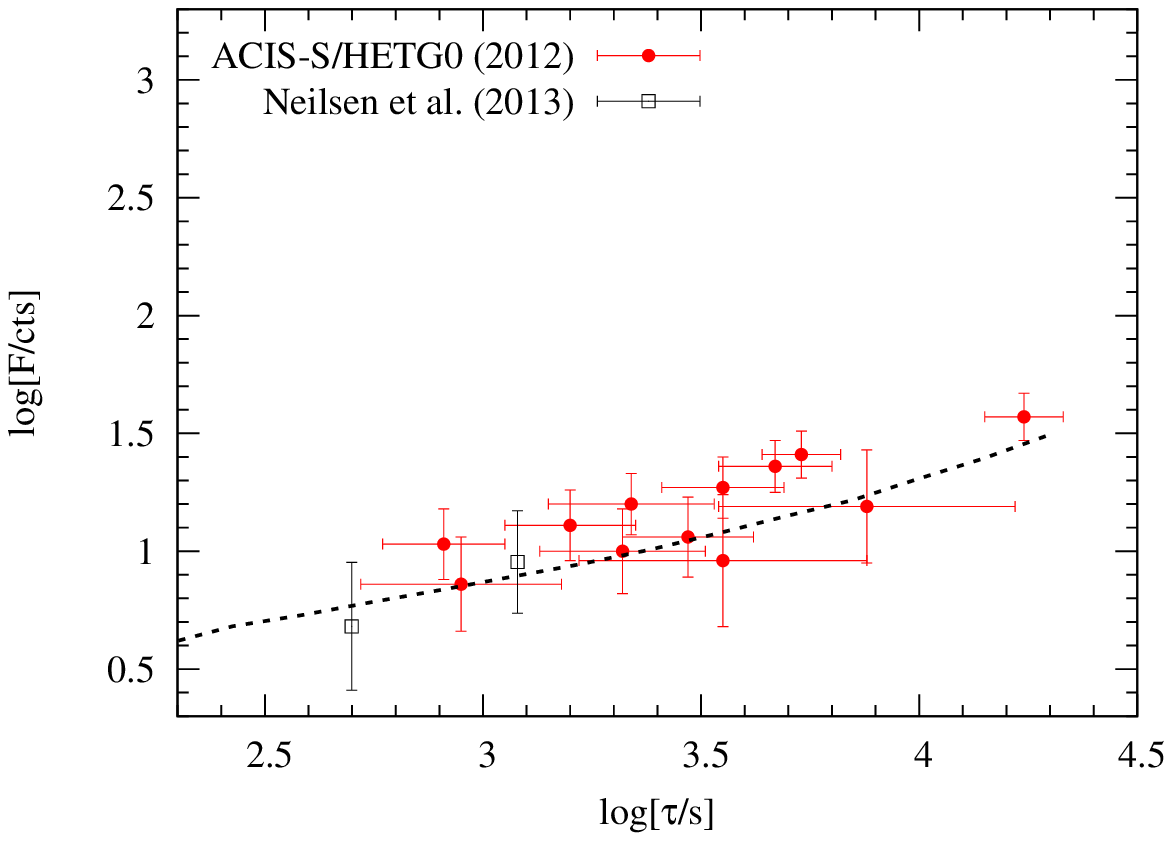}
\caption{Top panels: comparisons of the fluence-duration distributions 
of the flares commonly detected in the present work, as well as in P15 
for the ACIS-I observations (left) or N13 for the ACIS-S/HETG0 observations 
(right). Bottom panels: same as the top ones but for the flares detected 
only in our analysis or P15/N13 analysis. The dotted lines are the 50\% 
incompleteness limits as in Fig. \ref{fig:F_tau}.
}
\label{fig:F_tau_joint}
\end{figure*}

We now compare our results about the flares with those of P15 (for the 
ACIS-I data) and N13 (for the ACIS-S/HETG0 data). Based on the same 
ACIS-I observations, 19 flares are jointly detected in our analysis and 
in P15, which are shown in the top-left panel of Fig. \ref{fig:F_tau_joint}. 
There are quite a few apparent discrepancies in the flare parameters
between these two analyses. In particular, the fluences derived in P15 
seem to be systematically lower and the durations are systematically 
shorter than ours. Such differences may be expected from the different 
analysis methods and definitions of the flare parameters adopted in the 
analyses. In P15 the Bayesian block method \citep{1998ApJ...504..405S} 
was adopted to detect flares, while the likelihood fitting method is used 
in our analysis. The flare duration was defined as the first and last 
of the significant change points characterizing the flaring blocks, 
which is different from our definition of the 95\% emission window. 
Typically, the Bayesian block method tends to systematically underestimate 
the count rate and duration of a flare because only detected flaring blocks 
are counted. For flares with significant substructures (e.g., FlrID I3), 
however, the Bayesian block method tends to give longer durations than 
our definition. Compared with P15, we detect 14 more flares and miss 1 
flare (bottom-left panel of Fig. \ref{fig:F_tau_joint}). Most of these 
flares are weak and just above the detection threshold. 

For the ACIS-S/HETG observations, N13 detected 39 flares using the binned 
0th+1st order lightcurves and a different likelihood fitting method. 
Thirty-seven flares are common in their analysis and ours (top-right panel 
of Fig. \ref{fig:F_tau_joint}). Here the fluences of N13 flares have 
been multiplied by a factor of $\sim0.6$ to account for the difference of 
the data selection. Also the pileup effect on N13 flares has been 
corrected. The results of the common sample between these two analyses 
are mostly consistent with each other within the errorbars. Note that a 
few flares in the N13 sample have durations which are comparable to their 
adopted bin width (300 s) and are likely very uncertain. Furthermore, 
the duration of an N13 flare with subflares is defined as the total 
interval between their lowest and highest $2\sigma$ limits, in contrast 
to our definition as the $95\%$ emission-enclosed time interval.
Two N13 flares are not detected, whereas 12 more are detected in our 
re-analysis of the ACIS-S/HETG0 data. All these flares are faint (see 
the bottom-right panel of Fig. \ref{fig:F_tau_joint}), sensitive to the 
difference in the analysis methods. An alternative method, the Bayesian 
block algorithm, was tested in N13, which returned a set of 45 flares, 
whereas P15 detected 37 flares using the 0th order data of the 2012 
XVP observations with the same method.

In both analyses of the ACIS-I and -S/HETG0 data, we detect a number 
of new flares with relatively low fluences and long durations. The 
detection of these flares are justified with the expected sensitivity 
(the incompleteness curve) of our improved detection method. 

We also perform fittings to the fluence-duration correlation for the P15 
and N13 flare samples, with only the fluence errors considered, as shown
in Table \ref{table:fit}. The correlation slope $\beta$ is much smaller 
than the values from the fitting to our samples (\S4.1). This is mostly 
due to our inclusion of the uncertainties in the duration measurements, 
which affects the weights of individual flares in the $\chi^2$ fitting. 
\citet{2015ApJ...810...19L} studied the flare statistics through a joint 
fitting to the count rate distribution and structure function of the XVP
lightcurve, and found that $\beta>1.8$ at the 95\% confidence level, 
consistent with our results. 

P15 studied the flare rate in the past 15 years with the {\it Chandra}
and {\it XMM-Newton} observations. They found that the flare rate before 
2013 did not change significantly, which is consistent with our result. 
However, the bright flare rate showed evidence for a significant increase 
by a factor of $\sim10$ starting from August 2014, which happened to be 
about half year after the G2's pericenter passage \citep{2013ApJ...774...44G}. 
We can check if a similar enhancement occurred after the S2 star's 
pericenter passage. The 9 ACIS-I observations were taken in 2002 (with 
ObsID from 2951 to 3665), covering the period of the passage around 
2002.33. In total we find 12 flares in 545.6 ks exposure, which corresponds 
to a flare rate of $\sim1.9$ day$^{-1}$, well consistent with the 12 years' 
average. The number of bright flares (each with $>120$ cts, corresponding 
to the definition of bright flares for the ACIS-I observations in P15) 
in the 2002 observations is 2. Compared with the total number of bright 
flares of 7 in all the ACIS-I observations, no significant anomaly is shown 
in the bright flare rate, considering the exposure fraction (0.37) of the
2002 observations. The $3\sigma$ upper limit of the expected number of 
bright flares when detecting 2 flares is $10.9$, which corresponds to a 
rate of $1.7$ day$^{-1}$, about 4 times higher than the average one 
(7/1.47~Ms$^{-1}$~$\approx$~0.4~day$^{-1}$) in all the ACIS-I observations. 
A 10 times enhancement of the bright flare rate associated with the S2's 
pericenter passage, similar to that found in P15 during the G2's passage, 
can thus be excluded.

All these differences demonstrate the importance to enlarge the sample 
size and dynamic range of X-ray flares, and to carefully address their 
detection incompleteness and bias, as well as the parameter measurement 
uncertainties, as are achieved in the present work. As a result, we are 
now in a good position to provide a significantly improved assessment of 
the statistical properties of the flaring phenomenon of Sgr A$^\star$.

\section{Summary}

In this work we have systematically analyzed the {\it Chandra} lightcurves
of Sgr A$^\star$ from 1999 to 2012, including 46 ACIS-I and 38 ACIS-S/HETG0 
observations. Our analysis uses a combination of the unbinned maximum 
likelihood fitting algorithm and the MCMC method, which enables us to 
maximize the use of information in the data and to estimate the uncertainties
in all flare parameter measurements. This forward-fitting procedure also
allows us to account for the pileup effect directly in the lightcurve 
modeling. This removes a big uncertainty in such a correction when 
it is applied to a flare without accounting for its shape. We have further 
carried out simple variability analyses of the detected flare rate and the 
quiescent emission. Our major results and conclusions are as follows:

\begin{itemize}

\item We detect 33 (49) flares in the ACIS-I (-S/HETG0) data. The bulk of 
these flares overlap with those reported in existing studies (P15 and N13). 
We give not only improved measurements of the parameters (including the
first error estimates for the flare durations), but also a careful 
characterization of the detection incompleteness and bias. 
Our detections further reveal a number of faint flares (some of which 
have unusually long durations), which are missed in the existing studies. 
This discovery becomes possible due to the improved fitting method 
which allows for a more complete survey of the flare parameter space. 

\item Our analysis confirms the correlation of the flare fluence versus 
the duration, and gives new estimates of the mean relation $\log F=
\log(\alpha)+\beta\log(\tau)$, for both ACIS-I and -S/HETG0 flare samples. 
A direct fitting to the measurements, accounting for the uncertainties 
in both $\log F$ and $\log\tau$, gives $\beta\sim2.5-2.8$ for the two
samples, which is significantly larger than that obtained with the N13 
and P15 samples, but is consistent with that in \citet{2015ApJ...810...19L}. 
We further estimate an intrinsic dispersion of $\log(\tau)\sim 0.20-0.26$ 
around this relation. 

\item We do not find any significant long-term variation in the quiescent 
emission or the flare rate. In particular, no enhanced flare rate or 
quiescent emission is evident during the pericenter passage of the S2 
star in 2002 \citep{2002Natur.419..694S}. The $3\sigma$ upper limit
to the bright flare rate around S2's pericenter passage is about 4 times 
of the average rate for the entire ACIS-I flare sample, which rules out 
a factor of $\sim10$ increase in the bright flare rate similar to that 
observed after the G2's passage (P15). The intrinsic RMS variation of the 
quiescent emission among the ACIS-I (-S/HETG0) observations is 14\% (6\%), 
part of which is expected from the number fluctuation of weak flares below 
our detection limits. The mean quiescent emissions and flare rates in 
the ACIS-I and -S/HETG0 observations are consistent with 
each other, when their effective area difference is accounted for. 

\item The flares seem to cluster on the time scale of $20-70$ ks, 
particularly significant in the ACIS-S/HETG0 sample. This short-term 
clustering, as well as the non-variation of the long-term flare rate, 
suggests that the production of the flares may be described by a 
piecewise-deterministic Markov process (similar to that used to 
characterize earthquakes; \citealt{Davis1984}), which deserves further 
in-depth analysis and modeling.

\item The detection incompleteness and the redistribution matrix of 
the measured parameters are obtained through Monte Carlo simulations,
enabling further statistical studies of the flares. 

\end{itemize}

In a subsequent paper, we will present a rigorous statistical analysis
of the flares (e.g., their fluence-duration relation), based on the
characterization of the detection incompleteness and bias, as well as the 
error measurements of flare parameters as presented here. We will further 
examine the profiles (substructures and asymmetry) and phase-resolved 
spectra of the flares to shed light into their nature. 

\section*{Acknowledgments}

We thank Shawn Roberts for helpful comments on the manuscript, and
the referee for a careful reading and a useful review report.

\bibliographystyle{mn2e}
\bibliography{refs}

\begin{appendix}

\section{Lightcurves of all the detected flares}

\begin{figure*}
\centering
\includegraphics[width=0.34\columnwidth]{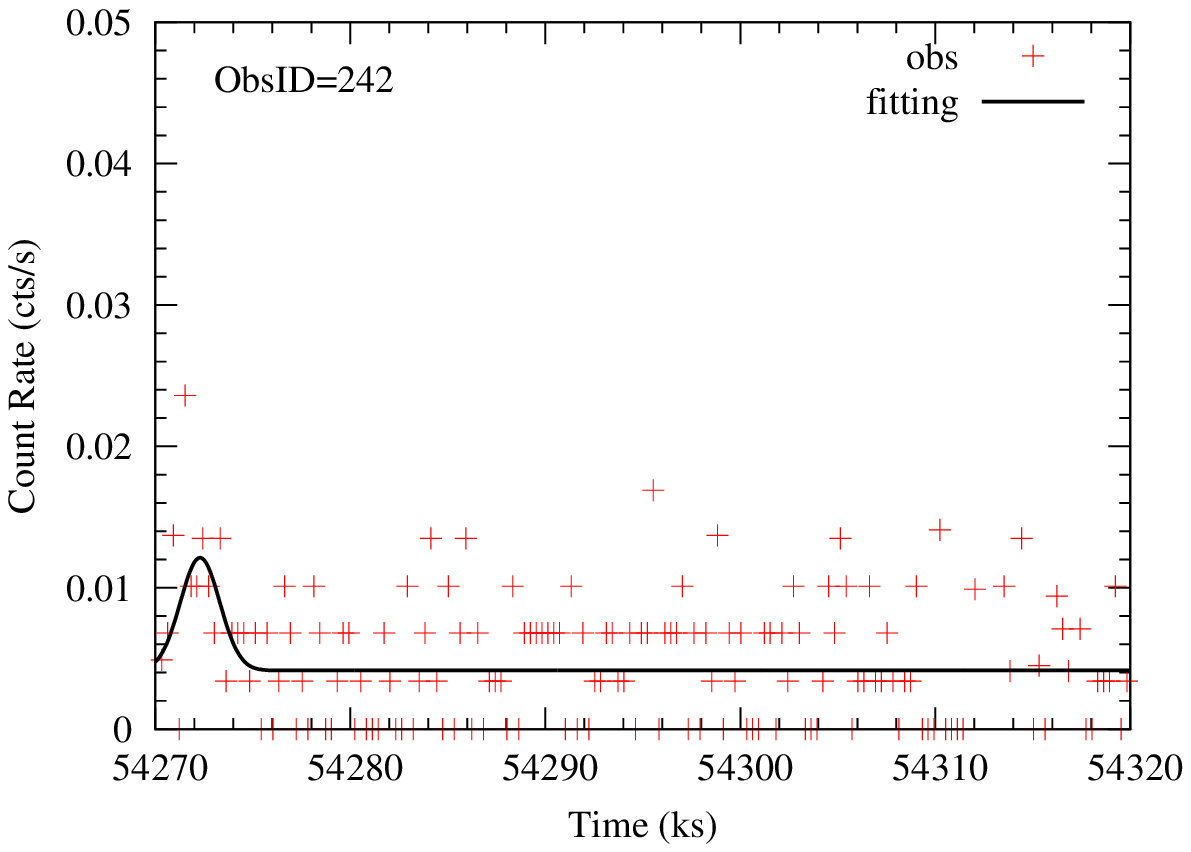}
\includegraphics[width=0.34\columnwidth]{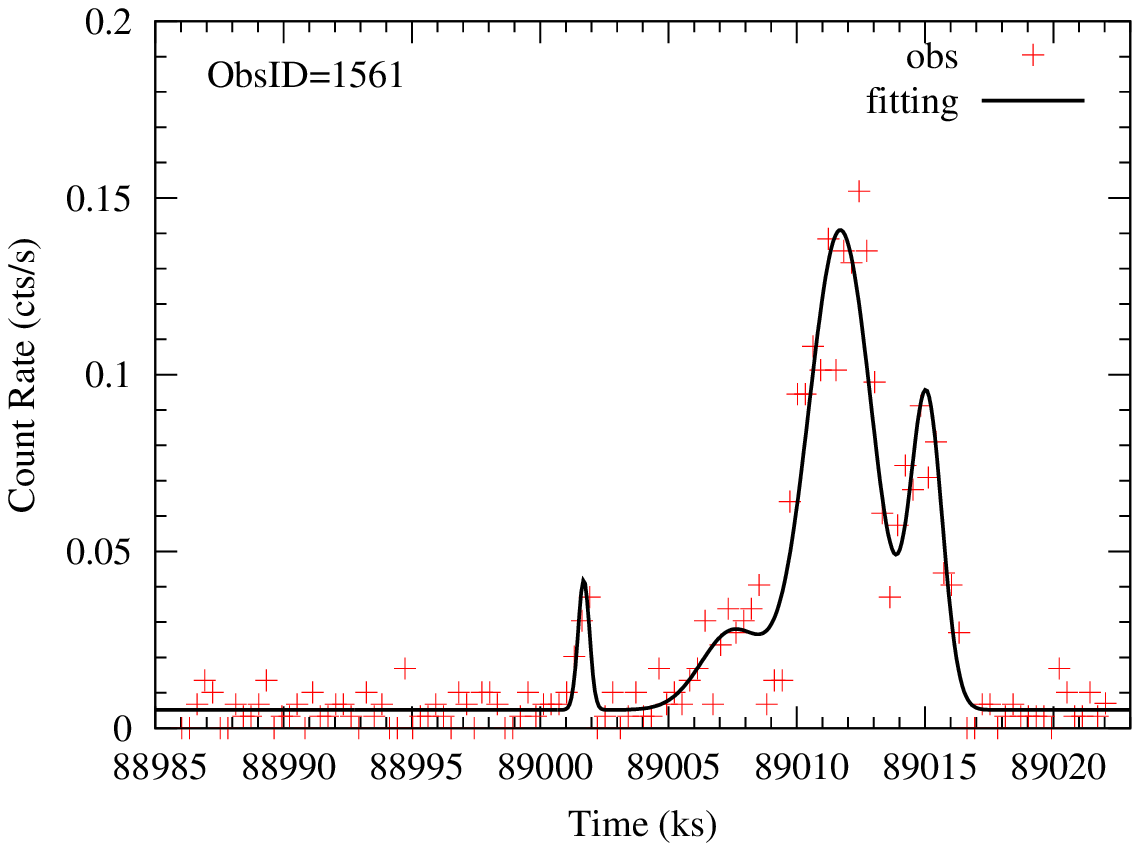}
\includegraphics[width=0.34\columnwidth]{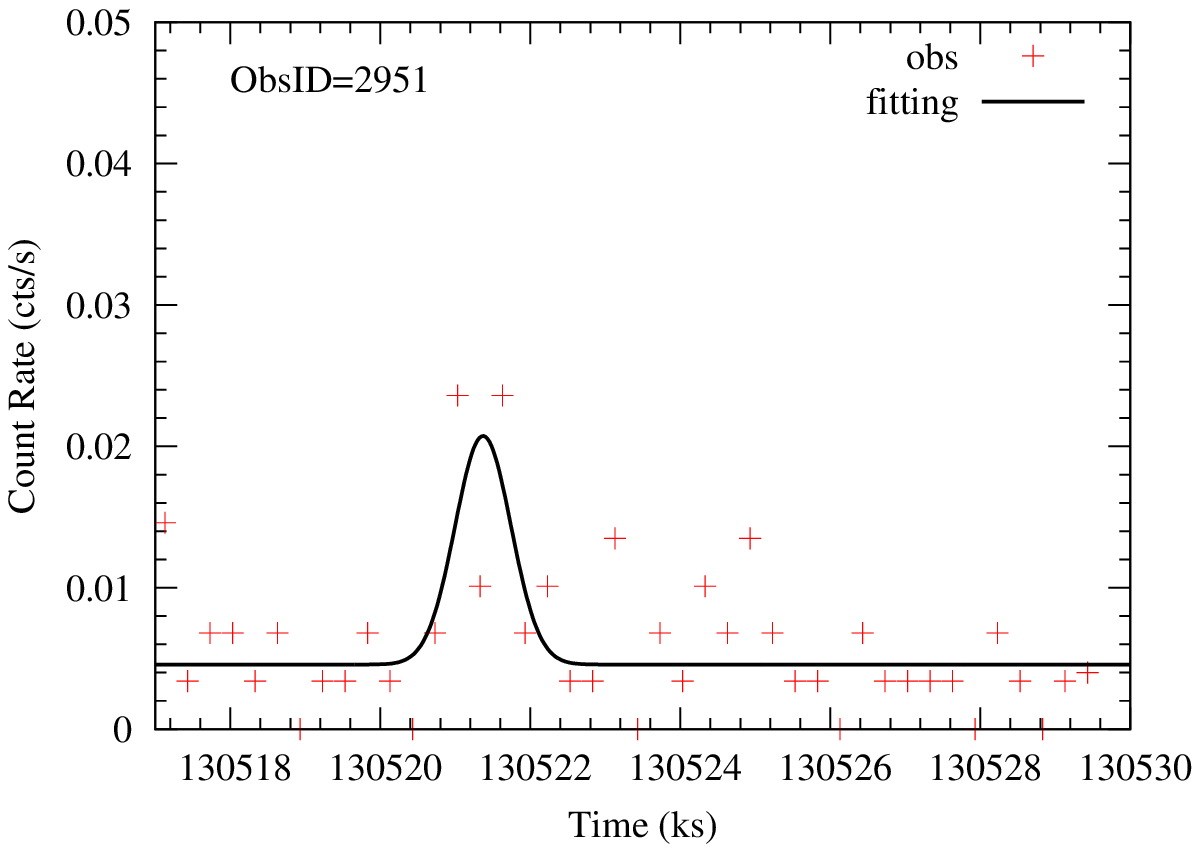}
\includegraphics[width=0.34\columnwidth]{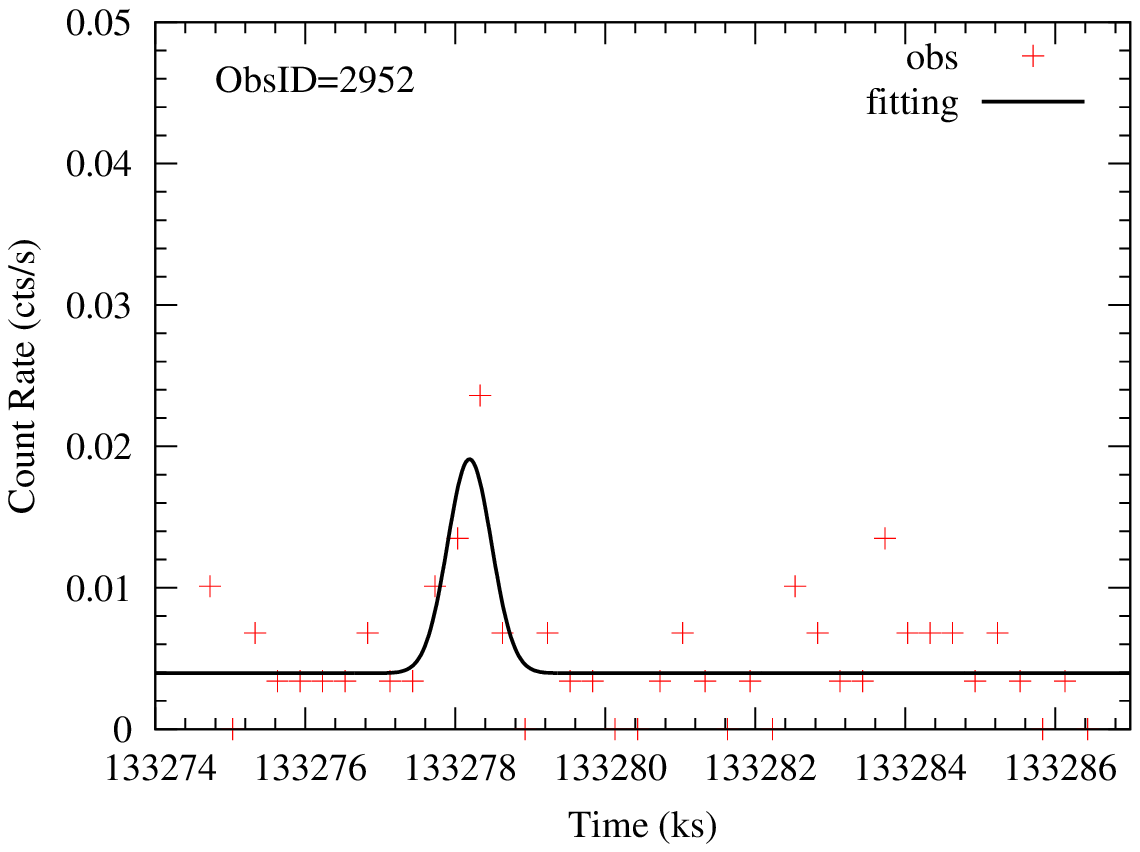}
\includegraphics[width=0.34\columnwidth]{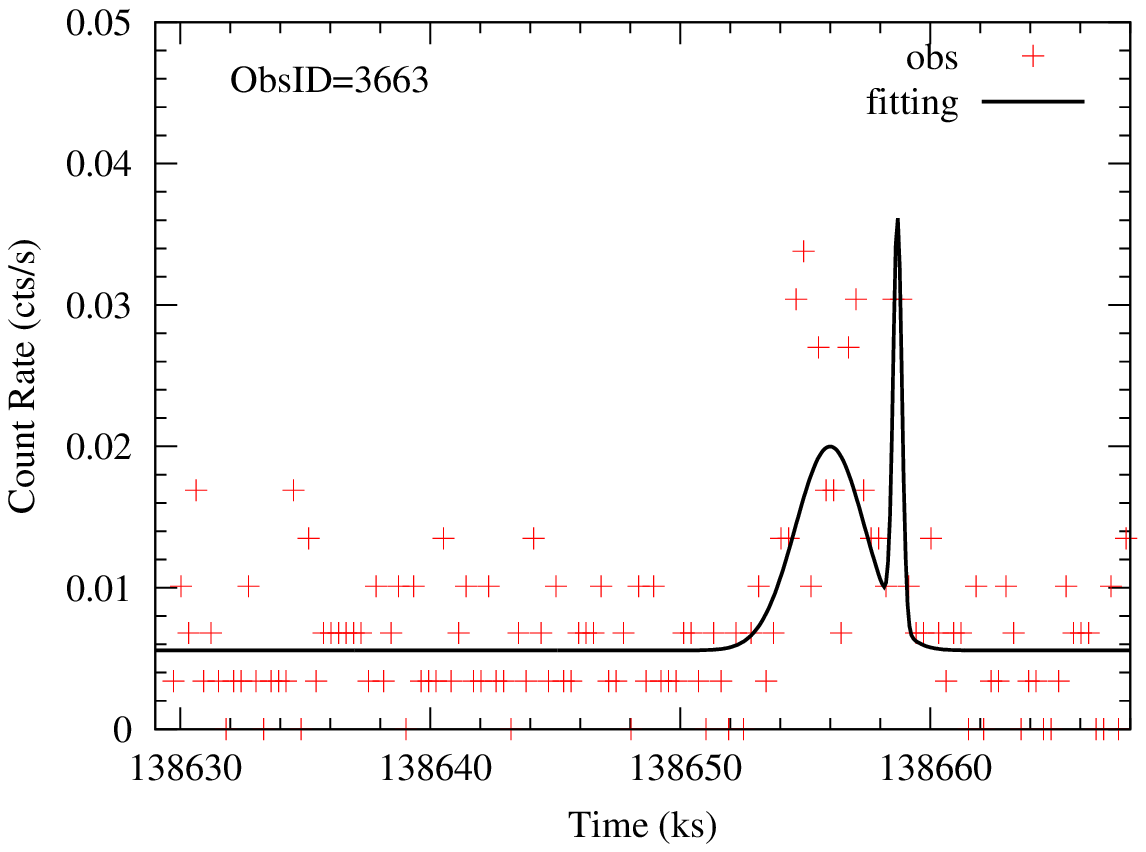}
\includegraphics[width=0.34\columnwidth]{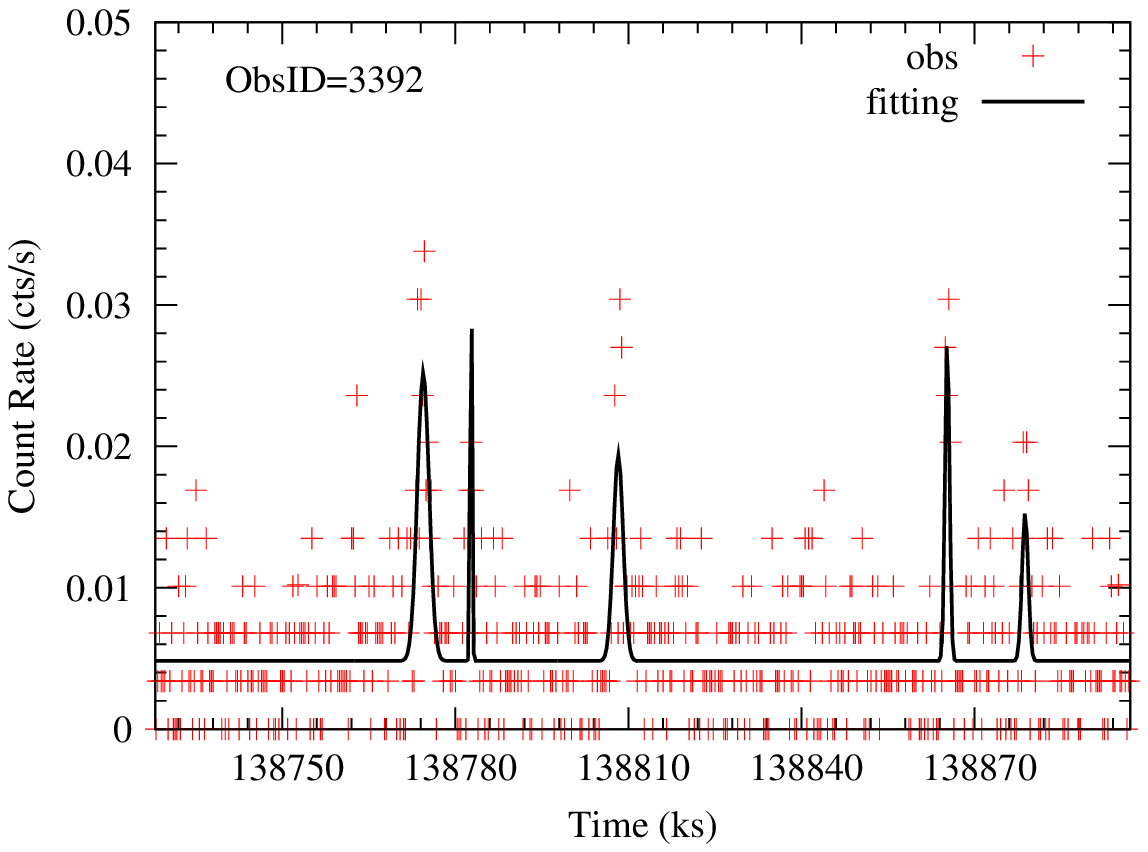}
\includegraphics[width=0.34\columnwidth]{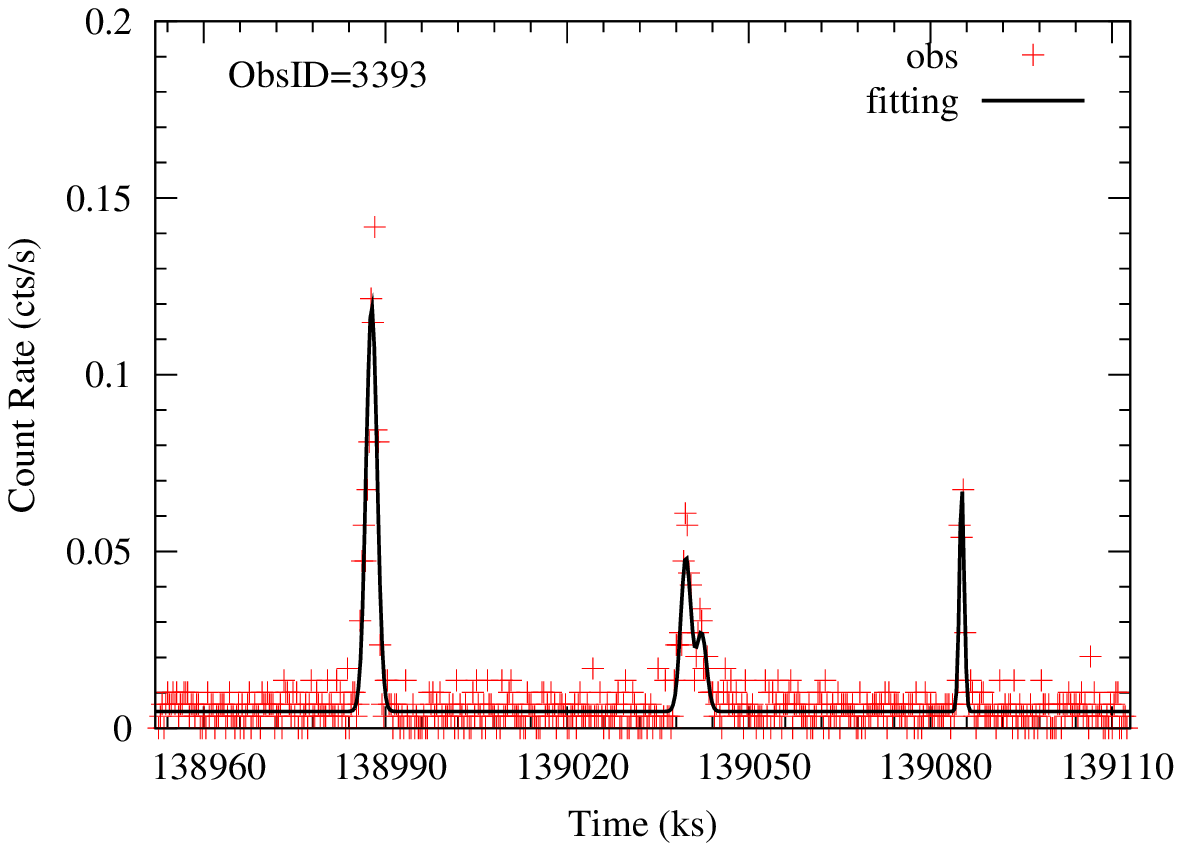}
\includegraphics[width=0.34\columnwidth]{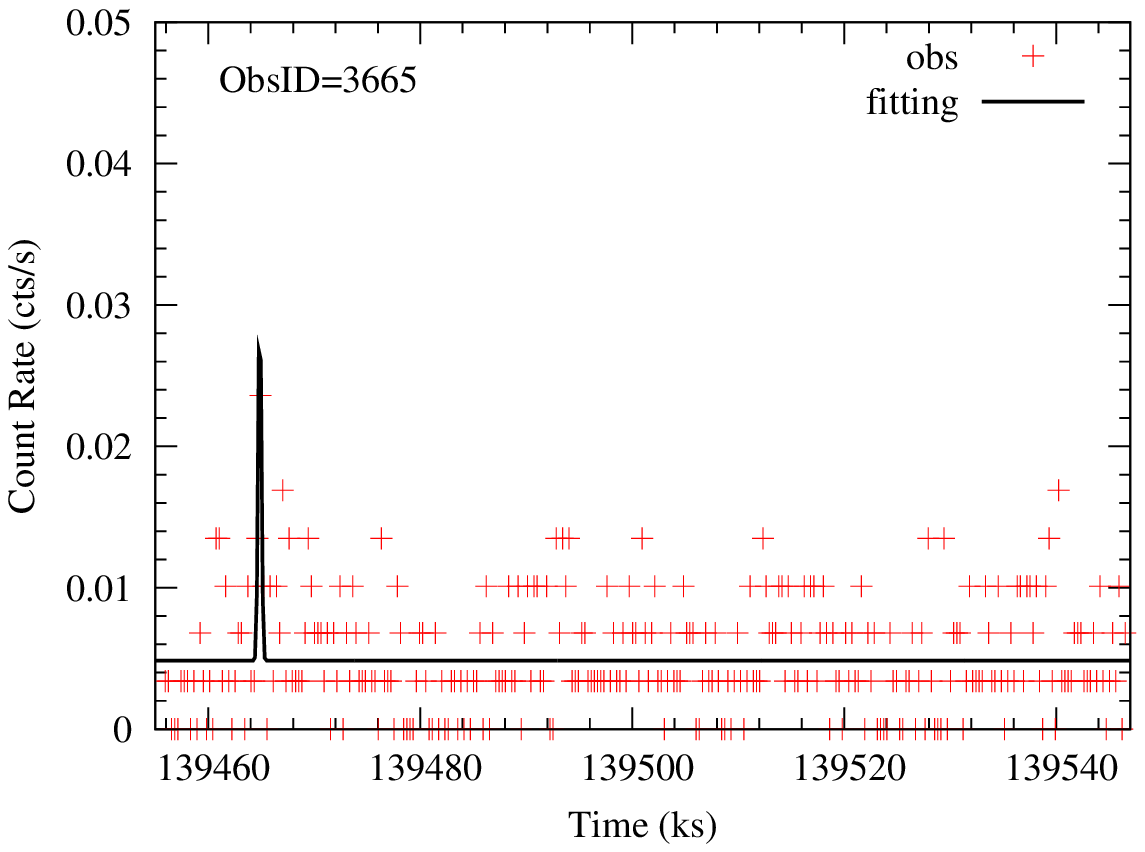}
\includegraphics[width=0.34\columnwidth]{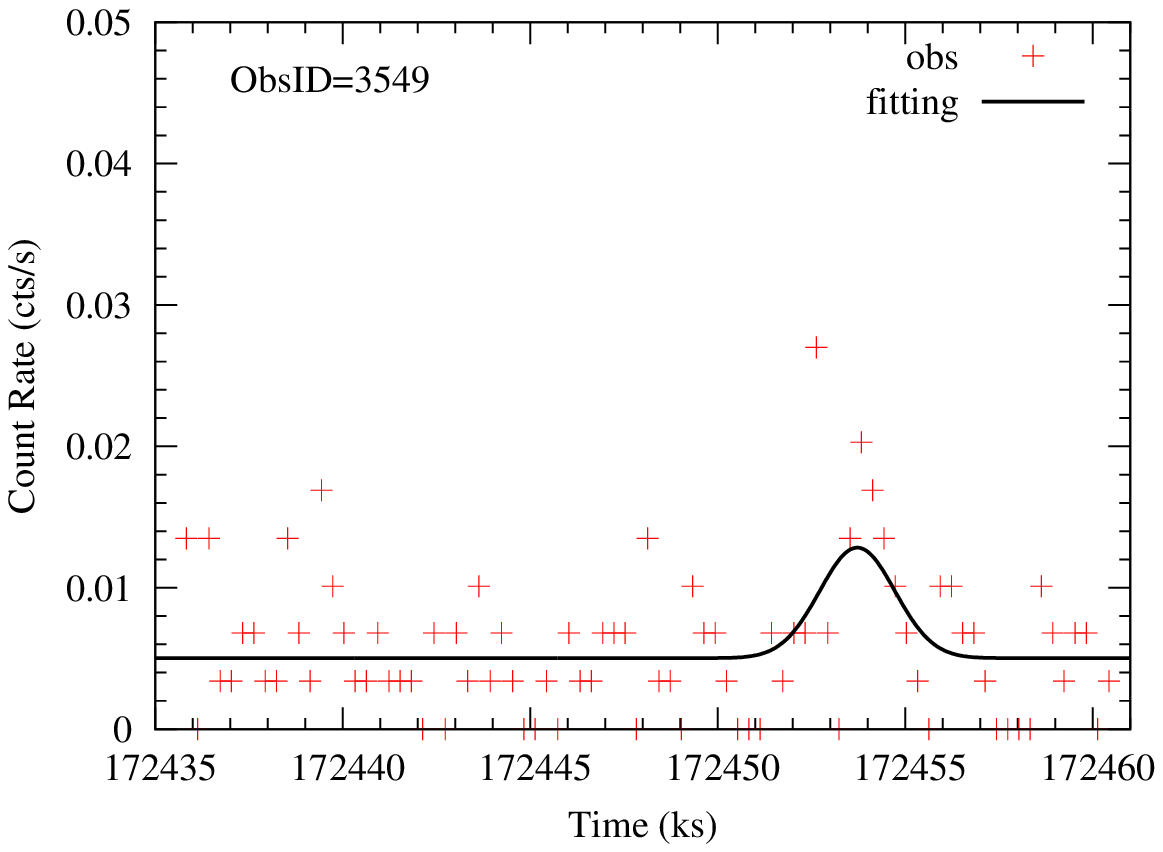}
\includegraphics[width=0.34\columnwidth]{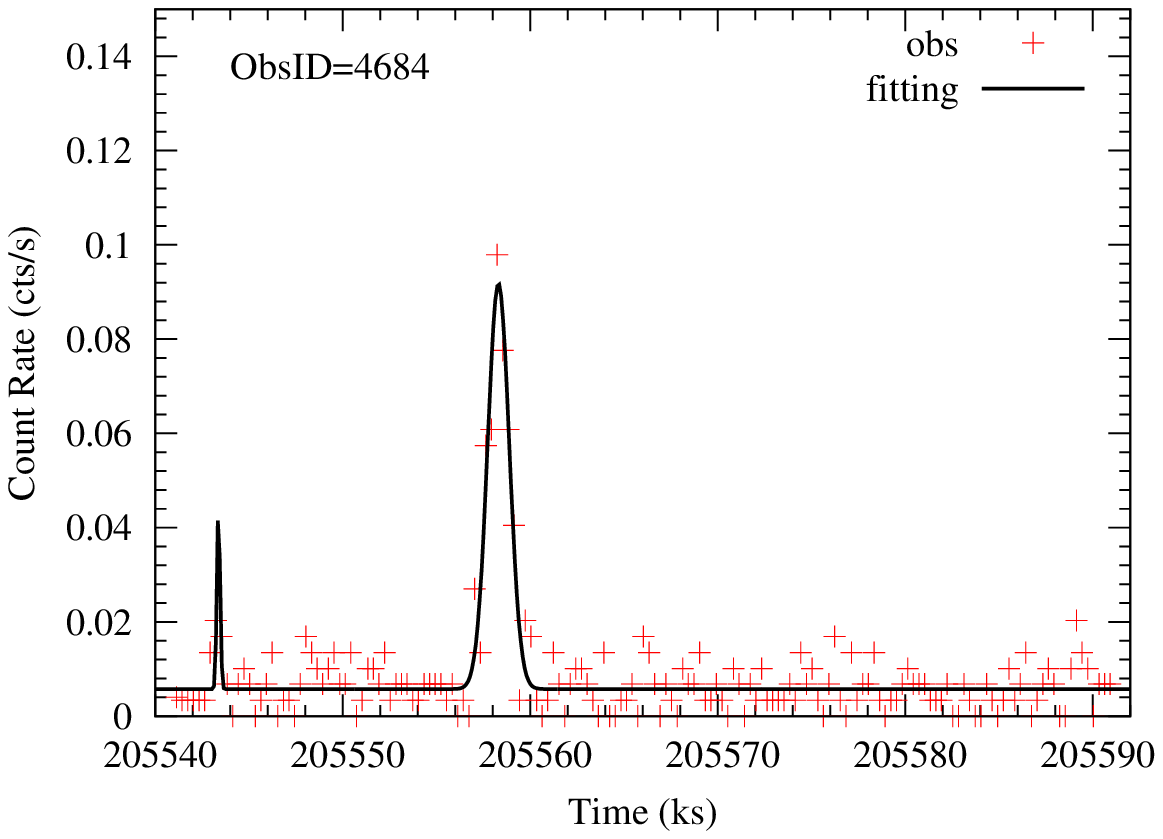}
\includegraphics[width=0.34\columnwidth]{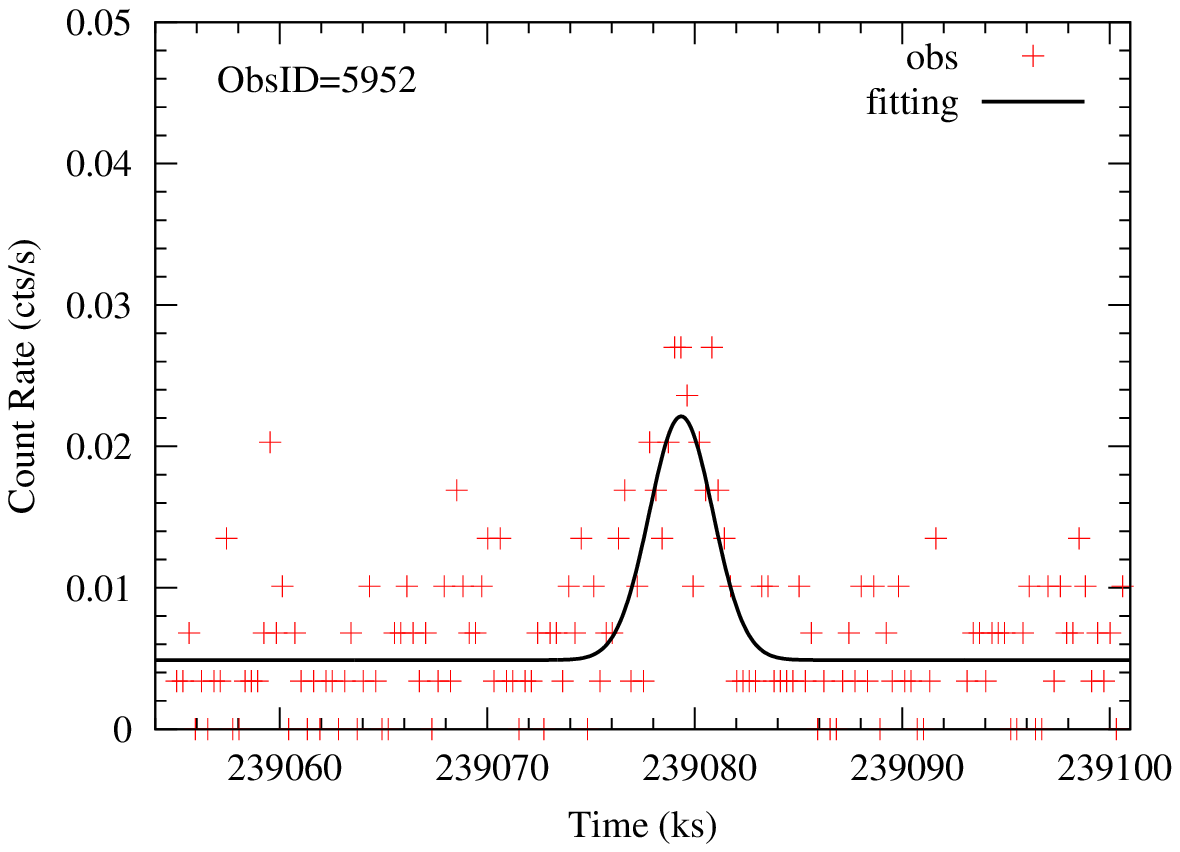}
\includegraphics[width=0.34\columnwidth]{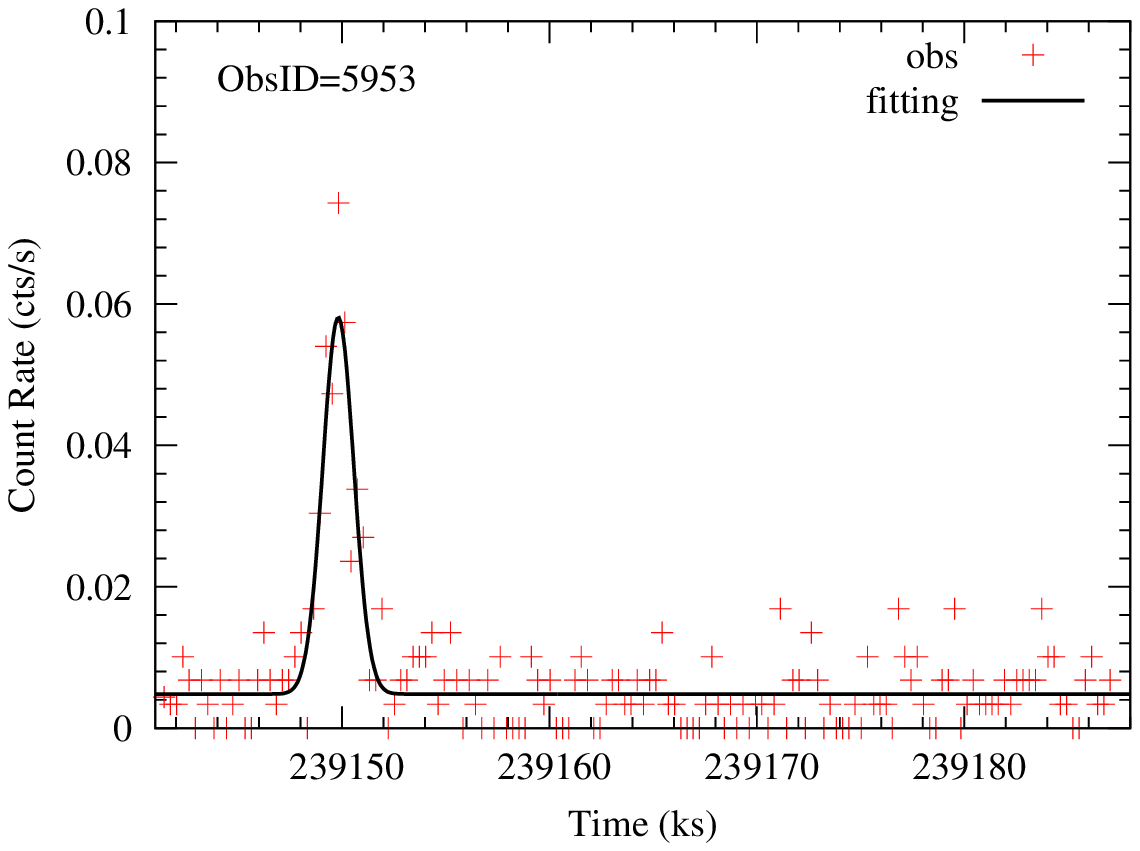}
\includegraphics[width=0.34\columnwidth]{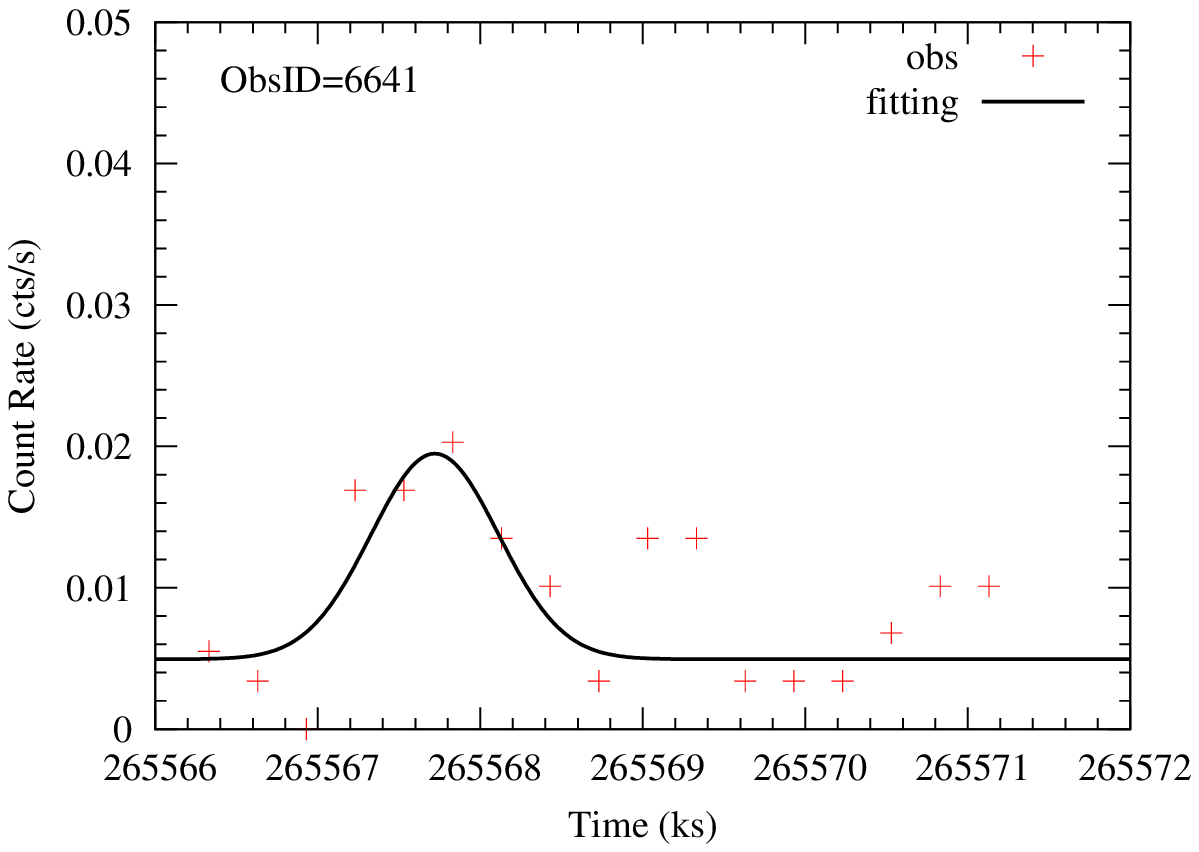}
\includegraphics[width=0.34\columnwidth]{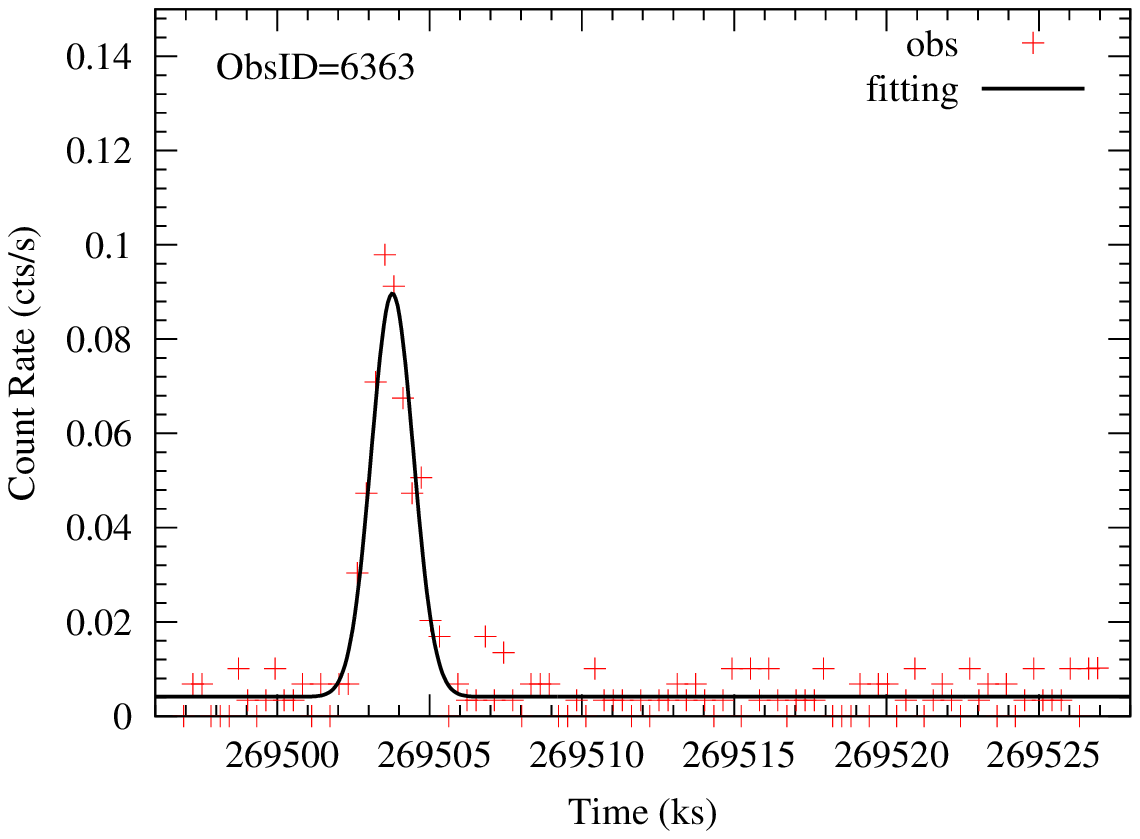}
\includegraphics[width=0.34\columnwidth]{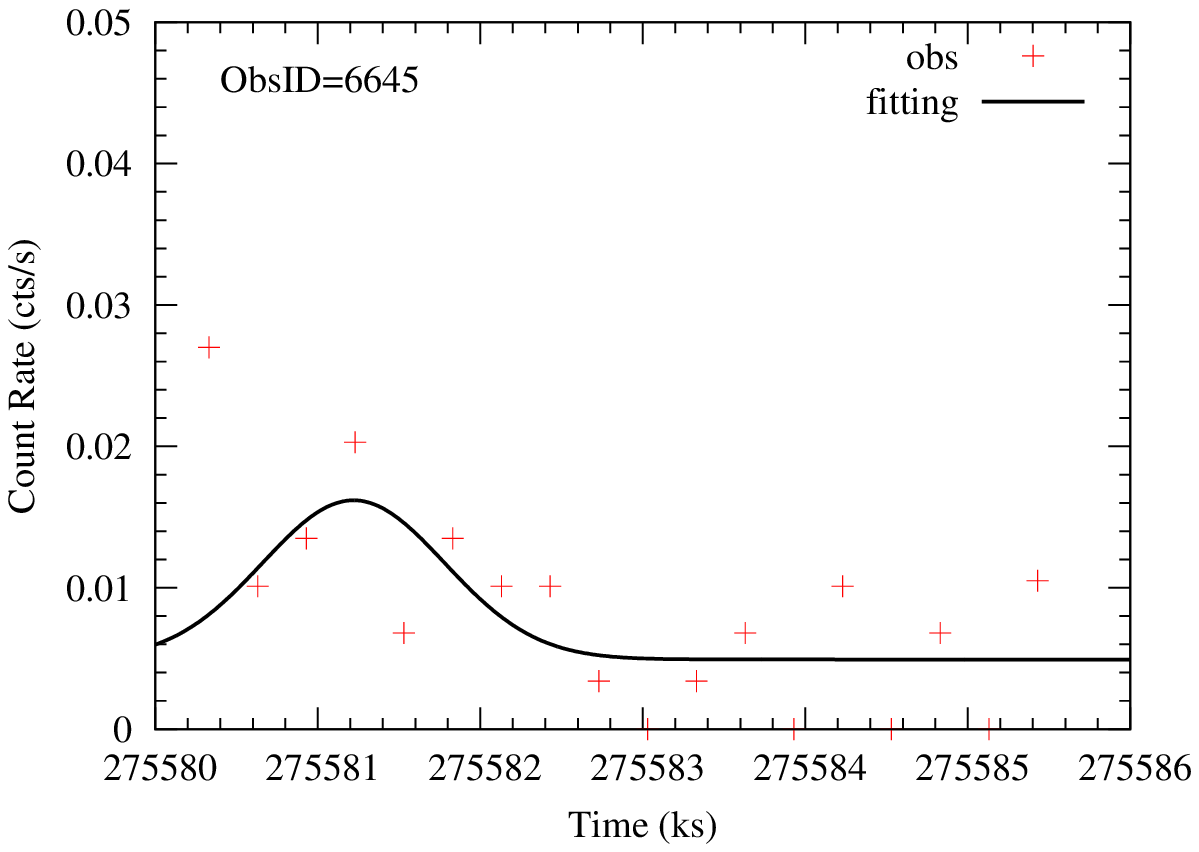}
\includegraphics[width=0.34\columnwidth]{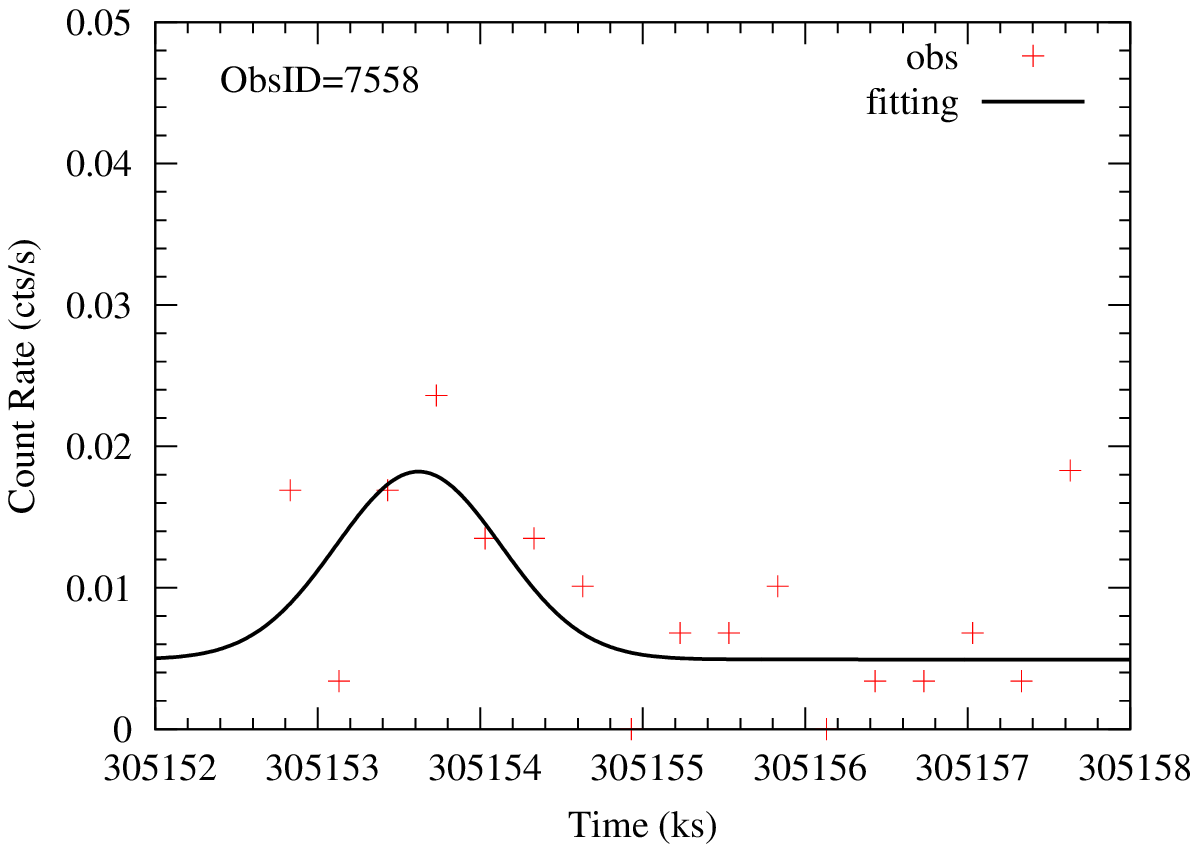}
\includegraphics[width=0.34\columnwidth]{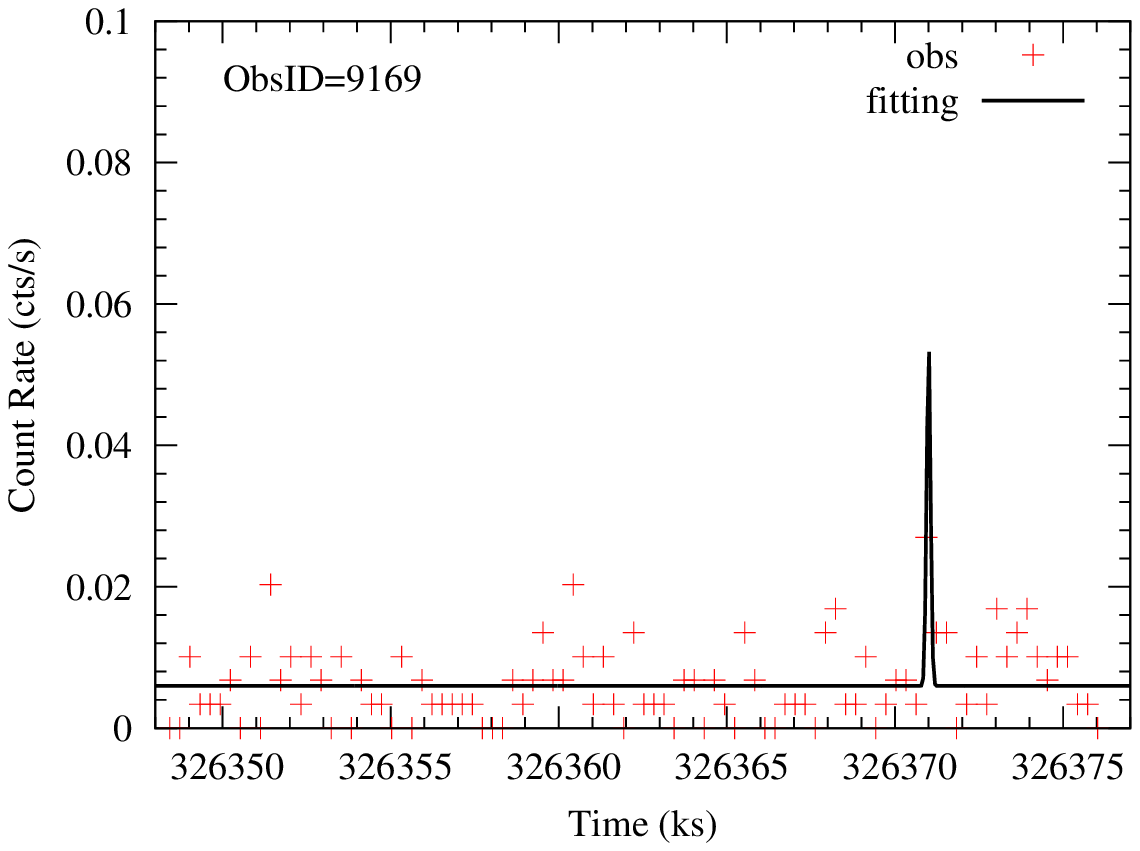}
\includegraphics[width=0.34\columnwidth]{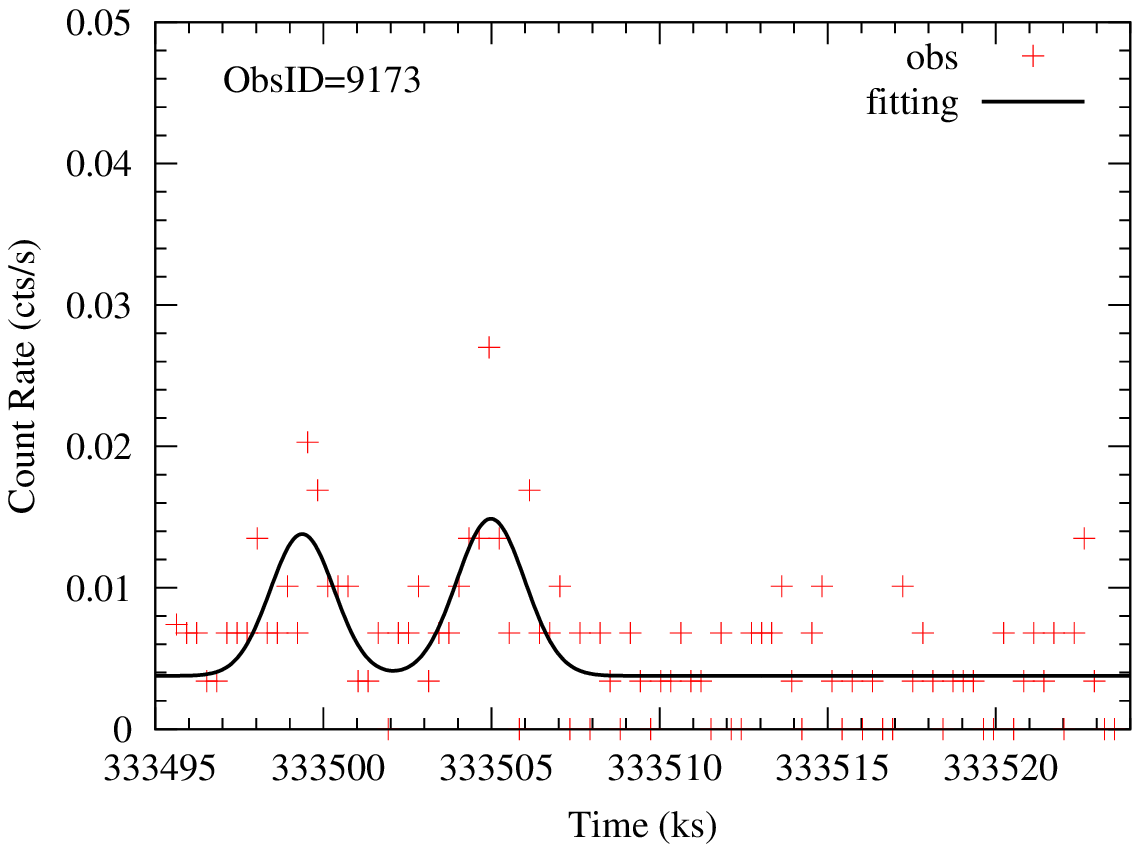}
\includegraphics[width=0.34\columnwidth]{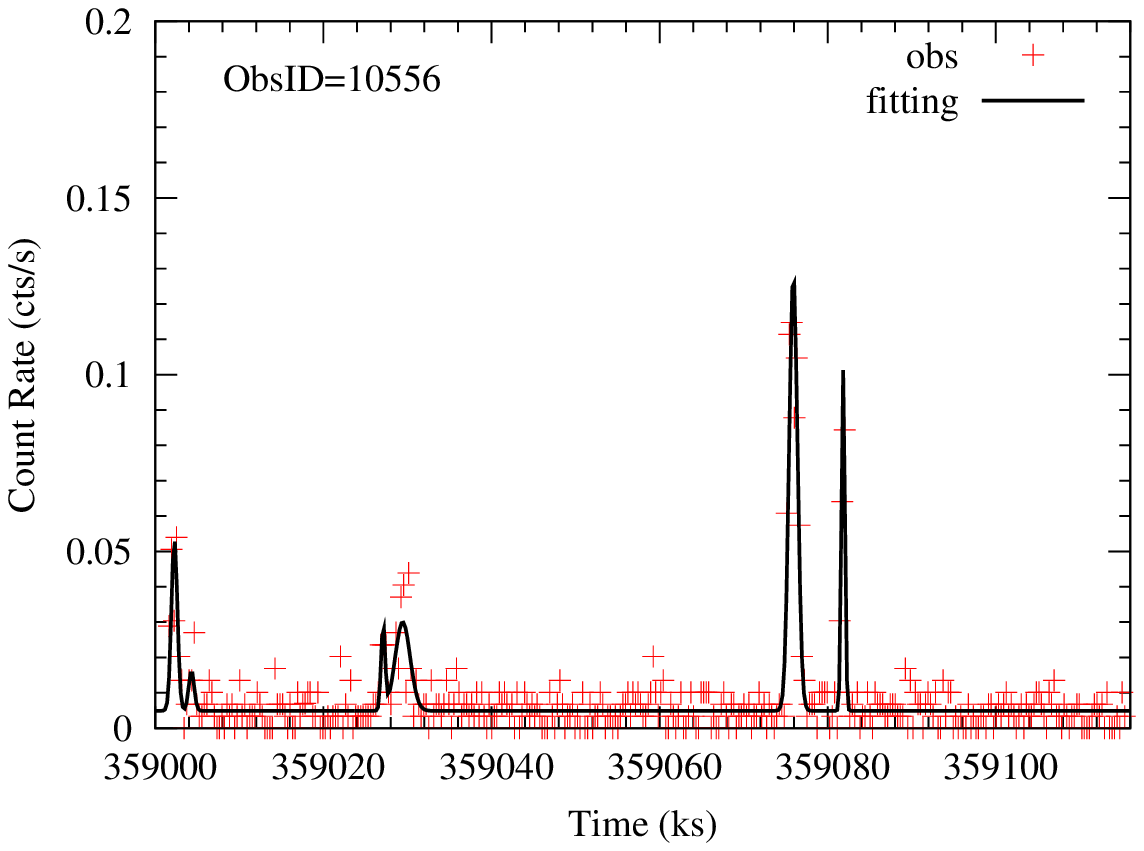}
\includegraphics[width=0.34\columnwidth]{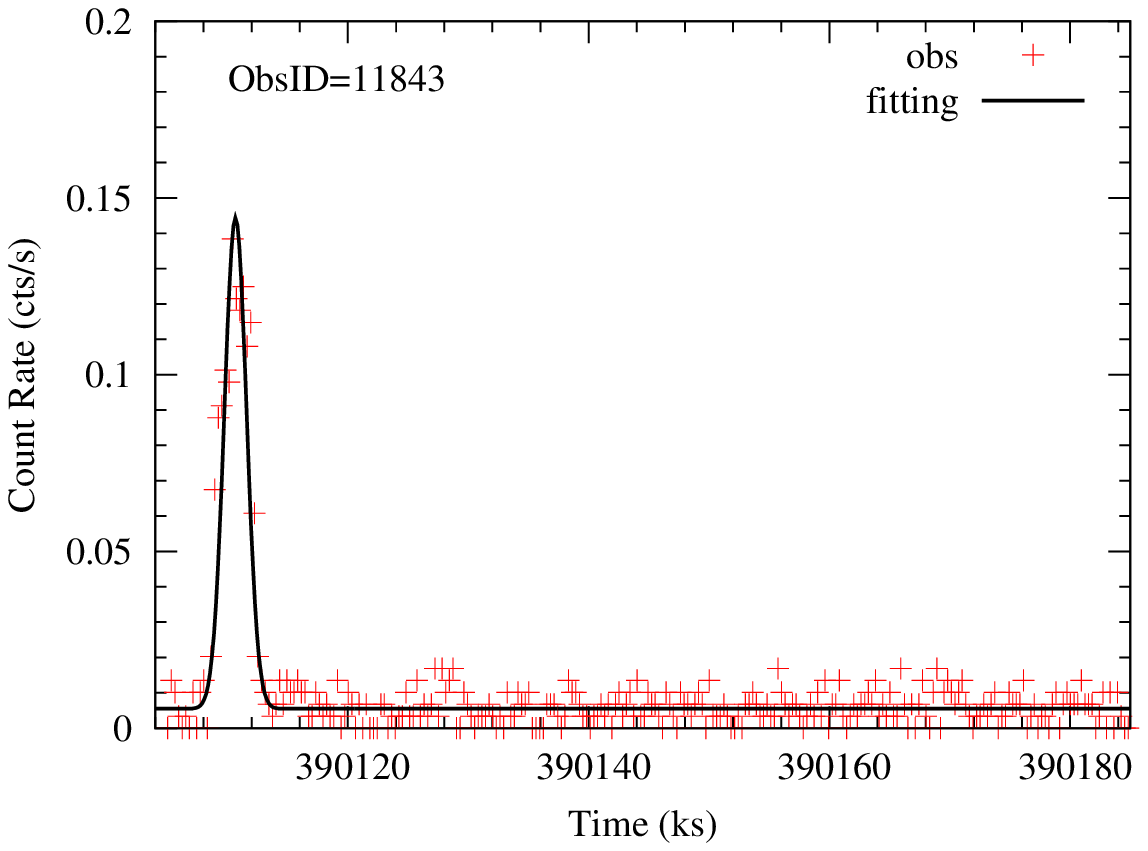}
\includegraphics[width=0.34\columnwidth]{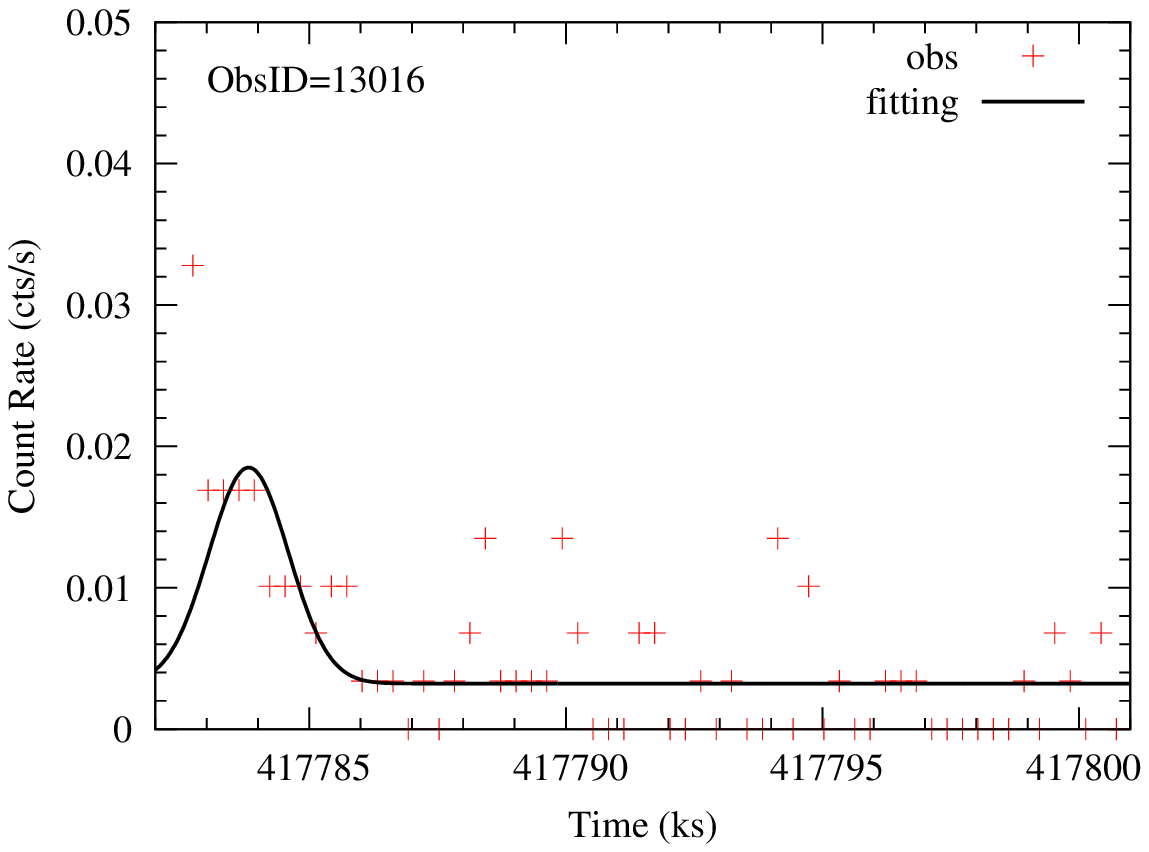}
\caption{Lightcurves of the detected flares in the {\it Chandra} ACIS-I 
observations, compared with the best-fitting results.
}
\end{figure*}

\begin{figure*}
\centering
\includegraphics[width=0.34\columnwidth]{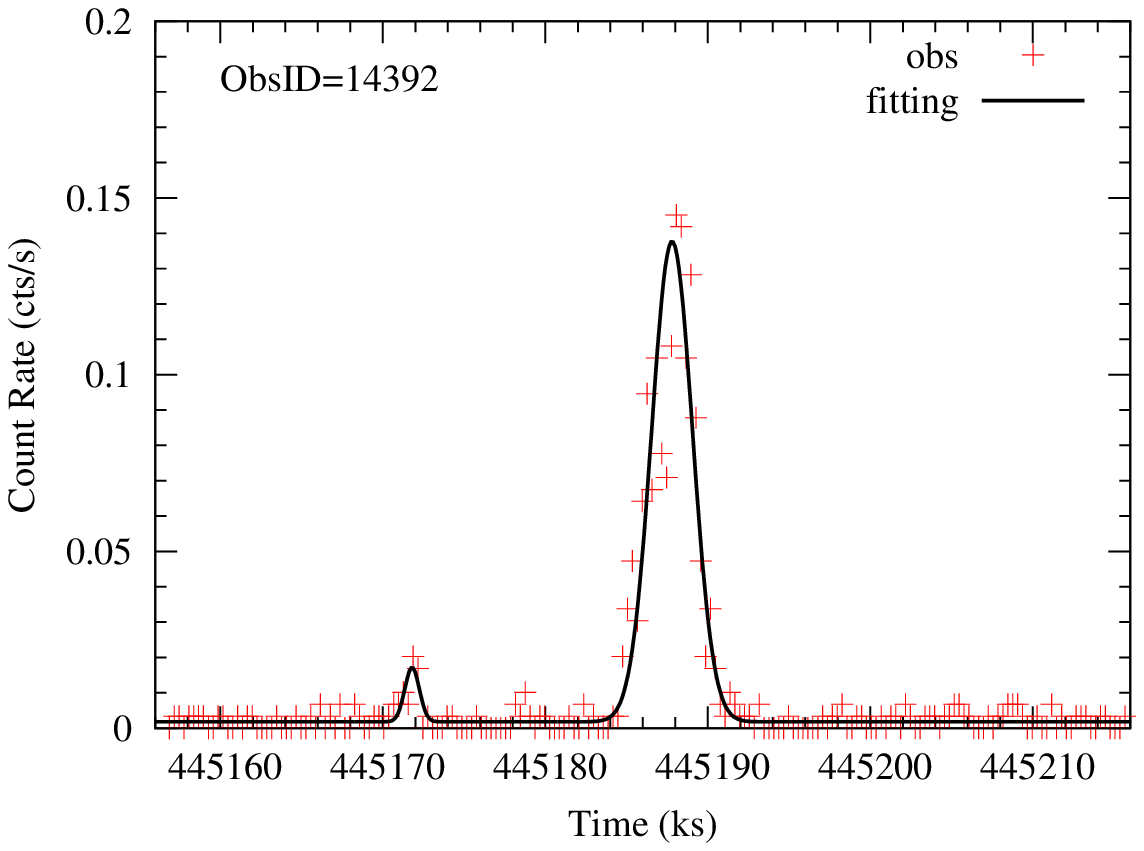}
\includegraphics[width=0.34\columnwidth]{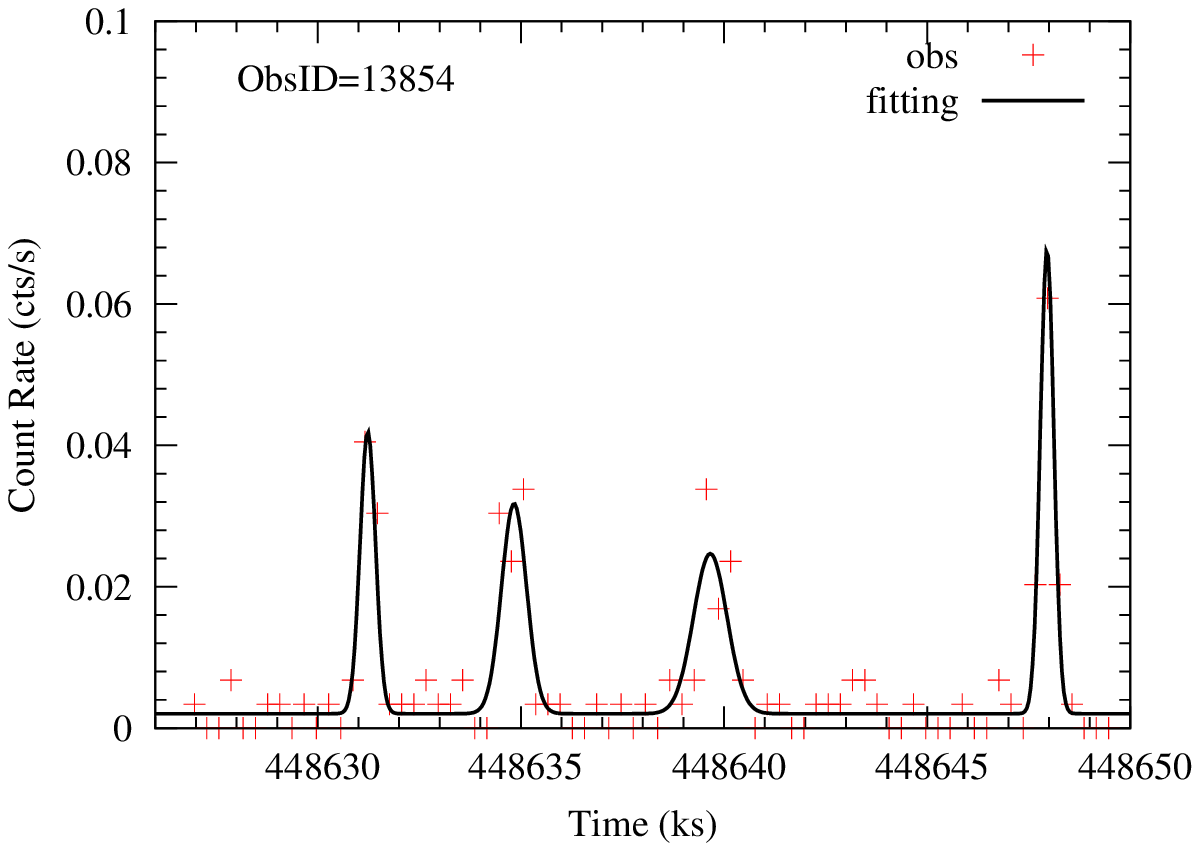}
\includegraphics[width=0.34\columnwidth]{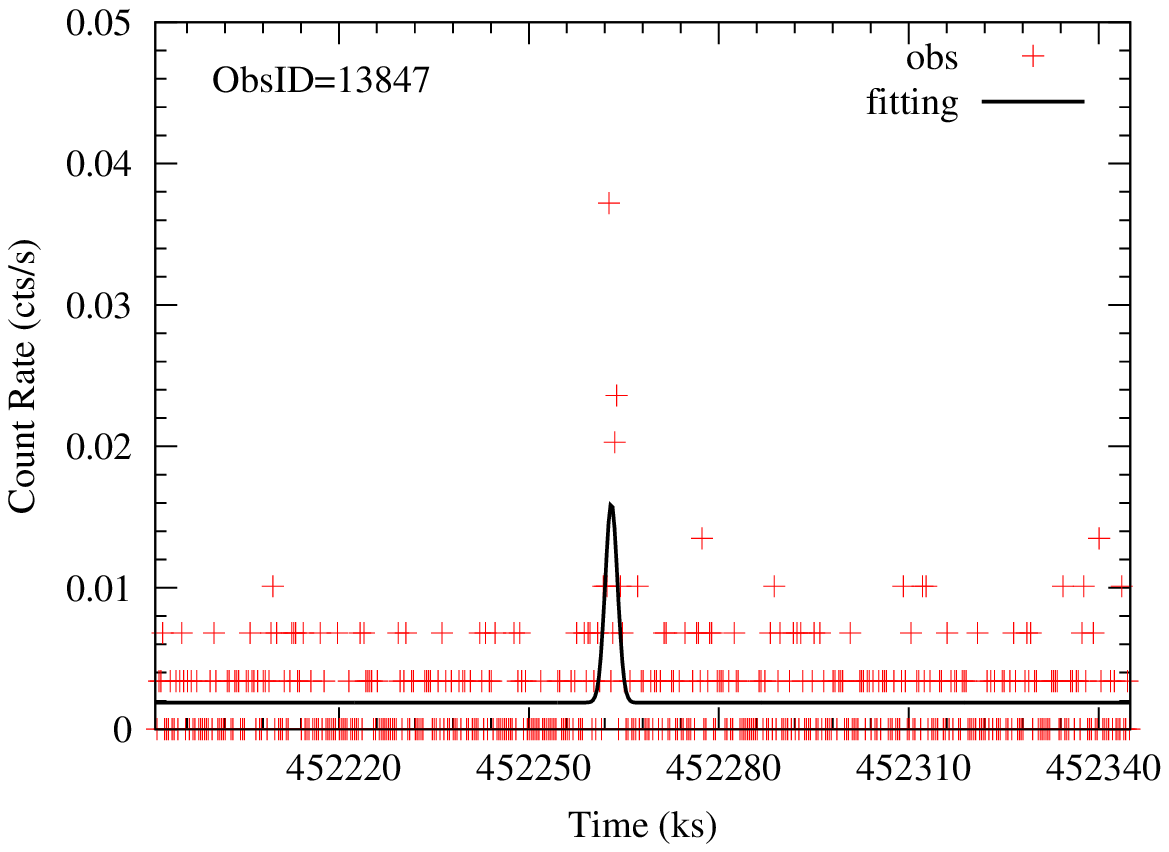}
\includegraphics[width=0.34\columnwidth]{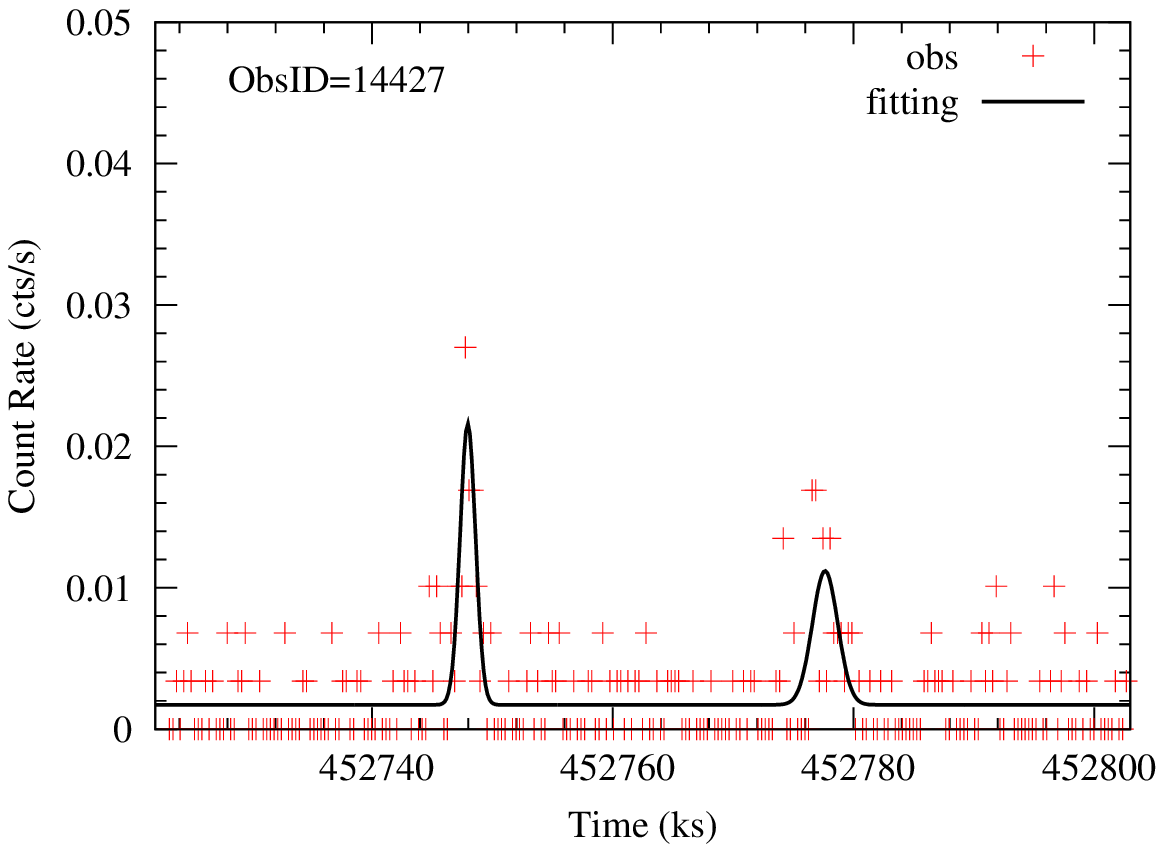}
\includegraphics[width=0.34\columnwidth]{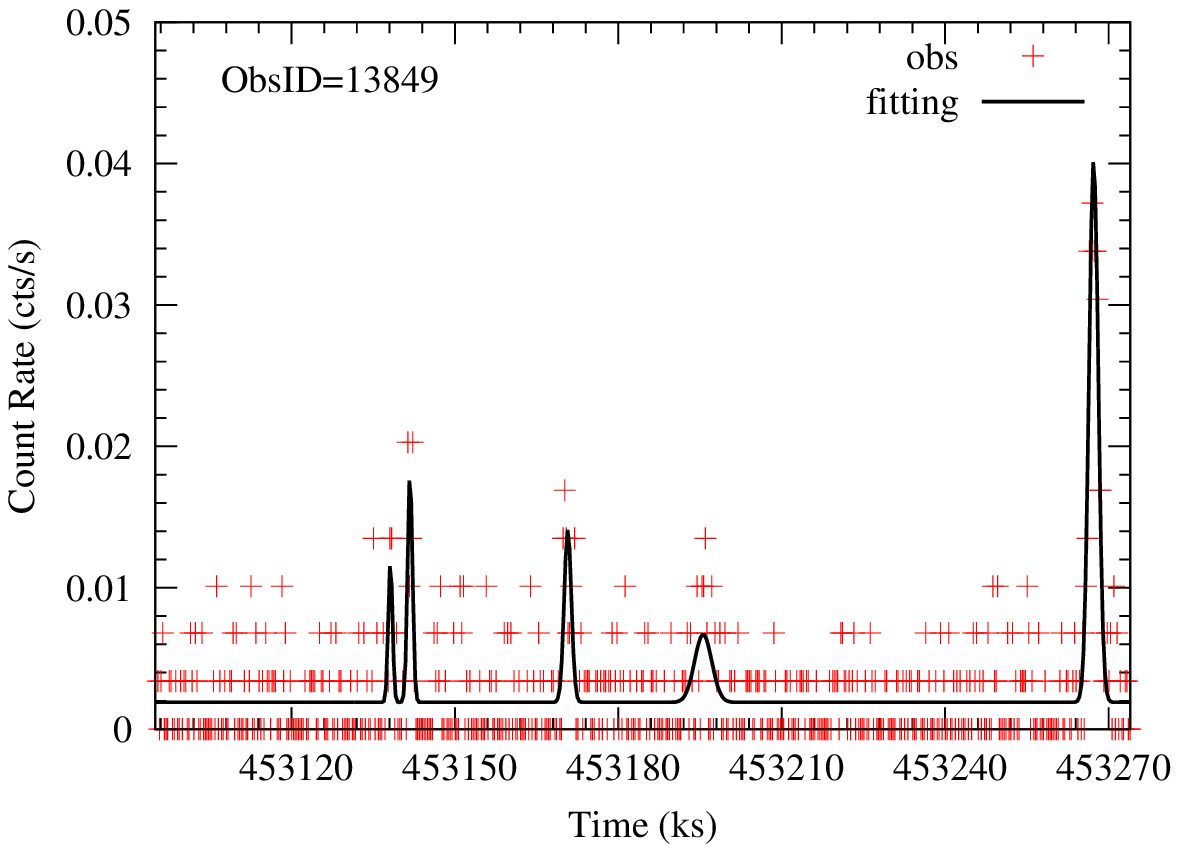}
\includegraphics[width=0.34\columnwidth]{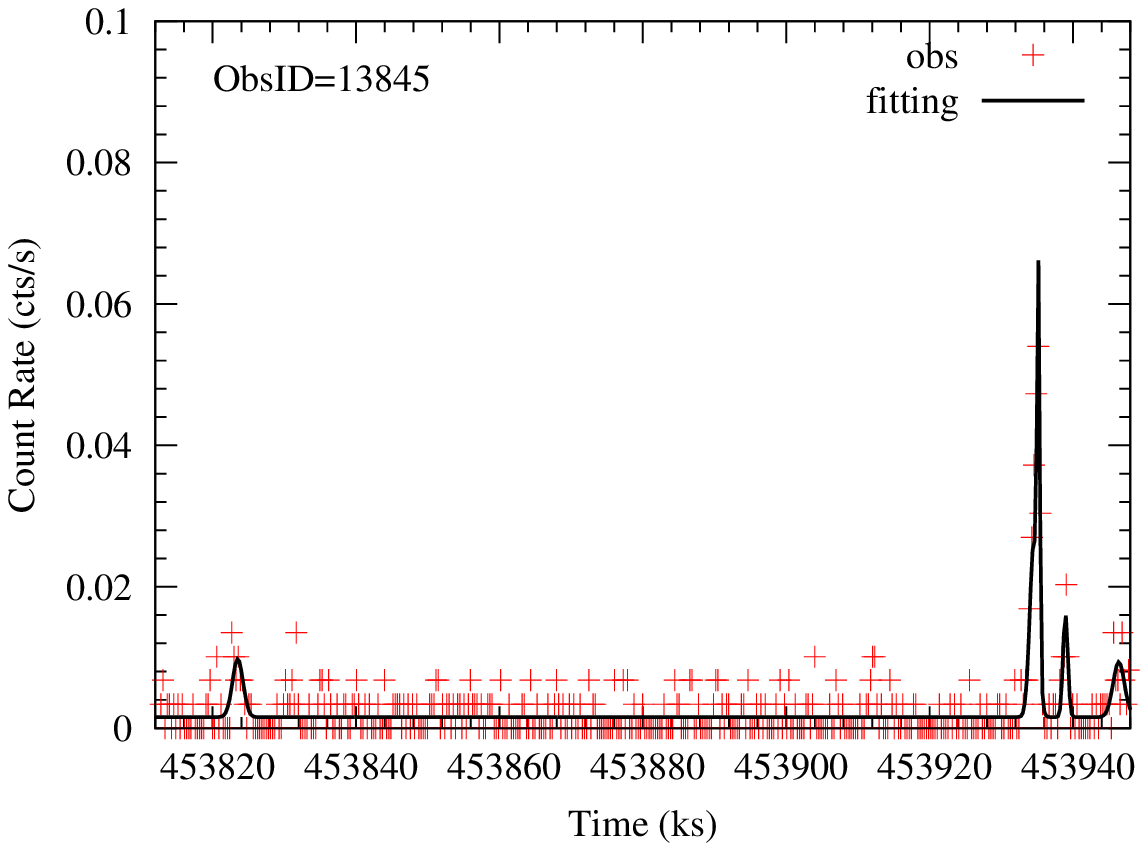}
\includegraphics[width=0.34\columnwidth]{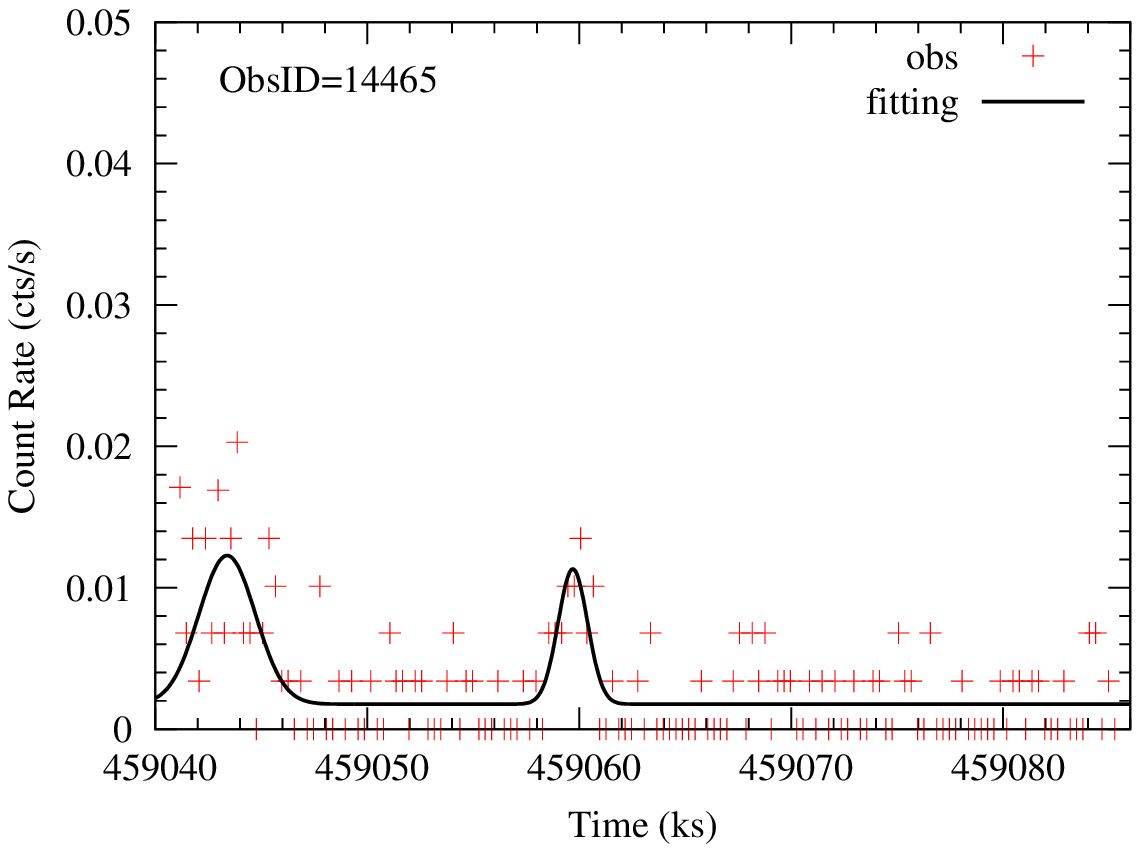}
\includegraphics[width=0.34\columnwidth]{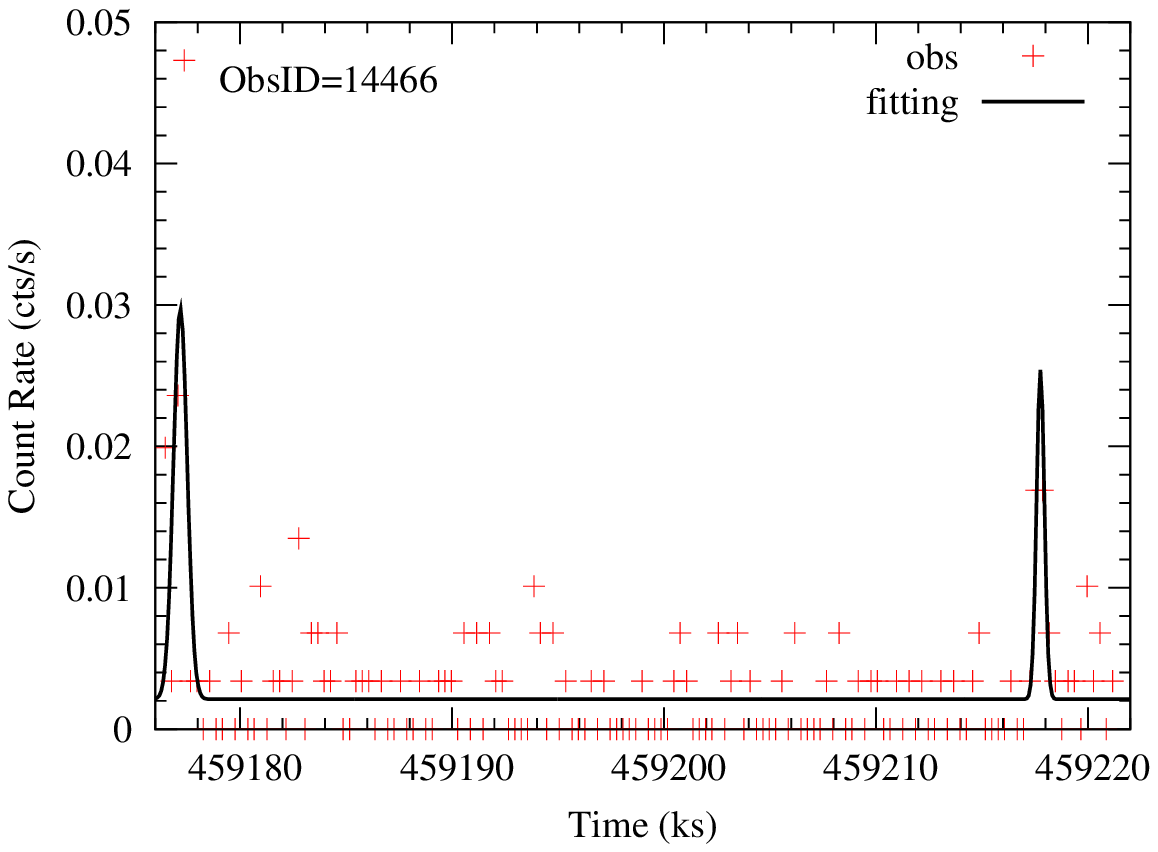}
\includegraphics[width=0.34\columnwidth]{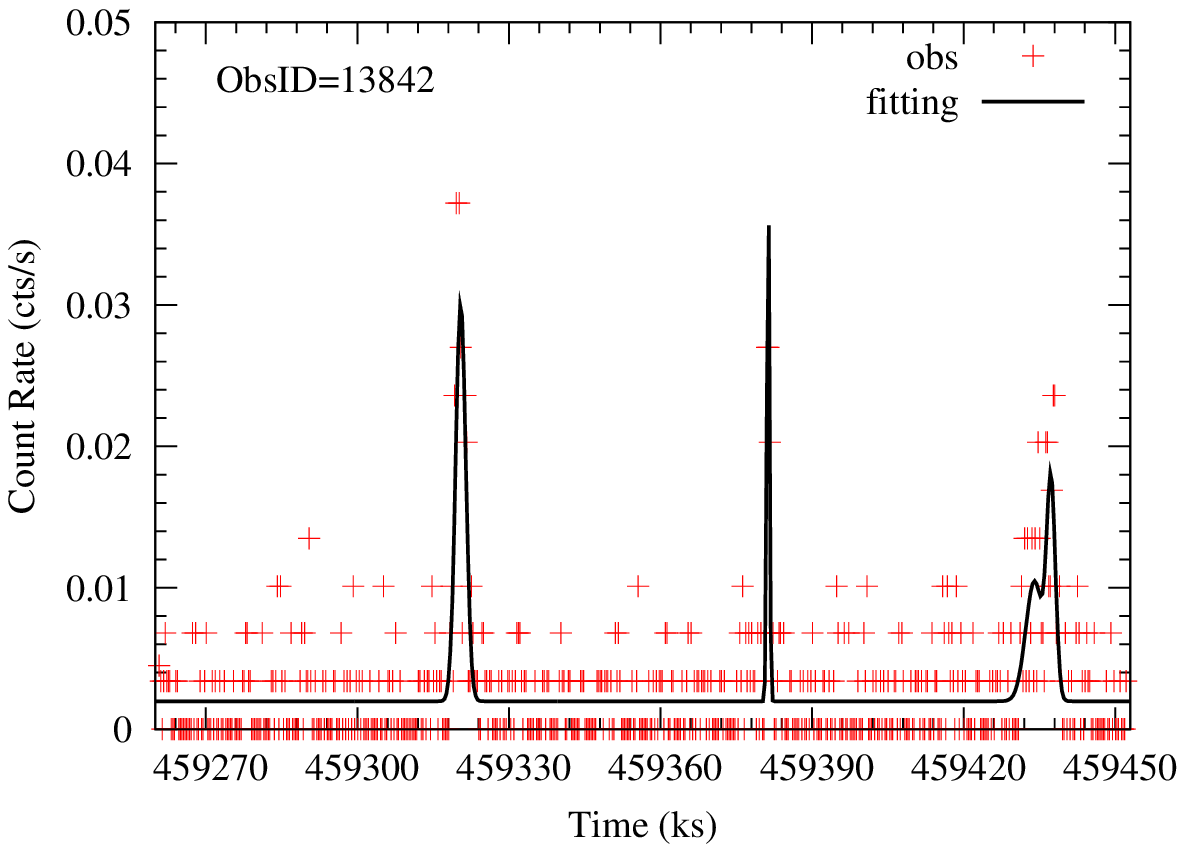}
\includegraphics[width=0.34\columnwidth]{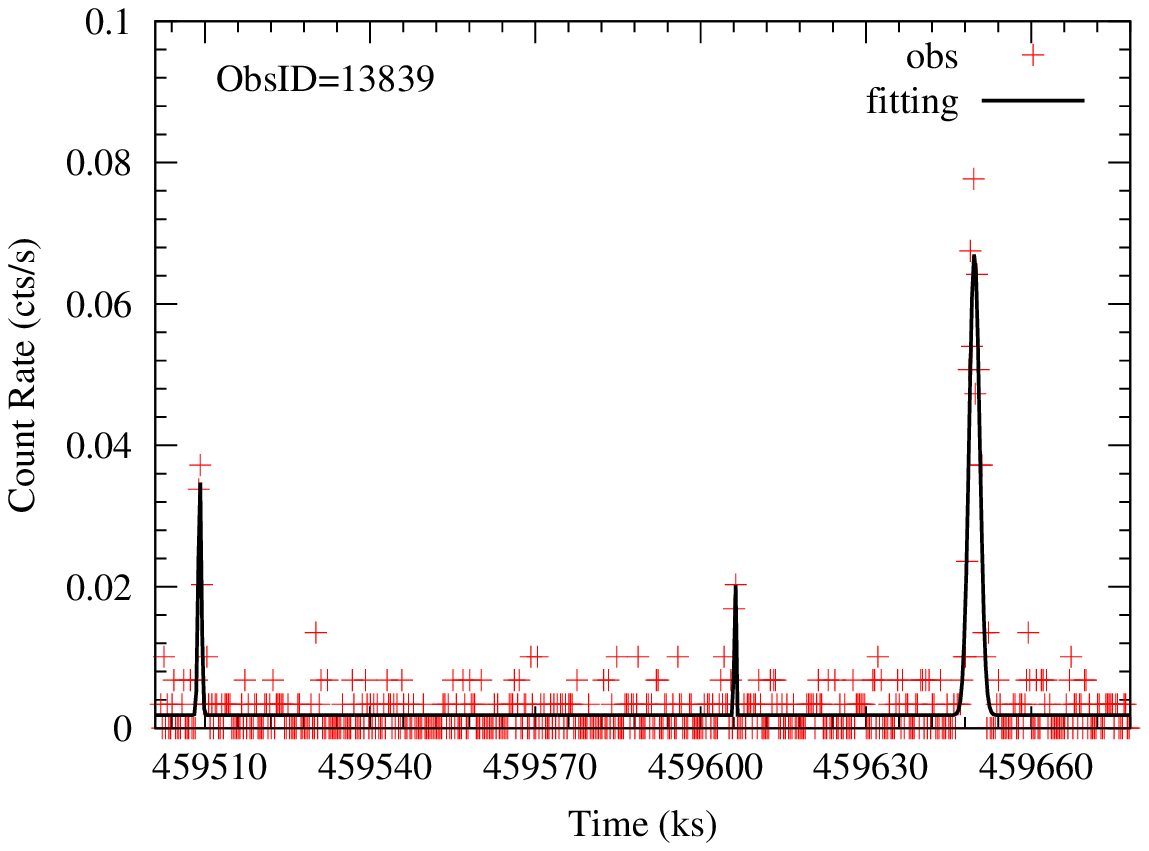}
\includegraphics[width=0.34\columnwidth]{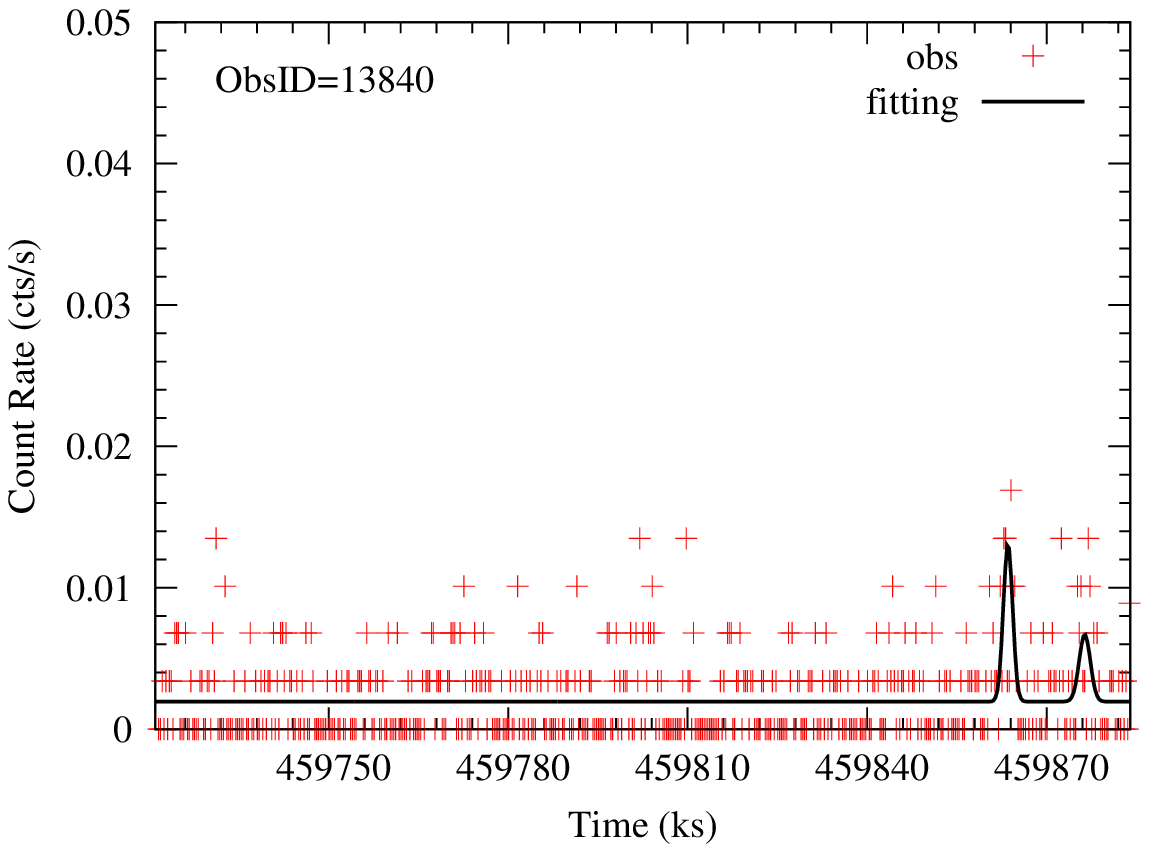}
\includegraphics[width=0.34\columnwidth]{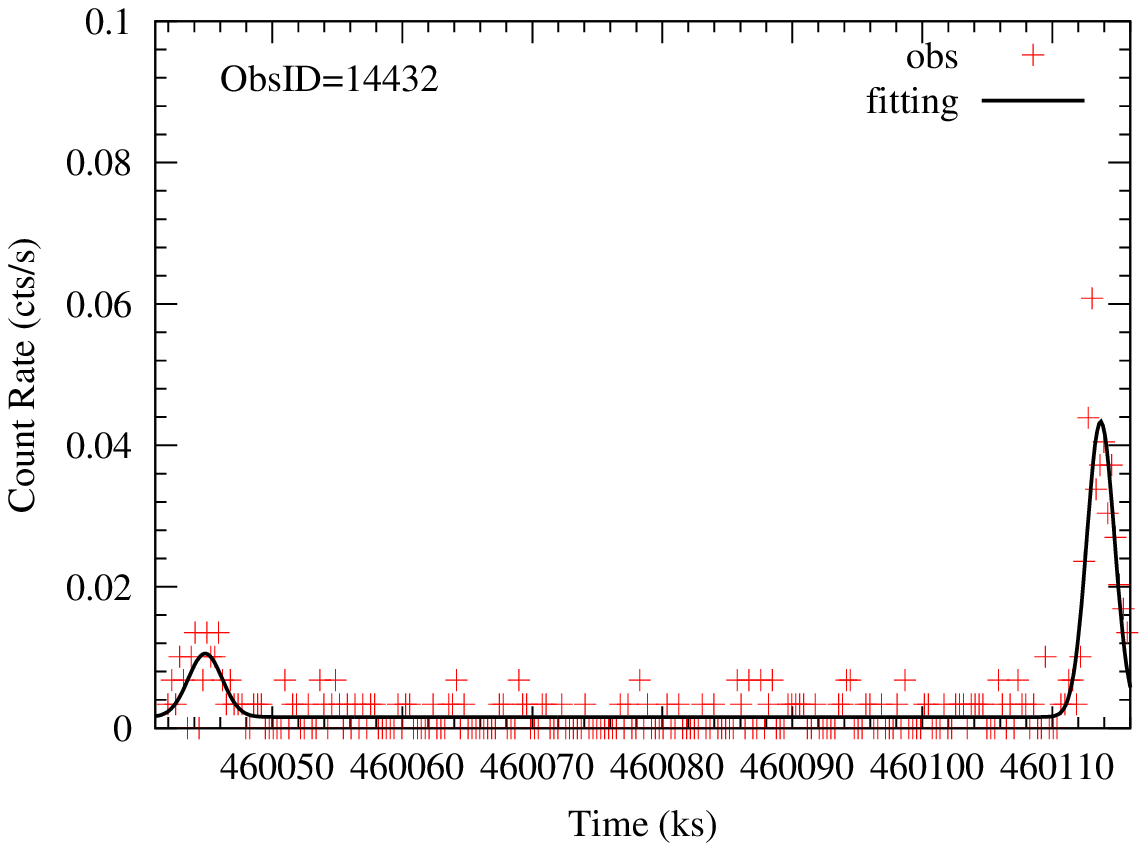}
\includegraphics[width=0.34\columnwidth]{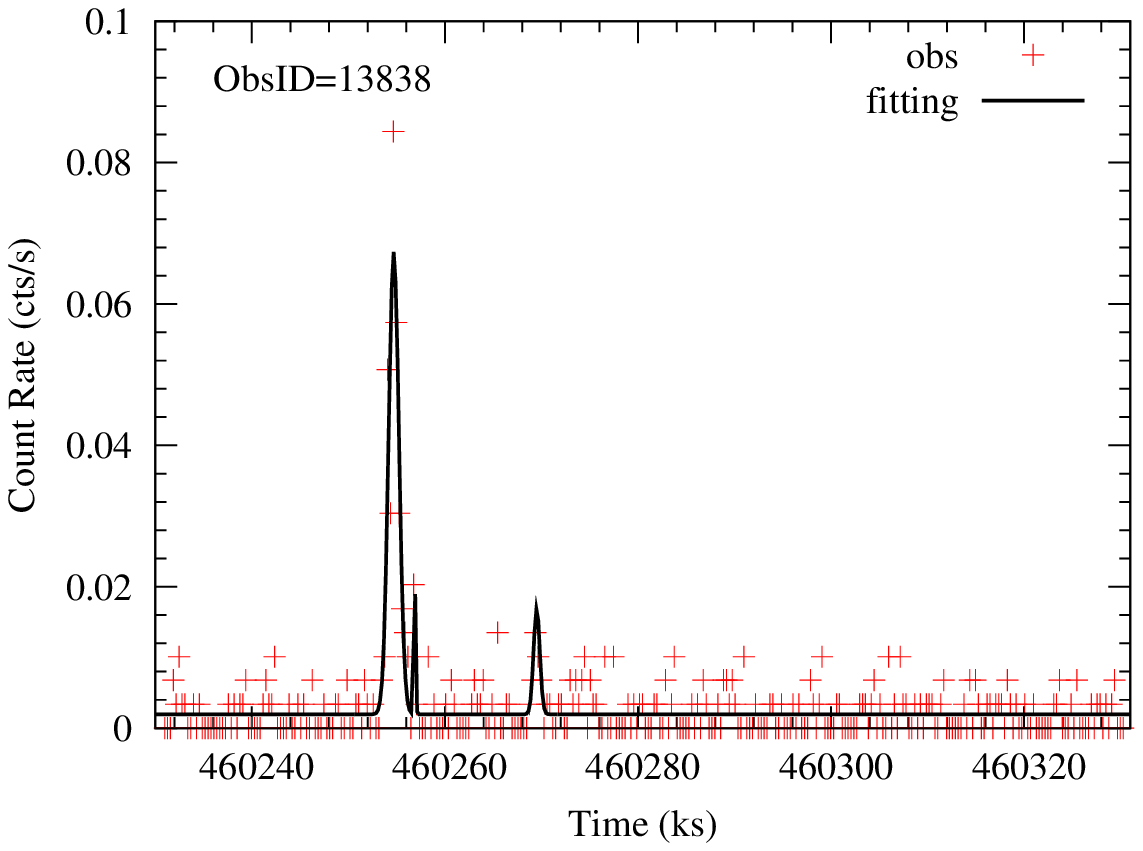}
\includegraphics[width=0.34\columnwidth]{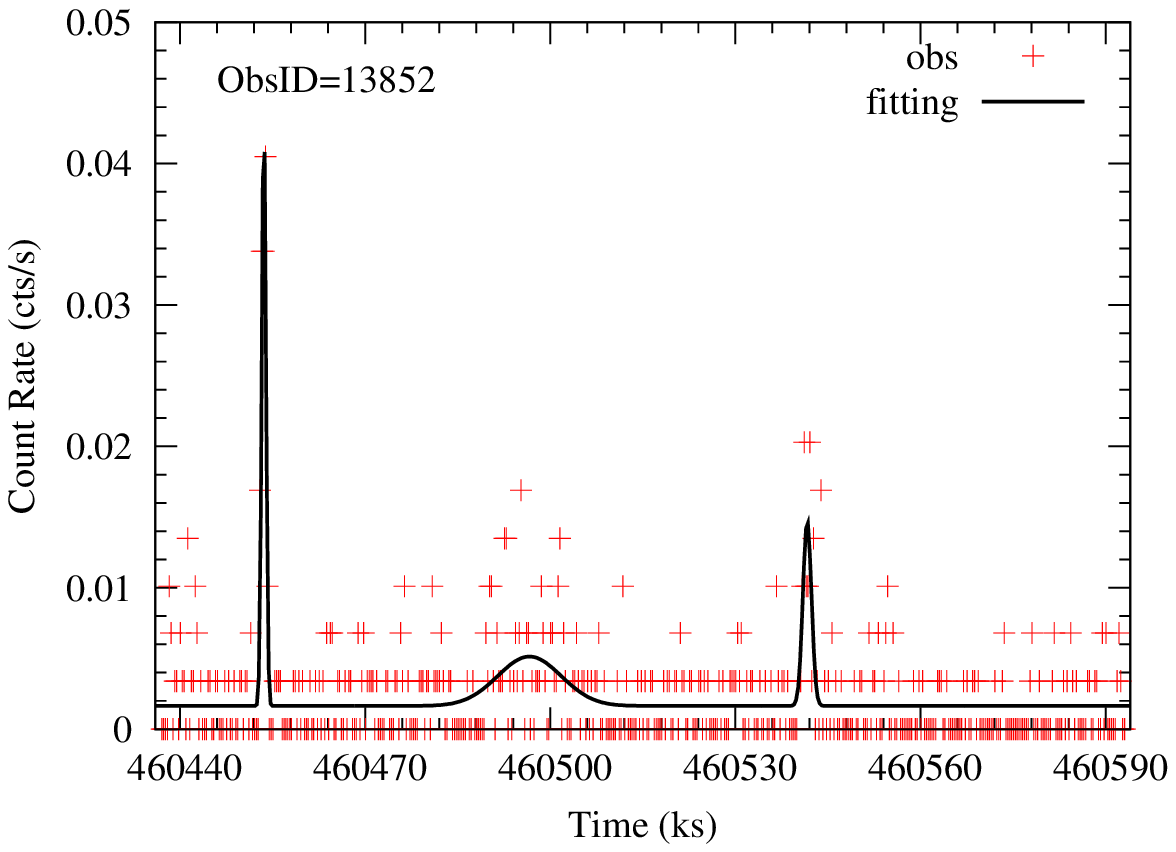}
\includegraphics[width=0.34\columnwidth]{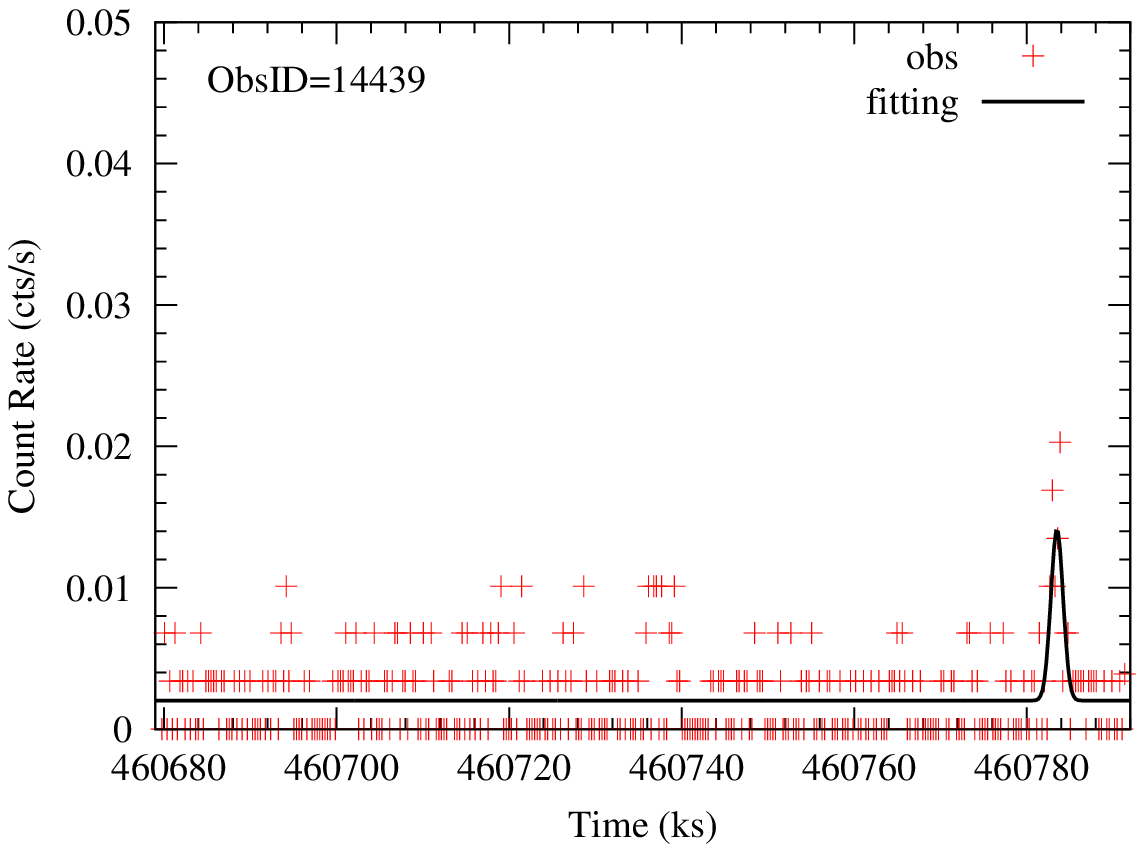}
\includegraphics[width=0.34\columnwidth]{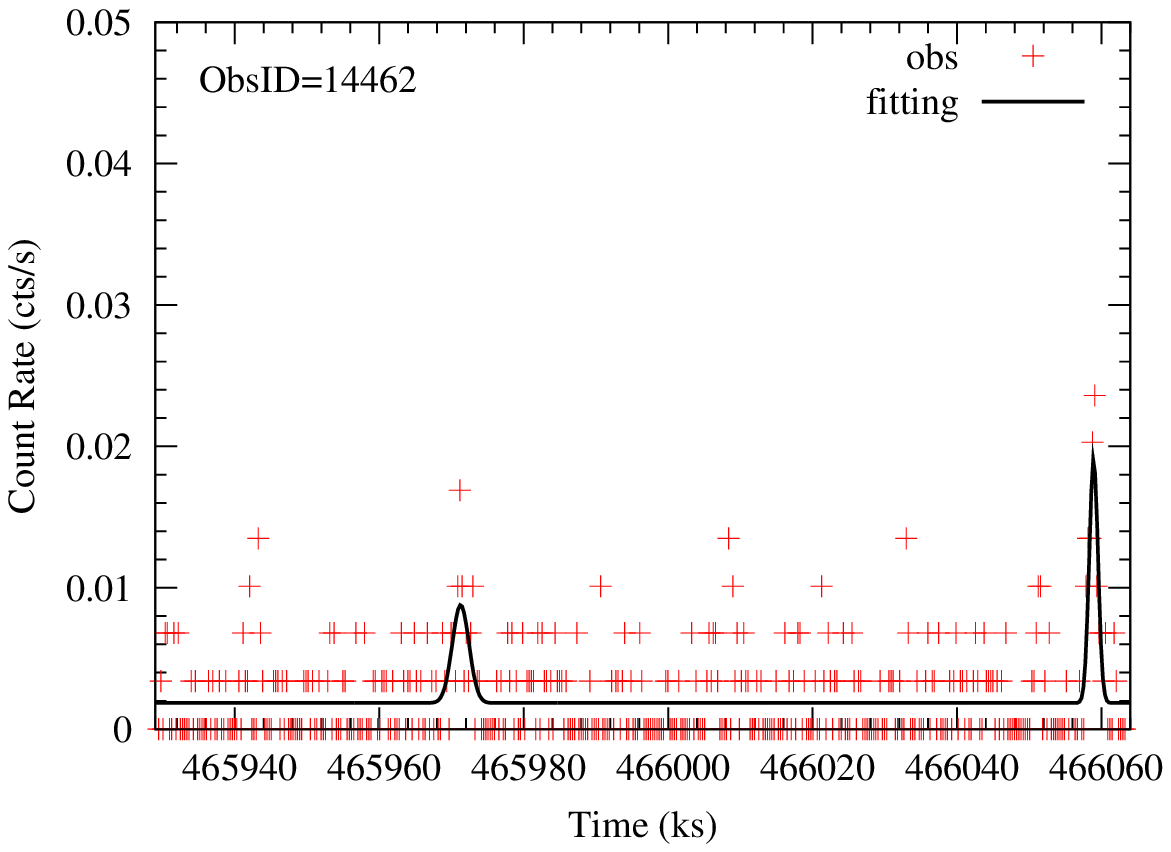}
\includegraphics[width=0.34\columnwidth]{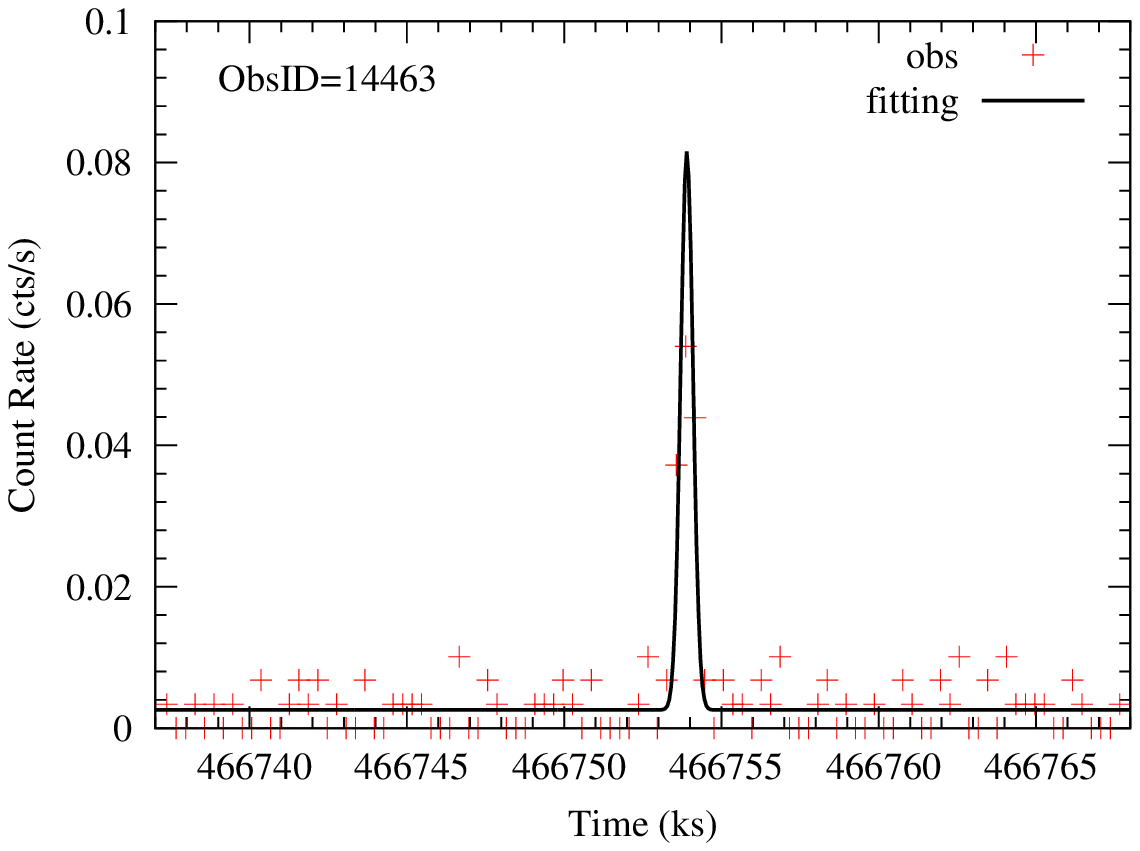}
\includegraphics[width=0.34\columnwidth]{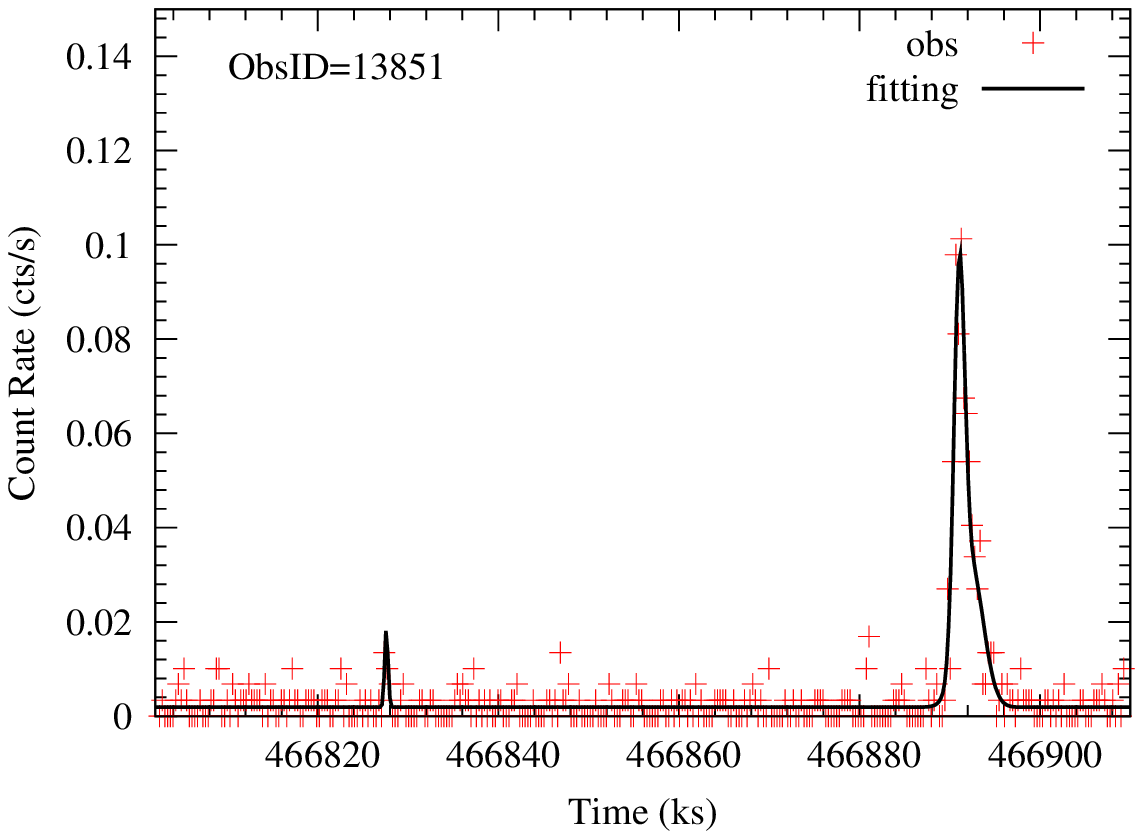}
\includegraphics[width=0.34\columnwidth]{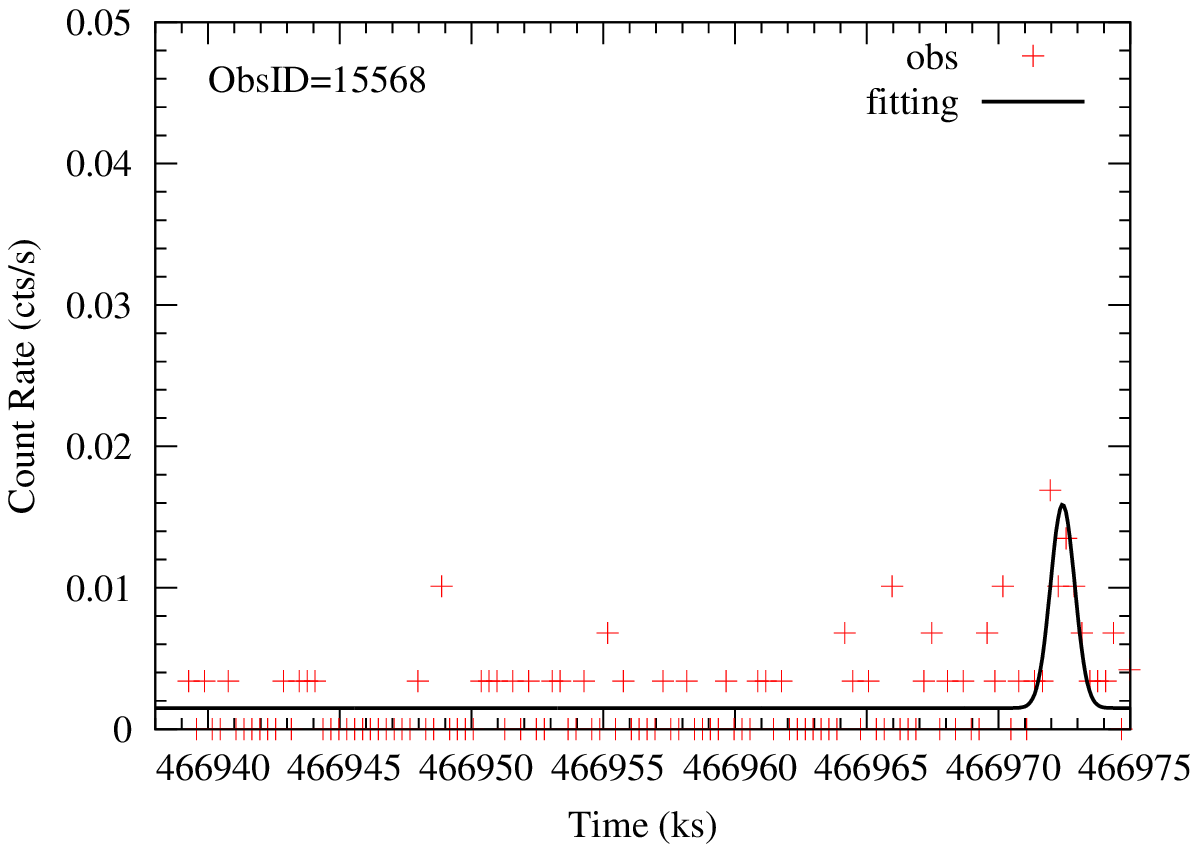}
\includegraphics[width=0.34\columnwidth]{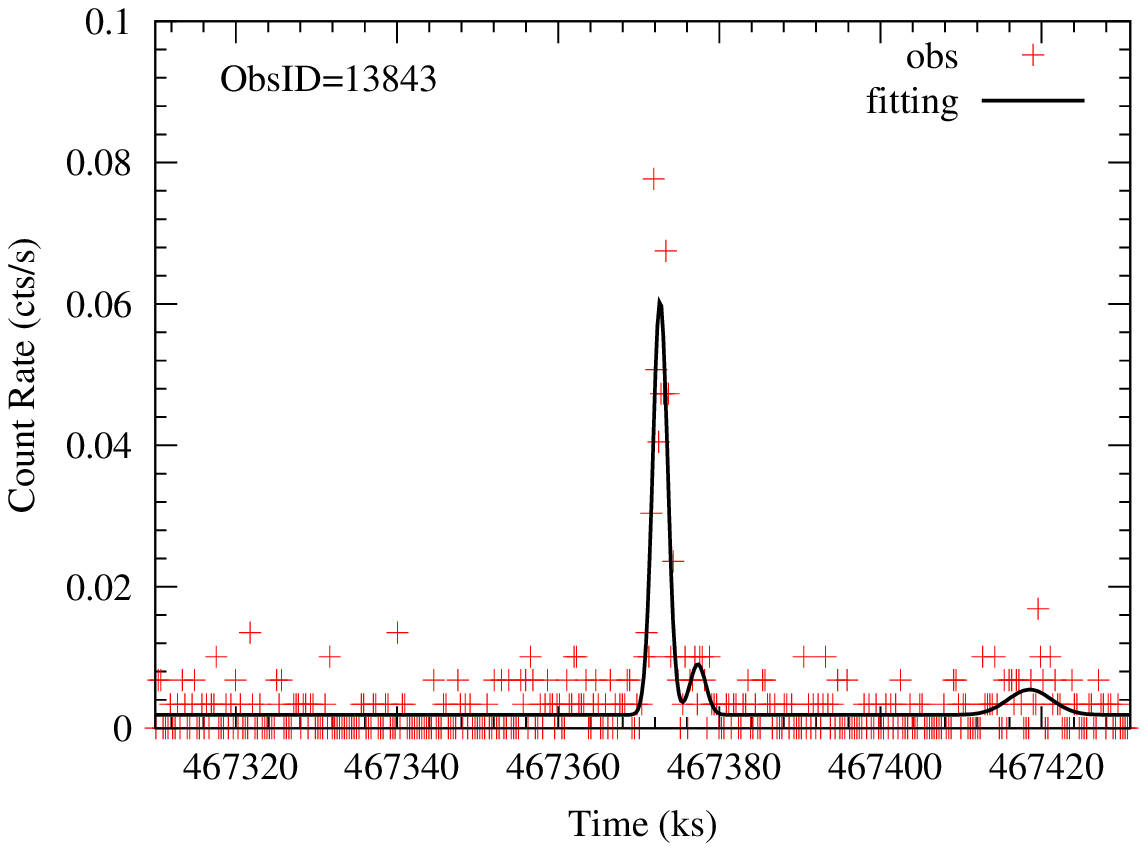}
\includegraphics[width=0.34\columnwidth]{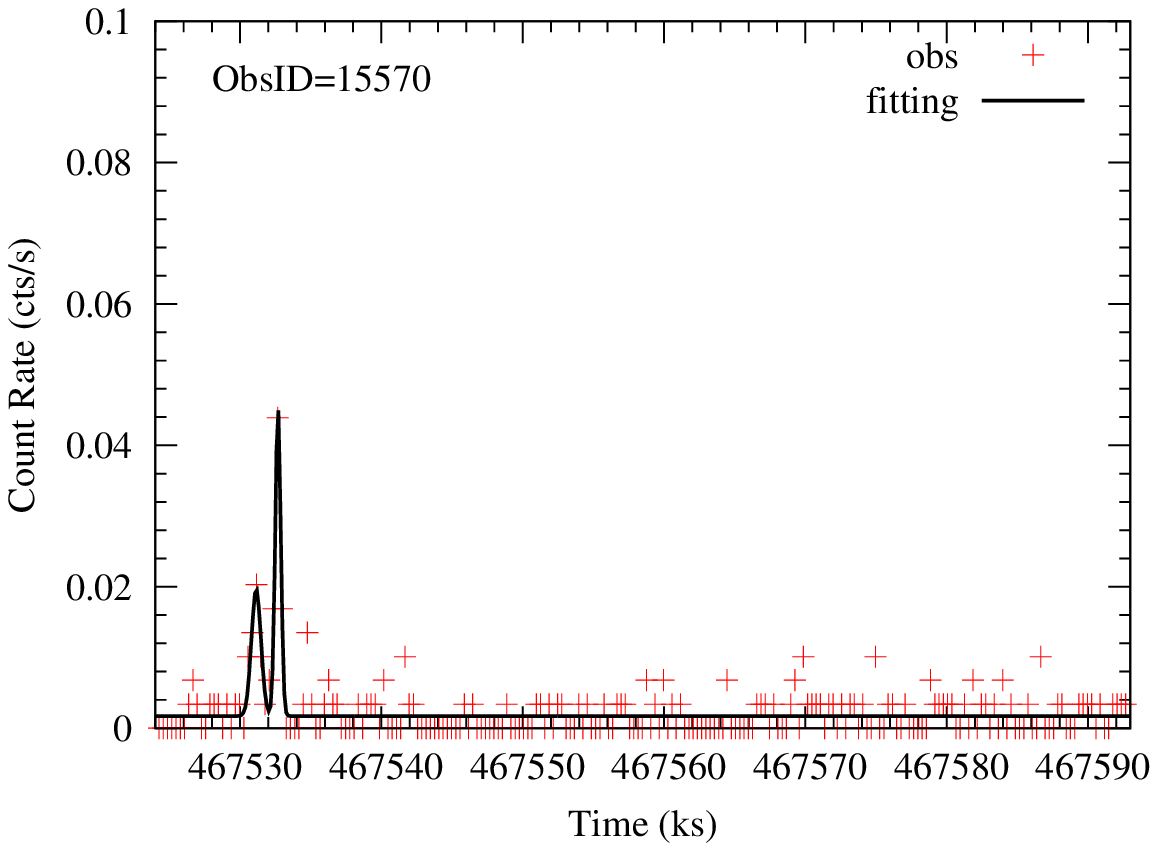}
\includegraphics[width=0.34\columnwidth]{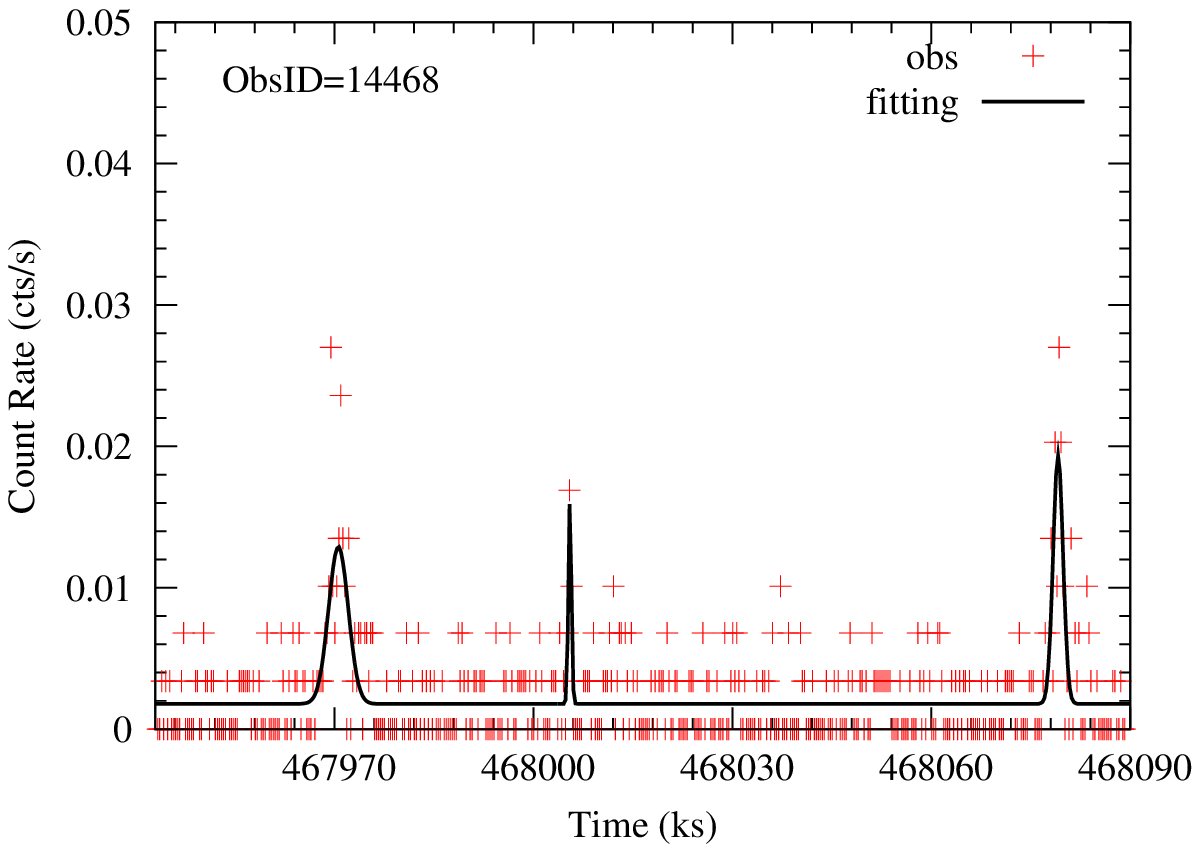}
\caption{Lightcurves of the detected flares in the {\it Chandra} ACIS-S/HETG0
observations, compared with the best-fitting results.
}
\end{figure*}

\end{appendix}

\end{document}